\documentclass{IEEEtran}

\usepackage{graphics,amssymb,array,cite,bm}
\usepackage{csquotes}
\usepackage[active]{srcltx}
\usepackage[english]{babel}
\usepackage[dvips]{graphicx}

\usepackage{hyperref}
\usepackage{amsfonts}
\usepackage{bbm}
\usepackage{mathrsfs}
\usepackage{xspace}
\usepackage{bm}
\usepackage{glossaries}

\usepackage{color}

\usepackage{algpseudocode,algorithm,algorithmicx}
\usepackage{enumitem}
\usepackage[latin1]{inputenc}

\makeglossaries

\newacronym{mac}{MAC}{multiple-access channel}
\newacronym{bc}{BC}{broadcast channel}
\newacronym{mimo}{MIMO}{multiple-input multiple-output}
\newacronym{siso}{SISO}{single-input single-output}
\newacronym{sc}{SC}{single-carrier}
\newacronym{mc}{MC}{multi-carrier}
\newacronym{ofdma}{OFDMA}{orthogonal frequency division multiple access}
\newacronym{af}{AF}{amplify-and-forward}
\newacronym{df}{DF}{decode-and-forward}
\newacronym{cf}{CF}{compress-and-forward}
\newacronym{mwrc}{MWRC}{multi-way relay channel}
\newacronym{pde}{PDE}{partial data exchange}
\newacronym{fde}{FDE}{full data exchange}
\newacronym{iid}{i.i.d.\@}{independent and identically distributed}
\newacronym{awgn}{AWGN}{additive white Gaussian noise}
\newacronym{awg}{AWG}{additive white Gaussian}
\newacronym{sic}{SIC}{successive interference cancellation}
\newacronym{snr}{SNR}{signal-to-noise ratio}
\newacronym{sinr}{SINR}{signal to interference plus noise ratio}
\newacronym{ber}{BER}{bit error rate}
\newacronym{zf}{ZF}{zero-forcing}
\newacronym{mmse}{MMSE}{minimum mean square error}
\newacronym{mse}{MMSE}{mean square error}
\newacronym{sud}{SUD}{single user decoding}
\newacronym{dof}{DoF}{degrees of freedom}
\newacronym{gdof}{GDoF}{generalized degrees of freedom}
\newacronym{nnc}{NNC}{noisy network coding}
\newacronym{dmn}{DMN}{discrete memoryless network}
\newacronym{csi}{CSI}{channel state information}
\newacronym{ee}{EE}{energy efficiency}
\newacronym{ian}{IAN}{treating interference as noise}
\newacronym{snd}{SND}{simultaneous non-unique decoding}
\newacronym{brd}{BRD}{best response dynamics}
\newacronym{br}{BR}{best response}
\newacronym{ne}{NE}{Nash equilibrium}
\newacronym{gne}{GNE}{generalized Nash equilibrium}
\newacronym{lhs}{LHS}{left-hand side}
\newacronym{rhs}{RHS}{right-hand side}
\newacronym{gee}{GEE}{global energy efficiency}
\newacronym{wsee}{WSEE}{weighted sum energy efficiency}
\newacronym{wpee}{WPEE}{weighted product energy efficiency}
\newacronym{wmee}{WMEE}{weighted minimum energy efficiency}
\newacronym{kkt}{KKT}{Karush Kuhn Tucker}
\newacronym{pc}{PC}{pseudo-concave}
\newacronym{qc}{QC}{quasi-concave}
\newacronym{ql}{QL}{quasi-linear}
\newacronym{pl}{PL}{pseudo-linear}
\newacronym{spc}{SPC}{strictly pseudo-concave}
\newacronym{sqc}{SQC}{strictly quasi-concave}
\newacronym{lfp}{LFP}{linear fractional problem}
\newacronym{clfp}{CLFP}{concave-linear fractional problem}
\newacronym{ccfp}{CCFP}{concave-convex fractional problem}
\newacronym{mmfp}{MMFP}{max-min fractional problem}
\newacronym{sorp}{SoRP}{sum-of-ratios problem}
\newacronym{porp}{PoRP}{product-of-ratios problem}
\newacronym{srp}{SRP}{single-ratio problem}
\newacronym{brb}{BRB}{branch-reduce-and-bound}
\newacronym{qos}{QoS}{quality-of-service}
\newacronym{comp}{CoMP}{cooperative multi-point}
\newacronym{ue}{UE}{user equipment}
\newacronym{bs}{BS}{base station}
\newacronym{mrc}{MRC}{maximum ratio combining}
\newacronym{d2d}{D2D}{device-to-device}
\newacronym{lmmse}{LMMSE}{linear minimum mean square error}
\newacronym{svd}{SVD}{singular value decomposition}
\newacronym{evd}{EVD}{eigen value decomposition}
\newacronym{ict}{ICT}{information communication technologies}
\newacronym{cagr}{CAGR}{Compound Annual Growth Rate}
\newacronym{urllc}{URLLC}{Ultra Reliable Low Latency Communications}
\newacronym{mbb}{mBB}{mobile broadband}
\newacronym{miot}{mIoT}{massive Internet of Things}
\newacronym{v2x}{V2X}{Vehicular-to-X}
\newacronym{ann}{ANN}{Artificial Neural Network}
\newacronym{dnn}{DNN}{Deep Neural Network}
\newacronym{fnn}{FNN}{\textbf{Feed-forward Neural Networks}}
\newacronym{rnn}{RNN}{\textbf{Recurrent Neural Networks}}
\newacronym{cnn}{CNN}{Convolutional Neural Network}
\newacronym{esn}{ESN}{Echo-State Network}
\newacronym{ai}{AI}{Artificial Intelligence}
\newacronym{gpu}{GPU}{Graphics Processing Unit}
\newacronym{etsi}{ETSI}{European Telecommunications Standards Institute}
\newacronym{nfv}{NFV}{Network Function Virtualization}
\newacronym{sdn}{SDN}{Software Defined Networking}
\newacronym{son}{SON}{Self-Organzing Networks}
\newacronym{ngmn}{NGMN}{Next Generation Mobile Networks}
\newacronym{ran}{RAN}{Radio Access Network}
\newacronym{relu}{ReLU}{Rectified Linear Unit}
\newacronym{sgd}{SGD}{Stochastic Gradient Descent}
\newacronym{mdp}{MDP}{Markov Decision Process}
\newacronym{itu}{ITU}{International Telecommunication Union}
\newacronym{uav}{UAV}{Unmanned Aerial Vehicles}

\newcommand{\eeq}{\end{equation}}

\newcommand{\br}{\mbox{\boldmath $r$}}

\newcommand{\bs}{\mbox{\boldmath $s$}}

\newcommand{\bb}{\mbox{\boldmath $b$}}

\newcommand{\ba}{\mbox{\boldmath $a$}}

\newcommand{\bp}{\mathbf{p}}

\newcommand{\bX}{\mbox{\boldmath $X$}}
\newcommand{\bh}{\mbox{\boldmath $h$}}

\newcommand{\bc}{\mbox{\boldmath $c$}}

\newcommand{\bgamma}{\mbox{\boldmath $\gamma$}}

\newcommand{\bV}{\mbox{\boldmath $V$}}

\newcommand{\bW}{\mbox{\boldmath $W$}}

\newcommand{\bpsi}{{\bf \Psi}}

\newcommand{\bmu}{\mbox{\boldmath $\mu$}}
\newcommand{\bsigma}{\mbox{\boldmath $\sigma$}}

\newcommand{\bx}{\mathbf{x}}
\newcommand{\bz}{\mbox{\boldmath $z$}}

\newcommand{\bv}{\mbox{\boldmath $v$}}
\newcommand{\bd}{\mbox{\boldmath $d$}}

\newcommand{\by}{\mbox{\boldmath $y$}}
\newcommand{\ds}{\displaystyle}
\newcommand{\bw}{\mbox{\boldmath $w$}}
\newcommand{\bY}{\mbox{\boldmath $Y$}}

\newcommand{\bbeta}{\mbox{\boldmath{$\beta$}}}

\newcommand{\bzero}{\mbox{\boldmath{$0$}}}

\newcommand{\beq}{\begin{equation}}

\newcommand{\thetab}      {{\bm \theta}}

\newcommand{\rhob}        {{\bm \rho}}

\newcommand{\phib}        {{\bm \phi}}

\newtheorem{remark}{Remark}

\begin{document}

\title{Wireless Networks Design in the Era of Deep Learning: Model-Based, AI-Based, or Both?}
\author{{Alessio Zappone, {\em Senior Member, IEEE}, Marco Di Renzo, {\em Senior Member, IEEE}, M\'erouane Debbah, {\em Fellow, IEEE}\\ (Invited Paper)}

\thanks{A. Zappone and M. Debbah are with the Large Networks and Systems Group, CentraleSupelec, Université Paris-Saclay, 3 rue Joliot-Curie, 91192 Gif-sur-Yvette, France,(alessio.zappone@l2s.centralesupelec.fr, merouane.debbah@l2s.centralesupelec.fr). M. Debbah is also with the Mathematical and Algorithmic Sciences Lab, Huawei France R\&D, Paris, France (merouane.debbah@huawei.com). The work of A. Zappone and M. Debbah has been supported by the H2020 MSCA IF BESMART, Grant 749336.

M. Di Renzo is with the Laboratory of Signals and Systems (CNRS - CentraleSupelec - Univ. Paris-Sud), Universit\'e Paris-Saclay, 3 rue Joliot-Curie, 91192 Gif-sur-Yvette, France, (marco.direnzo@l2s.centralesupelec.fr). 
}}
\maketitle

\begin{abstract}
This work deals with the use of emerging deep learning techniques in future wireless communication networks. It will be shown that data-driven approaches should not replace, but rather complement  traditional design techniques based on mathematical models. 

Extensive motivation is given for why deep learning based on artificial neural networks will be an indispensable tool for the design and operation of future wireless communication networks, and our vision of how artificial neural networks should be integrated into the architecture of future wireless communication networks is presented.  

A thorough description of deep learning methodologies is provided, starting with the general machine learning paradigm, followed by a more in-depth discussion about deep learning and artificial neural networks, covering the most widely-used artificial neural network architectures and their training methods. Deep learning will also be connected to other major learning frameworks such as reinforcement learning and transfer learning.

A thorough survey of the literature on deep learning for wireless communication networks is provided, followed by a detailed description of several novel case-studies wherein the use of deep learning proves extremely useful for network design. For each case-study, it will be shown how the use of (even approximate) mathematical models can significantly reduce the amount of live data that needs to be acquired/measured to implement data-driven approaches. 

Finally, concluding remarks describe those that in our opinion are the major directions for future research in this field.
\end{abstract}

\section{Introduction and Vision}
Our society is undergoing a digitization revolution, with a dramatic increase of both Internet users and connected devices. The fifth generation of wireless communication networks will be rolled out shortly, featuring innovative technologies such as infrastructure densification, antenna densification, use of frequency bands in the mmWave range, energy-efficient network management \cite{Buzzi5G,GEJSAC16,ZapNow15}, which promise to achieve the targets of 1000x higher data-rates and 2000x higher bit-per-Joule energy efficiency compared to the previous wireless generation \cite{5GNGMN}. However, as the 5G standardization phase is ongoing, it appears doubtful that a single 5G technology will be able to achieve the desired requirements. Indeed, it is widely believed that 5G will employ multiple technologies at the same time. This points towards extremely complex systems,  characterized by an infrastructure that becomes denser and denser to accommodate the exponentially increasing number of devices demanding connections. As a consequence, operational expenditures (OPEX) and capital expenditures (CAPEX), which are already a major challenge in present wireless networks \cite{Aliu2013}, will significantly increase.

Moreover, global IP traffic will continue increasing in the next years. Between 2020 and 2030, the \gls{cagr} will rise by 55\% annually, reaching 607 exabytes in 2025 and 5,016 exabytes  in 2030 \cite{ITUReport}. In addition, another critical challenge for future wireless networks is the extreme heterogeneity of the services to provide. Future wireless networks will have to support many innovative vertical services, each with its own specific requirements \cite{Euro5G}, e.g. 
\begin{itemize}
\item End-to-end latency of $1\,\textrm{ms}$ and reliability higher than $99.999\%$ for \gls{urllc}.
\item Terminal densities of $1$ million of terminals per square kilometer for \gls{miot} applications. 
\item Per-user data-rate larger than $50\;\textrm{Mb/s}$ for \gls{mbb} applications. 
\item Terminal location accuracy of the order of $0.1\,\textrm{m}$ for \gls{v2x} communications. 
\end{itemize}
These numbers are beyond what 5G networks have been designed to handle, and the integration of such diverse vertical services into the same network architecture calls for an extremely flexible and adaptive architecture, which clashes against today's \enquote{one-size-fits-all} paradigm. Therefore, new approaches to increase the network flexibility have recently started attracting research attention, such as software networks and the use of \glspl{uav}.

Software networks are primarily based on the network slicing paradigm, which proposes to logically separate the control and data plane, thus effectively slicing the physical network into multiple virtual networks co-existing over a common shared physical infrastructure. Each network slice constitutes a logically separate virtual network that can be customized to meet the specific requirements of a specific vertical service, by using techniques like \gls{sdn} \cite{KreutzSoftwareNets} and \gls{nfv} \cite{Abdelwahab16}. Network slicing applies to both the core and access network segments and paves the way for a new generation of programmable and software-oriented wireless networks, that are able to support flexible and on-demand network resources provisioning, allowing service providers to tailor the use of resources to the specific needs of the different classes of services to be provided. 

Besides increasing the flexibility of the network through network slicing and reprogrammability, the use of \glspl{uav} is meant to increase the flexibility of the physical network infrastructure. \glspl{uav} like drones and other flying objects will act as flying access points, that can be redeployed based on heterogeneous traffic conditions to support on-demand connectivity requests \cite{YanikomerogluWCL2017}. 

Thus, \textbf{future wireless networks will be characterized by an unprecedented level of complexity, which makes traditional approaches to network deployment, design, and operation no longer adequate}. Every aspect of past and present wireless communication networks is regulated by \emph{mathematical models}, that are either derived from theoretical considerations, or from field measurement campaigns. Mathematical models are used for initial network planning and deployment, for network resource management, as well as for network maintenance and control. However, any model is always characterized by an inherent trade-off between their accuracy and their tractability. Very complex scenarios like those of future wireless networks are unlikely to admit a mathematical description that is at the same time accurate and tractable. In other words, we are rapidly reaching the point at which the quality and heterogeneity of the services we demand of communication systems will exceed the capabilities and applicability of present modeling and design approaches. 

In order to face this \textbf{complexity crunch} challenge, for the first time since the inception of wireless communications, it is not enough to simply devise a more performing transmission technology. Being simply able to transmit data at a faster rate does not ensure the flexibility required to accommodate diverse classes of users with extremely heterogeneous service requirements. Besides developing faster transmission technologies, future research efforts should be aimed also at improving the network infrastructure itself, making it intelligent enough to flexibly and automatically adapt to sudden wireless scenario changes and rapid traffic evolutions. In order to provide end-users with a perceived seamless and limitless connectivity, the re-configuration of network resources and/or the re-deployment of network nodes in response to new data demands, as well as to connectivity problems and/or failures of hardware components, must be prompt and timely. To this end, it is necessary to make the network fully self-organizing, automating all management, operation, and maintenance tasks, limiting direct human intervention as much as possible. This is the concept of \gls{son}, which is not new to wireless networks, as it was introduced by the \gls{ngmn} alliance, and even standardized by 3GPP for LTE networks. However, despite having garnered much attention since its inception, \gls{son} failed to achieve the expected end-goal of fully automated networks. It was employed primarily for specific \gls{ran}  applications, but without providing a true end-to-end solution. In our opinion, this is mainly due to the lack of intelligence and cognition in past and present networks. In order to enable truly self-organizing networks, it is essential to have an infrastructure capable of cognitive behavior. Intelligence must be spread across all network segments, making network nodes self-aware, self-organizing, and self-healing,  by sensing the surrounding environment and processing the acquired data. These requirements have recently given rise to the concept of smart radio environments, which is discussed in detail in \cite{EURASIP_RIS}.
It is estimated that a fully  automated and self-aware network, with self-configuration and self-healing capabilities would reduce CAPEX and OPEX by a factor 5 relative to 2010 levels \cite{HuaweiSONWhite}, i.e. relative to a period when the complexity and expected performance of wireless networks were quite lower than today. Therefore, the gain compared to the extremely more complex networks of the future is expected to be significant. 

\subsection{AI-Based Wireless Networks}
The need for an intelligent wireless network motivates to endow each network segment with  \textbf{\gls{ai}} capabilities and to employ a \textbf{data-driven} paradigm in which network nodes are able to determine the best policy to employ based on the experience obtained by processing previous  data. On the one hand, this clearly reduces the reliance on mathematical models as far as network design and operation is concerned, but, on the other hand, it does not necessarily imply that traditional mathematical-oriented models and approaches should be dismissed. In fact, it is our opinion that there is much to be gained by the joint use of model-based and \gls{ai}-based techniques and we envision future wireless networks where model-based and \gls{ai}-based techniques are used in synergy. A major goal of this work is to support this point, and indeed Section \ref{Sec:Applications} will present specific approaches for cross-fertilization between these two seemingly contrasting approaches, together with the related quantitative analysis.

But how to develop artificially intelligent wireless networks? A framework that enables this is \emph{machine learning}, in particular through one of its techniques, namely \textbf{deep learning}. Machine learning provides several techniques that endow computers with the ability to learn from data,  instead of being explicitly programmed \cite{BishopBook2017}. Machine learning techniques are not new to communication systems, and indeed several machine learning approaches have been developed and proposed to aid the design and operation of communication systems, e.g. support vector machines, decision-tree learning, Bayesian networks, genetic algorithms, rule-based learning, and inductive logical programming, among others. Detailed surveys and tutorials about machine learning and its applications to wireless networks can be found in \cite{SimeoneNowPub,Simeone2018,Shalev-ShwartzNowPub,Bkassiny2013,TembineBook2011}, and its use to enable \gls{son} networks has been proposed in \cite{Moysen2018}. However, deep learning \cite{Bengio2016,BengioNowPub,DengNowPub}, which is the most popular machine learning technique in many fields of science, has started attracting the attention of the communication community only very recently. 

Deep learning is a particular machine learning technique that implements the learning process elaborating the data through \glspl{ann}. As it will be explained in more detail in Section \ref{Sec:MLandDL}, the use of \glspl{ann} is the key factor that makes deep learning more performing than other machine learning schemes, especially when a large amount of data is available. This has made deep learning the first among the top ten \gls{ai} technology trends of 2018 \cite{TopTenAI}, and the leading machine learning technique in many scientific fields such as image classification, text recognition, speech recognition, audio and language processing, robotics. Despite all this, as already said, its use in communication systems has been envisioned only very recently \cite{Chen2017}, and its potential is at the moment almost untapped. In our opinion, this is mainly due to the fact that, unlike other fields of science, communication engineers could traditionally rely on mathematical models for system design, thereby making the use of data-driven approaches not strictly necessary. However, as we have described, this fundamental postulate is going to be weakened in the near future, which puts forth the need for deep learning in communication systems. Moreover, recent technological advancements make deep learning a viable technology for application to future communication networks. More precisely:
\begin{itemize}
\item In order to gain the most out of deep learning algorithms, it is necessary to process large datasets. At present, exactly the exponential increase of wireless devices results in a corresponding growth of traffic data \cite{Imran2014,Bi2015,Cheng2017}. 
\item Modern advancements in computing capacity makes it possible to execute larger and more complex algorithms much faster. In particular, \glspl{gpu} can be repurposed to execute deep learning algorithms at speeds many times faster than traditional processor chips. 
\end{itemize}
Recently, several leading telecommunication companies have supported the use of deep learning for communications \cite{HuaweiAI,QualcommAI}. Moreover, initial steps towards the standardization of intelligent wireless communication systems have already been taken. \gls{etsi} activated an Industry Specification Group named \emph{Experiential Network Intelligence}, with the purpose to define a \emph{cognitive network management} architecture capable of using \gls{ai} techniques and context-aware policies to adjust the services that are offered, based on changes in user needs, environmental conditions, and business goals. Such a paradigm is referred to as the \emph{observe-orient-decide-act}  control paradigm and represents the first standardization step towards the definition of an \emph{experiential} system, i.e. a system that learns from previous experience to improve its knowledge of how to act in the future. This is anticipated to help operators automate their network configuration and monitoring processes, thereby reducing their operational expenditure and improving the use and maintenance of their networks. Similarly, a standardization initiative for machine learning in future mobile networks has been activated by the \gls{itu}, with the aim of specifying an architectural framework for machine learning in future networks, defining the integration of machine learning functionalities into the architecture of future mobile networks, as well as identifying techniques for network management in future wireless environments. More specifically, the recently approved \enquote{ITU-T Y.3172 architectural framework for machine learning in future networks including IMT-2020} \cite{ITU_ML}, constitutes another important component for the adoption of machine learning to operate and optimize wireless networks.

On the other hand, in order to make the vision of \gls{ai}-based wireless networks true, there are also some challenges that must be overcome. In particular, two challenges appear today as the most relevant ones:
\begin{itemize}
\item \textbf{Data acquisition.} As already mentioned, in order to get the most out of deep learning algorithms, a large amount of data is required. As stated above, this is nowadays possible since the increase of traffic provides a huge amount of data that can be collected and exploited. However, the question remains of how to acquire the necessary amount of data in a practical and cost-effective way, e.g., by taking into account the overhead, time, and energy costs, especially in scenarios with high mobility and fast varying network conditions. In our opinion, the first half of the solution lies in the pervasive use of new, \emph{intelligent}, materials, known as \textbf{meta-materials}, which have communication as well as data storage and processing abilities. As detailed in Section \ref{Sec:AIforDeployment}, meta-materials can provide the fabric for \gls{ai}-enabled wireless networks. 
As for the second half of the solution, in our opinion it lies in the cross-fertilization between \gls{ai}-based and model-based techniques, which, as detailed in Section \ref{Sec:Applications}, can significantly reduce the amount of data that needs to be physically acquired through field measurement campaigns. 
\item \textbf{\gls{ai} integration into communication networks.} While it appears clear that future communication networks will have to rely on \gls{ai}, it is not clear how \glspl{ann} should be integrated into the architecture of communication networks. Should the acquired data be stored at a centralized location, where a single \gls{ann} manages a large network domain, or should each network device store its own data and run a local \gls{ann}? Our answer to this question is provided in Section \ref{Sec:DLforRM}, where it is argued that a decentralized paradigm is to be preferred, and two possible approaches are described.
\end{itemize}

Before concluding this section, we believe it is important to emphasize that machine learning is anticipated to be a game-changing technology not only for mainstream wireless communication networks, but also for emerging communication technologies that are being investigated as a way to complement traditional wireless approaches in specific scenarios. Among others, we mention \emph{optical wireless communications} \cite{PathakVLC,KarunatilakaLED}, which promise very high data rates by communicating over the visible spectrum, and \emph{molecular communications}, which are not based on electromagnetic waves but exploit chemical signals as information carriers, thus enabling communication through media where electromagnetic signals do not propagate well, such as water, inside human bodies or the walls of buildings \cite{Nakano12,Farsad16}. Both technologies have garnered much interest in recent years, but they share the main drawback of being difficult to be accurately described by tractable mathematical models. Therefore, model-less, AI-driven approaches can provide a decisive contribution to the practical implementation of wireless optical and molecular communication systems, as, for example, observed in  \cite{Hager2018}, which employs deep learning to solve Schr\"odinger equations in fiber-optic communications.

\subsection{Contributions and Organization}
The vast majority of survey contributions on machine learning focus on different fields than  communication networks, e.g. \cite{BengioNowPub,DengNowPub,Bengio2016,Shalev-ShwartzNowPub,BishopBook2017,LeCun2015,Schmidhuber2014}. As far as communications are concerned, most previous surveys discuss general machine learning techniques \cite{SimeoneNowPub,Bkassiny2013,Jiang_2017,Alsheikh14,Klaine2017,TembineBook2011}, without providing a dedicated analysis of deep learning. Only a few very recent overview works focus specifically on deep learning and \glspl{ann} for wireless communications \cite{Chen2017,Kasnesis17,Zhang2018}. All these three previous contributions envision the use of deep learning in future wireless networks, identifying \gls{ai} as the key technology of the future and identifying many use-cases and scenarios in which deep learning has the potential of simplifying the design and improving the performance. In addition, none of the above works provides at the same time an in-depth quantitative analysis of several applications of deep learning for the design of wireless networks, 
an extensive overview of wireless applications of deep learning, as well as a self-contained mathematical treatment of deep learning by \glspl{ann} that discusses the main types of \glspl{ann} and the related training algorithms. Moreover, none of the above works addresses possible approaches for cross-fertilization between deep learning techniques and traditional mathematical modeling design approaches. In this context, our work provides the following five major contributions (\textbf{C.1}-\textbf{C.5}): 
\begin{enumerate}
\item[\textbf{(C.1)}] The connection between model-based and data-driven methodologies is elaborated. A systematic framework to embed the prior knowledge contained in available mathematical models into deep learning techniques is described, and is shown to significantly reduce the amount of training data that is needed to achieve good communication performance. 
\item[\textbf{(C.2)}] A possible network architecture based on the use of the emerging technology of meta-materials is put forth. It is shown in particular that it facilitates the acquisition of the data required to train \glspl{ann}. Also, the issue of managing and operating an \gls{ai}-based communication networks based on meta-materials is discussed. 
\item[\textbf{(C.3)}] Several case-studies where deep learning is proved to be useful are described. For each considered case-study, the mathematical formulation of the problem, the specific \gls{ann} architecture that is used, and the corresponding analysis and numerical results are discussed.
\item[\textbf{(C.4)}] A solid and self-contained description of the theoretical foundations of deep learning, the most relevant \glspl{ann} architectures and training methods, as well as the most widely-used guidelines for hyper-parameters tuning are given. 
\item[\textbf{(C.5)}] The connection between deep learning and other machine learning frameworks, such as \textbf{deep reinforcement learning}, \textbf{deep federated learning}, and \textbf{deep transfer learning} are discussed. Several case-studies where these learning frameworks are jointly used are quantitatively analyzed. Moreover, the approach of \textbf{deep unfolding} is proposed as a way to map iterative algorithms to \glspl{ann} architectures. 
\end{enumerate}
The rest of this work is organized as follows:
\begin{itemize}
\item The rest of this section elaborates on contribution \textbf{C.2}, by discussing the potential and advantages of \gls{ai}-based wireless networks, for application to network deployment and planning, resource management, and maintenance and operation. Furthermore, our vision on data gathering and management in \gls{ai}-based networks is presented.
\item Section \ref{Sec:MLandDL} discusses in detail the connection between machine learning and deep learning. First, the fundamental paradigms of supervised learning, unsupervised learning, and reinforcement learning are introduced, and then the role of deep learning and \gls{ann} in this general framework is explained. 
\item Section \ref{Sec:Basics} is focused, together with Section \ref{Sec:MLandDL}, on contribution \textbf{C.4}, providing the theoretical description of deep learning, introducing the basic components of \glspl{ann}, the most widely-used \gls{ann} architectures and training methods. In addition, the connection between deep learning, reinforcement learning, transfer learning, and deep unfolding are explained, providing Contribution \textbf{C.5}.
\item Contributions \textbf{C.1} and \textbf{C.3} are addressed in Section \ref{Sec:Applications}. First, a detailed overview of the applications and research contributions of deep learning to wireless communications is provided. Next, several examples and use-cases of practical interest are presented, in which the joint use of mathematical models and deep learning methods are shown to yield significant gains compared to state-of-the-art approaches. For each use-case, a quantitative analysis is explicitly carried out, by describing the design of an \gls{ann} to tackle the problem and discussing the resulting performance. 
\item Finally Section \ref{Sec:Conclusions} concludes this paper by outlining the major challenges to overcome in order to fully enable the rise of \gls{ai}-based wireless communication networks.
\end{itemize}

\subsection{Deep Learning for Network Deployment and Planning}\label{Sec:AIforDeployment}
Future wireless networks will be more than allowing people, mobile devices, and objects to communicate with each other \cite{1}. Future wireless networks will be turned into a distributed intelligent wireless communication, sensing, and computing platform, which, besides communications, will be capable of sensing the environment to realize the vision of smart living in smart cities by providing context-awareness capabilities, of locally storing and processing information in order to accommodate the time critical, ultra-reliable, and energy efficient delivery of data, of accurately localizing people and objects in environments and scenarios where the global positioning system is not an option. Future wireless networks will have to fulfill the challenging requirement of interconnecting the physical and digital worlds in a seamless and sustainable manner \cite{2}, \cite{3}.

To fulfill these challenging requirements, we think that it is not sufficient anymore to rely solely on wireless networks whose \textit{logical} operation is software-controlled and optimized \cite{4}. The \textit{wireless environment} itself needs to be turned into an intelligent software-reconfigurable entity \cite{5}, whose operation is optimized to enable uninterrupted connectivity. Future wireless networks need a smart environment, i.e., a wireless environment that is turned into a reconfigurable space that plays an active role in transferring and processing information. We refer to this emerging wireless future as \enquote{smart radio environment} \cite{EURASIP_RIS}.

\begin{figure}[!t]
\centering
\includegraphics[width=\columnwidth]{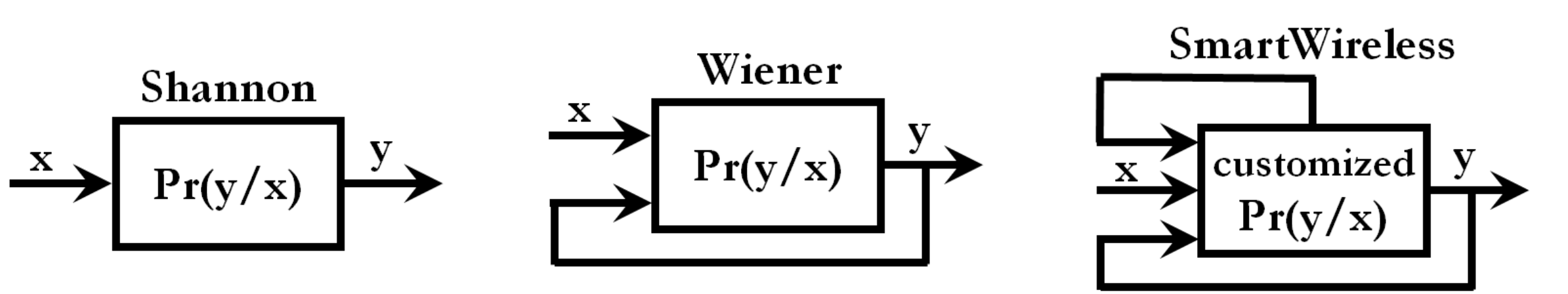}
\caption{Current networks vs. a smart radio environment (or smart wireless).}
\label{SmartWireless_BlockDiagram}
\end{figure}
To better elucidate our notion of reconfigurable and programmable wireless environment, let us consider the block diagram illustrated in Fig. \ref{SmartWireless_BlockDiagram}. Conceptually, the difference between current wireless networks and a smart radio environment can be summarized as follows. According to Shannon \cite{24}, the system model is given and is formulated in terms of transition probabilities (i.e., $\Pr \left\{ {{y \mathord{\left| {\vphantom {y x}} \right. \kern-\nulldelimiterspace} x}} \right\}$). According to Wiener \cite{25}, the system model is still given, but its output is feedback to the input, which is optimized by taking the output into account. For example, the channel state is sent from a receiver back to a transmitter for channel-aware beamforming. In a smart radio environment, the environmental objects are capable of sensing the system's response to the radio waves (the physical world) and feed it back to the input (the digital world). Based on the sensed data, the input signal and the response of the environmental objects to the radio waves are jointly optimized and configured through a software controller, respectively. For example, the input signal is steered towards a given environmental object, which reflects it towards the receiver by suitably-optimized phase shifts. In turn, the receiver is also steered towards the incoming signal. \\

Different solutions towards realizing the vision of smart radio environments are currently emerging \cite{6}-\cite{14}. Among them, the use of reconfigurable meta-surfaces constitutes a promising and enabling solution to fulfill the challenging requirements of future wireless networks \cite{15}. Meta-surfaces are thin meta-material layers that are capable of modifying the propagation of the radio waves in fully customizable ways \cite{16}, thus having the potential of making the transfer and processing of information more reliable \cite{17}. Also, they constitute a suitable distributed platform to perform low-energy and low-complexity sensing \cite{18}, storage \cite{19}, and analog computing \cite{20}. In \cite{14}, in particular, the authors have put forth a network scenario where every environmental object is coated with reconfigurable meta-surfaces, whose response to the radio waves is software-programmed  by capitalizing on the enabling technology and hardware platform currently being developed in \cite{21}.

\begin{figure}[!t]
\centering
\includegraphics[width=\columnwidth]{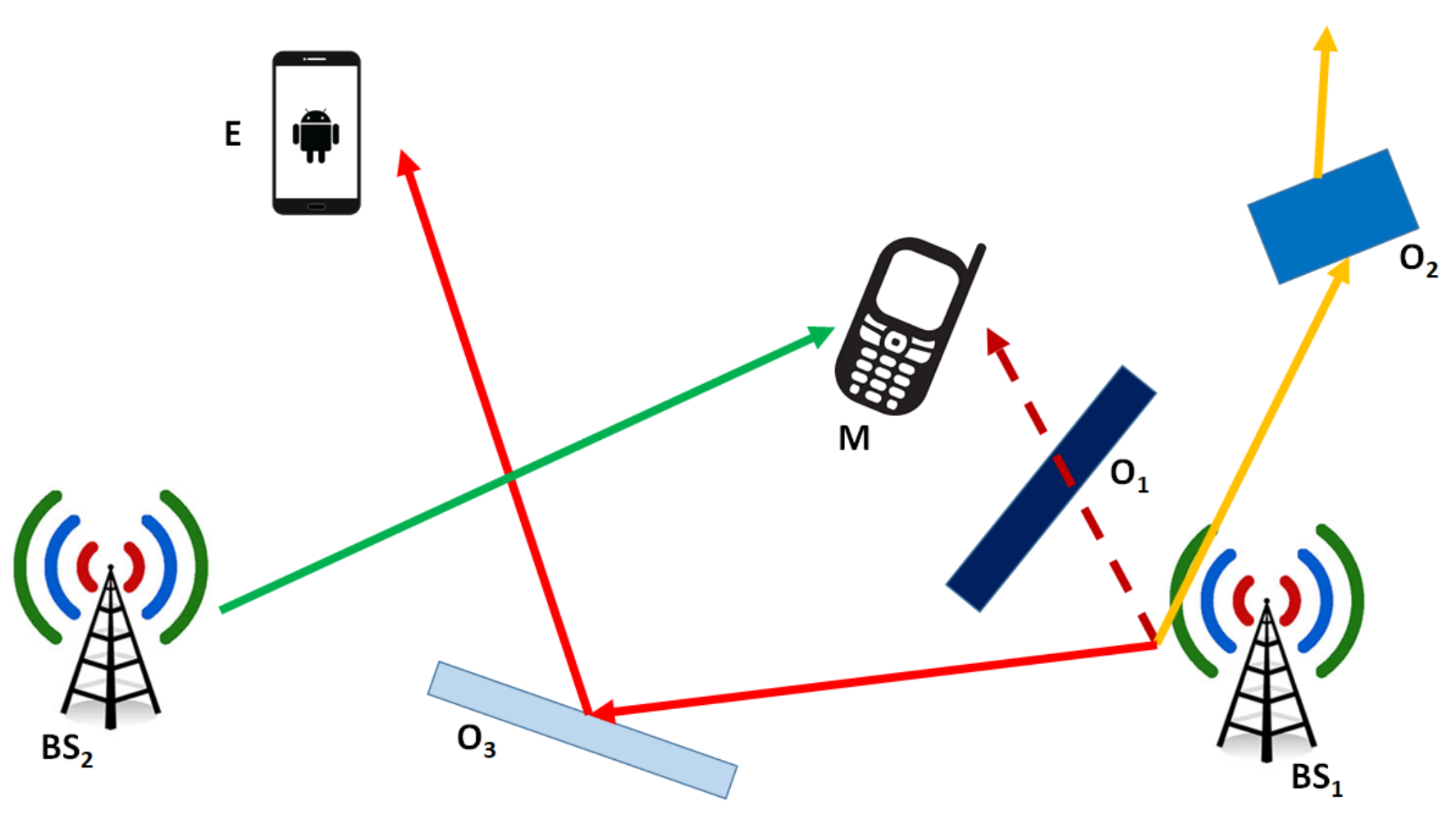}
\caption{Current cellular networks operation.}
\label{CellularScenario_Today}
\end{figure}
\begin{figure}[!t]
\centering
\includegraphics[width=\columnwidth]{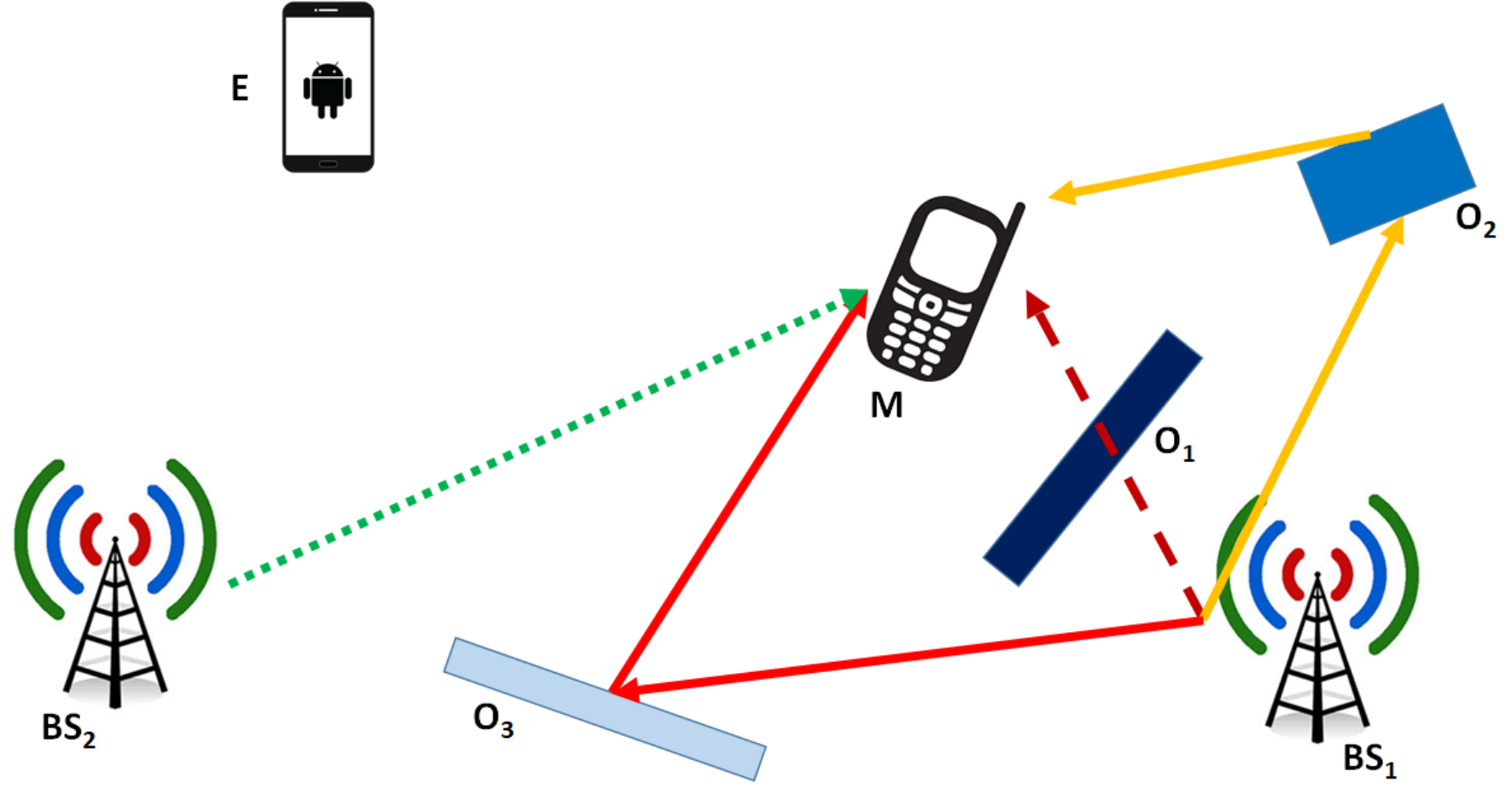}
\caption{Cellular networks operation in a smart radio environment.}
\label{CellularScenario_Tomorrow}
\end{figure}

An example of using reconfigurable meta-surfaces in a cellular network scenario is sketched in Figs. \ref{CellularScenario_Today} and \ref{CellularScenario_Tomorrow}. In Fig. \ref{CellularScenario_Today}, a mobile terminal (M) wants to connect to the Internet via a cellular network. In the absence of environmental objects (O1, O2, O3), BS1 is the base station that provides the best signal to M. Due to the blocking object O1, however, the received signal from BS1 is not sufficiently strong, and M connects to the Internet via BS2, while BS1 is kept active to serve other users. Since BS2 is far from M, its received signal is not sufficient for high rate transmission. Because of the refractive object O2, the signal emitted by BS1 generates strong interfering signals in other locations. Also, the reflective object O3 generates a strong reflected signal towards a malicious user (E) that can intercept the signal from BS1. In Fig. \ref{CellularScenario_Tomorrow}, by contrast, we illustrate the operation of cellular networks in a smart radio environment. The objects O1, O2, O3 are now coated with reconfigurable meta-surfaces that modify the radio waves according to the generalized laws of reflection and refraction \cite{16}. Figure \ref{CellularScenario_Tomorrow} shows how the operation of wireless networks changes fundamentally. The link BS1-M is still obstructed by O1. The responses of the reconfigurable meta-surfaces on O2 and O3 are, however, appropriately controlled and optimized: O2 refracts the signal from BS1 towards M and avoids interfering other users. O3 reflects the signal towards M and protects BS1 against E. In contrast to Fig. \ref{CellularScenario_Today}, the reflected and refracted signals at M allow it to reliably connect to the Internet. Now, BS2 serves other users at, e.g., a higher speed.

Current research efforts towards realizing the vision of smart radio environments are primarily focused on implementing hardware testbeds, e.g., reflect-arrays and meta-surfaces, and on realizing point-to-point experimental tests \cite{6}-\cite{14}. To the best of the authors knowledge, on the other hand, there exist no theoretic and algorithmic methodologies that provide one with the ultimate performance limits of this emerging wireless future, and with the algorithms and protocols for achieving those limits. We argue, in addition, that the design of smart radio environments is unlikely to be possible by relying solely on conventional methods. We believe, on the other hand, that deep learning and \gls{ai} will play a major role in this context. In the following two sections, we will first discuss in deeper details the difference and potential advantages of smart radio environments against current wireless network solutions, and then discuss the importance of deep learning in this context.

\subsubsection{Current Networks vs. Future Smart Radio Environments}
To better elucidate the difference and significance of smart radio environments with respect to the most advanced technologies employed in wireless networks at present, let us consider, as an example, a typical cellular network.

The distinguishable feature of cellular networks lies in the users' mobility. The locations of the base stations cannot, in general, be modified according to the user's locations. Some exceptions, however, exist \cite{26}, \cite{27}, and we will elaborate on this below. The mobility of the users throughout a location-static deployment of base stations renders the user distribution uneven throughout the network, which results in some base stations to be severely overloaded and some others to be under-utilized. This is a well-known issue in cellular networks, and is tackled in different ways, among which load balancing  \cite{28} and the densification of base stations (ultra-dense networks). Network densification is certainly a promising approach, but has its own limitations \cite{29}, \cite{105}. It is known, e.g., that network densification increases the network power consumption as the number of base stations per square kilometer increases. This is exacerbated even more with the advent of the Internet of Things (IoT), where the circuit power consumption increases with the number of users per square kilometer \cite{TWC_Paper2018}, \cite{135}. Ultra-dense network deployments, in addition, enhance the level of interference, which needs to be appropriately controlled in order to achieve good performance. Furthermore, each base station necessitates a backhaul connection, which may not be always available. Other solutions based on massive Multiple-Input-Multiple-Output (MIMO) schemes could be employed, but they usually necessitate a large number of individually controllable radio transmitters and advanced signal processing algorithms \cite{30}. Similar comments (i.e., power consumption, hardware complexity, blocking of links, etc.) apply to using millimeter-wave communications \cite{31}, \cite{32}. It is worth mentioning that millimeter-wave systems can take advantage of the presence of reconfigurable meta-surfaces as a source of controllable reflectors that can overcome non-line-of-sight propagation conditions, and enable the otherwise impossible communication between the devices \cite{9}. Without pretending to be exhaustive, other relevant solutions that are typically employed in wireless encompass retransmission methods that negatively impact the network spectral efficiency, the optimized deployment of specific network elements, e.g., relays, which increase the network power consumption as they are made of active elements (e.g., power amplifiers), and that either reduce the achievable link rate if operated in half-duplex mode or are subject to severe self-interference if operated in full-duplex mode \cite{Relay_1}-\cite{Relay_3}.

Meta-surfaces-enabled smart radio environments are fundamentally different. The meta-surfaces are made of low-cost passive elements that do not require any active power sources for transmission \cite{3}. Their circuitries can be powered with energy harvesting modules, too \cite{34}. They do not apply any sophisticated signal processing algorithms (coding, decoding, etc.), but primarily rely on the programmability and re-configurability of the meta-surfaces and on their capability of shaping the radio waves impinging upon them \cite{35}. They can operate in full-duplex mode without significant or any self-interference, and do not need any backhaul connections. Even more importantly, the meta-surfaces are deployed where the issue naturally arises: where the environmental objects, which, in current wireless networks, reflect, refract, distort, etc. the radio waves in undesirable and uncontrollable ways, are located. Since the input-output response of the meta-surfaces is not subject to conventional Snell's laws anymore, the locations of the objects that assist a pair of transmitter and receiver to communicate, and the functions that they apply on the received signals can be chosen to minimize the impact of multi-hop-like signal attenuation. In addition, the phase of the many atomic elements (i.e. the scattering particles), that constitute the meta-surfaces can be optimized to coherently focus the waves towards the intended destination, thus obtaining a substantial beamforming gain without using active elements. These functionalities, in addition, are transparent to the users, as there is no need to change the hardware and software of the devices. Furthermore, the number of environmental objects can potentially exceed the number of antennas at the endpoint radios, which implies that the number of parameters for system optimization will exceed that of current wireless network deployments \cite{11}. The freedom of controlling the response of each meta-surface and choosing their location via a software-programmable interface makes, in addition, the optimization of wireless networks agnostic to the underlying physics of wireless propagation and meta-materials. Despite the practical challenges of deploying robotic (terrestrial) base stations capable of autonomously moving throughout a given region \cite{26}, \cite{27}, experimental results conducted in an airport environment, where the base stations were deployed on a rail located in the ceiling of a terminal building \cite{36}, showed promising gains. The possibility to deploy mobile reconfigurable meta-surfaces is, on the contrary, practically viable. The meta-surfaces can be easily attached to and removed from objects (e.g., facades of buildings, indoor walls and ceilings, advertising displays), respectively, thus yielding a high flexibility for their deployment. The position of small-size meta-surfaces on large-size objects, e.g., walls, can be adaptively optimized as an additional degree of freedom for system optimization: Thanks to their 2D structure, the meta-surfaces can be mechanically displaced, e.g., along a discrete set of possible locations (moving grid) on a given wall. It is apparent, therefore, that the concept of smart radio environment can potentially impact wireless networks immensely. First contributions that investigate the use of meta-surfaces for the design of wireless networks have appeared in \cite{ZapLIS2019,HuangICASSP}.

\subsubsection{The role of deep learning in smart radio environments}
As discussed, the concept of smart radio environment is a fundamental paradigm shift compared to the current designs of wireless networks. 
But what is the interplay between smart radio environments and \gls{ai}-based communication networks? We believe the two paradigms are intertwined, at the same time enabling and being enabled by each other.  As already mentioned, besides the ability of improving the communication performance, meta-surfaces are expected to be equipped with sensors that allow them to estimate the current conditions of the environment. This equips them with the capability of acquiring lots of data that can be locally stored and processed, and/or sent to fusion centers. Thus, meta-surfaces provide the fabric of future \gls{ai}-based wireless networks. Thanks to the pervasive use of meta-surfaces, smart radio environments will be naturally able to acquire and harness a large amount of data that travels over communication networks and that is required to maximize the performance of deep learning algorithms based on \glspl{ann}. In this sense, smart radio environments constitute an enabler for the implementation of \gls{ai}-based communication networks. 

On the other hand, the massive use of meta-surfaces, reconfigurable reflect-arrays, reconfigurable large-intelligent surfaces, provides a large number of degrees of freedom whose optimization entails a large computational complexity. 
By direct inspection of Fig. \ref{SmartWireless_BlockDiagram}, it is apparent that smart radio environments are much more difficult to optimize than current wireless networks. In a smart radio environment, the operation of each environmental object may be optimized, besides the operation of the transmitter and receiver (the end points of the network). Accurately modeling such an emerging network scenario and optimizing it in real time and at a low complexity is an open issue. Indeed, it is very challenging to devise a model that is sufficiently accurate to account for customizable reflections, refractions, blocking, displacements of the surfaces, etc. Moreover, even if such a model could be developed, it would be very unlikely amenable to optimization due to the large number of variables to optimize and the complexity of the resulting utility functions. Compared with current network models, in addition, Fig. \ref{SmartWireless_BlockDiagram} highlights that smart radio environments need much more context-aware information for configuring and optimizing the operation of all the environmental objects, which results in a larger feedback overhead that has a strong impact in applications with high mobility. Unfortunately, in order to optimize such a complex system, with so many degrees of freedom, typical optimization-oriented approaches are not feasible, as they would require a too high complexity overhead. Luckily, as discussed in the 
the coming subsection \ref{Sec:DLforRM}, deep learning can be used to significantly simplify the resource management task. In this sense, \gls{ai} by deep learning and \glspl{ann} makes smart radio environments practically implementable, especially when model-based and \gls{ai}-based approaches are used jointly, as discussed in detail in Section \ref{Sec:Applications}.


\subsection{Deep Learning for Network Resource Management}\label{Sec:DLforRM}
The goal of resource management is to allocate the available network resources in order to maximize one or more performance metrics. Transmit powers, beamforming vectors, receive filters, frequency chunks, computing power, memory space, etc., can be scheduled among the network terminals based on traffic demands, propagation channel conditions, terminals requirements, so as to optimize the network throughput, the communication latency, the energy efficiency, while at the same time ensuring that all end-users experience the guaranteed \gls{qos}.  Formally speaking, denoted by $f$ the performance function to maximize and by $\bx\in{\cal S}$ the resource to allocate, with ${\cal S}$ the set containing the admissible values of $\bx$, the resource allocation problem can be cast as the optimization program
\begin{align}\label{Eq:GenResProb}
\ds\max_{\bx\in{\cal S}}f(\bx)\;.
\end{align}
Thus, the conventional approach to resource management is based on the use of traditional optimization theory techniques. However, as already mentioned, this approach only works if one is able to come up with a suitable mathematical model of the problem, i.e. with tractable, but accurate, formulas describing the objective $f$ and the feasible set ${\cal S}$. This is typically not the case in interference-limited systems, where the presence of multi-user interference makes most relevant radio resource allocation problems NP-hard. For example, power allocation for sum-rate maximization is known to be NP-hard in interference-limited networks \cite{LuoNP2008}, which implies that also beamforming problems and energy efficiency maximization problems are NP-hard \cite{ZapNow15} as well.
Moreover, even if we could solve NP-hard problem with affordable complexity, the optimal resource allocation will inevitably depend on the system parameters, e.g. the users' positions, the number of connected users, the slow-fading or fast fading channel realizations. Anytime one of these parameters  changes, which occurs quite frequently in mobile environments, the optimization problem needs to be solved anew. This causes a significant complexity overhead, that limits the real-time implementation of available optimization frameworks, especially in large and complex systems like future wireless communication networks. Clearly, all of these issues become even more prominent in smart radio environments where the number of variables to optimize will far exceed conventional numbers. In this context, the use of deep learning techniques based on \glspl{ann} can significantly reduce the burden of system design, enabling true online resource management. A first contribution that demonstrates the use of deep learning for the design of a meta-surface-enabled wireless network  has appeared in \cite{HuangSMDL2019}.

Our proposed approach to solve resource allocation problems by deep learning is based on the observation that the general resource allocation problem in \eqref{Eq:GenResProb} can be regarded as an unknown function mapping from the ensemble of all network parameters of interest, denoted by $\bc\in\mathbb{R}^{N}$, with $N$ the number of system parameters of interest, to the corresponding optimal resource allocation $\bx^{*}\in{\cal S}$. Formally, we can view Problem \eqref{Eq:GenResProb} as the non-linear map
\beq\label{Eq:Fintro}
{\cal F}:\bc\in\mathbb{R}^{N}\to \bx^{*}\in{\cal S}\subseteq\mathbb{R}^{N} \;.
\eeq
Thus, our proposal is to convert Problem \eqref{Eq:GenResProb} into learning the unknown map \eqref{Eq:Fintro}, a task that \glspl{ann} are able to tackle. Indeed, as it will be discussed in Section \ref{Sec:MLandDL}, \glspl{ann} are, under very mild assumptions, universal approximators, i.e., if properly trained, they are able to learn the input-output relation between the system parameters and the desired resource allocation to use, thus emulating the function ${\cal F}$ in \eqref{Eq:Fintro}. This means that we can optimize a desired performance function for given system parameters without explicitly having to solve any optimization problem, but rather letting an \gls{ann} compute the resource allocation for us. A detailed analysis of this approach will be presented in Section \ref{Sec:Applications}.

With this in mind, the natural question that arises is how to integrate \gls{ann}-based resource management into the topology and architecture of a wireless network. Where should we store the data required by the \gls{ann} tasked with network resource management, and where should the related computations be executed? Ideally, the optimal approach would be to have a cloud-based approach in which an \enquote{artificial brain} placed in a single point oversees all tasks related to resource management across the whole network or at least a network segment. All available data should be collected and stored in this artificial brain which is tasked with executing all required computations and with feeding back the resulting optimal resource allocation policy to all other network terminals. Unfortunately, such a centralized approach is not compatible with future wireless networks due to at least three major reasons:
\begin{enumerate}
\item \textbf{Latency.} Some vertical sectors of future wireless networks, e.g. \gls{urllc}, require strict end-to-end communication latency requirements, lower than a millisecond. Thus, for these applications, it is not possible to wait for the cloud to perform the computations and then feed the results back to the end-users. Instead, it would be more convenient to perform the computations locally at the users' terminals.
\item \textbf{Privacy.} Unlike previous wireless networks generations, future wireless networks will not be simply limited to realizing faster mobile network or to providing richer functions in smartphones. The integration of  innovative vertical services aims at making the vision of the \enquote{everything connected world} true, but this comes with critical privacy and security requirements. Accordingly, for some vertical applications it is not desirable to share information with the cloud, which makes cloud-based deep learning not a convenient approach. In this context, it should be mentioned that, even if network security methods exist and provide us with privacy, integrity, and authentication, their use represents an overhead in terms of additional complexity and additional data to transmit \cite{StallingsBook}. Indeed, commercial solutions to privacy and/or authentication require the use of specific cryptographic algorithms such as \emph{Advanced Encryption Standard} (AES) and \emph{Rivest-Shamir-Adleman} (RSA), which run on top of the physical layer and require to execute finite fields operations on each block of transmitted data. Moreover, data integrity is typically guaranteed by the use of Hash codes, which also require the execution of specific operations to generate the Hash code for each packet of transmitted bits. Clearly, this results in overheads that might significantly reduce the communication performance of large-scale networks. Moreover, the perceived level of trust by the end-users will be inherently higher if no sensible data needs to be transmitted. 
\item \textbf{Connectivity.} Future wireless networks promise ubiquitous service delivery. This means that a user terminal should be able to operate also in areas or times in which a poor connection to the cloud exists. This requirement is not compatible with a pure cloud-based implementation, but instead each user device should have some \enquote{local intelligence} to be able to operate in these scenarios, too.
\end{enumerate}
Therefore, in order to make deep learning compatible with future wireless communication networks, the intelligence can not be concentrated only in a centralized network brain. Instead, some intelligence should be distributed across the network mobile devices, implementing a \emph{\textbf{Mobile \gls{ai}}} architecture. It is interesting to observe that this approach resembles the way in which human knowledge is developed: like human societies in which there is a collective intelligence that belongs to everybody, and an individual intelligence, the mobile \gls{ai} paradigm envisions both a \emph{cloud intelligence}, which every node of the network can access by connecting to the cloud, and a \emph{device intelligence} specific to each network device.

In order to implement this mobile \gls{ai} paradigm, a first natural approach that we put forth is to regard each device in the network as a rational and independent decision-maker, which acquires its own local dataset and uses it to build its own local \gls{ann} model. This technique does not require any interaction between the network infrastructure and the edge users, as far as data sharing and processing are concerned, and has the potential of enabling the 5G vision of distributed, self-managing networks true. On the other hand, due to limited storage and processing capabilities, mobile devices might not be able to develop accurate models on their own and the resulting performance gap must be analyzed. Moreover, the self-organizing nature of the devices poses questions about reaching a stable network operating point and about the efficiency of such a point. The Noble-prize-winner framework of game theory appears as the natural way to answer at least the last points, as it provides sophisticated mathematical tools to analyze the interactions among independent decision-makers \cite{Fudenberg1993,Basar-Olsder,myerson1997game}. Game theory has been already extensively used for resource management in wireless communication networks \cite{TembineBook2011,Han2011,mackenzie2006game}, although never in connection with deep learning. 

A second approach that we envision is based on the use of the so-called \emph{federated learning} technique \cite{McMahan2017,Kon2017}. The main idea of federated learning is to distribute the data and computation tasks among a federation of local devices that are coordinated by a central server. The server owns a global \gls{ann} model that is built by appropriately combining the local models from the devices, which  are developed based on local datasets. The server, on the other hand, is updated only with the updates of the global model, without the need of collecting and processing the datasets themselves. By leveraging this approach, the individual intelligence owned by each device contributes to the collective intelligence of the whole federation of devices, which is maintained by the server. As a refinement of this approach, \cite{Chen2018} proposes not to exchange the updates of the model, but rather the updates of  the algorithm that is used to compute the model. In other words, each local model is computed by processing the local dataset by some algorithm, and the devices do not communicate the model to the server, but instead send only an update of the parameters of the algorithm that is used to compute the global model.

Regardless of the specific approach that is employed, the mobile \gls{ai} paradigm comes with several  fundamental open problems. In a scenario where each wireless node has cognitive abilities (i.e. its own  \gls{ann}), and whose behavior is influenced by its own local experience (i.e. local data), different wireless devices will learn how to behave based on datasets that might differ in both quantity (different nodes might have different measurement and storage capabilities) and quality (different nodes might experience different data perturbations due, for example, to the non-ideality of the measurement sensors). This could potentially lead to instabilities and, in the worst case, could cause the communication network to collapse. Hence, new control mechanisms are necessary in order to ensure the correct evolution over time of \gls{ai}-based communication networks.

\subsection{Deep Learning for Network Operation and Maintenance}
Maintenance and operation of a wireless network is a broad field that involves many different tasks, such as users' localization, channel estimation, quality-of-service monitoring, fault and anomaly detection, hand-over execution, intrusion detection, etc. Although seemingly quite diverse, operation and maintenance tasks have a common denominator, as they both involve the acquisition of some measurable data, from which the desired information must be extracted. Formally speaking, all above tasks can be formulated as the task of guessing the realization of some random vector $\bx$ based on the observation of another random vector $\by$, that is somehow correlated to $\bx$, i.e. that was generated from $\bx$ through some unknown transformation. Such a problem can be cast into the framework of classical decision and estimation theory, but classical detection and estimation methods require the conditional distribution $f(\bx|\by)$ and the prior distribution $f(\bx)$, whose availability is strongly related to the availability of a tractable model for the specific problem at hand. Even in present wireless applications, this is an unrealistic assumption for several operation and maintenance tasks. A notable example is that of hand-overs of users moving along the boundary of two cells, a crucial problem in cellular networks. This is typically (and heuristically) handled by comparing the users' \gls{snr} towards the neighboring cells over a given time window. However, deriving a statistical model for this scenario that accounts for the users' mobility patterns is quite challenging, and indeed the optimization of the thresholds for hand-overs is an open problem even in present cellular networks. Given the foreseen complexity increase in future wireless communication networks, statistical approaches will become less and less practical.

A suitable way of coping with the lack of models and statistical information about the random vectors $\bx$ and $\by$ is represented by machine learning. Indeed, operation and maintenance is probably the field of wireless communications in which machine learning approaches have been used first. Recent surveys on applications of machine learning for maintenance tasks have appeared in \cite{Wei2017,Han2016Survey,Zhang2012,Granjal2015}, and have shown how machine learning performs well even without any statistical distribution information. Specifically, available solutions assume that a training set containing examples of correct matches between the realizations of $\bx$ and $\by$ is available, e.g. based on observing and storing previous traffic data. By processing the training set according to specific procedures called \emph{training algorithms}, machine learning methods are able to learn a rule for predicting the value of $\bx$ corresponding to unobserved values of $\by$.

As far as the integration of deep learning for network maintenance into future wireless architectures is concerned, it is our opinion that it could be carried out following a more centralized approach than for the resource management scenario described in Section \ref{Sec:DLforRM}. Indeed, most operation and maintenance tasks (e.g. fault and anomaly detection, hand-overs, intrusion detection) are inherently centralized in the sense that all computations are executed by network infrastructure nodes and do not require any specific information exchange with edge-users. On the other hand, in case of very large datasets and very demanding computations to perform, we envision the use of a distributed or federated learning approach, but only among dedicated network nodes. More in detail, a suitable approach consists of  sharing storage and computation tasks among a cluster of fixed infrastructure nodes connected by high-speed links and deployed in different points of the network. In this case, each node of the cluster could either be tasked with operating and maintaining only a specific part of the network, or the data and computing power of each cluster node could be jointly exploited via a federated learning approach.

\section{Machine Learning and Deep Learning}\label{Sec:MLandDL}
The term \emph{machine learning} broadly refers to algorithmic techniques able to perform a given task  without running a fixed computer program explicitly written and designed for the problem at hand, but instead processing available data and progressively learning from it. Formally speaking, a computer program is said to learn from experience \textbf{E} with respect to a task \textbf{T} and performance measure \textbf{P}, if its performance at task \textbf{T}, as measured by \textbf{P}, improves with experience \textbf{E} \cite{Mitchell1997}.

The tasks that can be solved by machine learning are very diverse. In general, machine learning techniques  prove extremely useful to execute tasks for which no explicit and/or viable programming approach exists to date, e.g. classification, regressions, pattern recognition, automatic language translation, anomaly detection, etc. As diverse as the task to perform may be, a machine learning algorithm can be mathematically described by the map 
\beq\label{Eq:MLTechnique}
{\cal F}:\bx\in{\cal X}\subseteq\mathbb{R}^{n}\rightarrow \by\in{\cal Y}\subseteq\mathbb{R}^{m}\;,
\eeq
wherein $\bx$ is a data vector whose components are the \emph{features} describing the task to be solved, $\by$ is the output produced by the machine learning algorithm representing the answer to the problem at hand, ${\cal X}$ and ${\cal Y}$ are the sets in which $\bx$ and $\by$ may vary. It is important not to confuse the task performed by a machine learning technique with the action of learning. The former is the final objective of the algorithm, while the latter is the method that is used to carry out the task.

In order to evaluate the ability of a machine learning algorithm to solve the assigned task, i.e. to produce  output vectors close to the desired ones, a performance criterion \textbf{P} must be defined. Several performance measures can be considered and typically the best choice is application-dependent. Typical choices are the mean squared error (MSE) or the cross-entropy functions, which will be formally introduced in Section  \ref{Sec:TrainingANN}, where the training procedure for \glspl{ann} is  described.

The last component of a machine learning algorithm to be introduced is the experience \textbf{E}, i.e. the knowledge and data that the algorithm can exploit to carry out the task. Machine learning algorithms typically experience a set of data points ${\cal S}_{TR}$, called \textbf{training set}. Depending on the information contained in ${\cal S}$, machine learning algorithms can be grouped into two main categories:
\begin{itemize}
\item \textbf{Unsupervised learning}: the experienced data training set ${\cal S}_{TR}$ contains only input features, i.e. ${\cal S}_{TR}=\{\bx_{1},\ldots,\bx_{N}\}$. Based on ${\cal S}_{TR}$, the machine learning algorithm must be able to extrapolate the statistical structure of the input or any other information needed to carry out the desired task. 
\item \textbf{Supervised learning}: the experienced data training set ${\cal S}_{TR}$ contains both input features and the corresponding desired outputs, referred to as \emph{labels} or \emph{targets}, i.e. ${\cal S}=\{(\bx_{1},\by_{1}),\ldots,(\bx_{N},\by_{N})\}$. Thus, in supervised learning, the training set  provides a series of examples to instruct the algorithm how to behave when some specific inputs are considered. 
\end{itemize}
In both supervised and unsupervised learning, the available dataset is fixed. This models a scenario in which the algorithm does not directly interact with the environment where it operates. Instead, a different machine learning paradigm that does not fall in the categorization above is that of \textbf{reinforcement learning} \cite{SuttonBookRL}. The approach of reinforcement learning is to enable a feedback loop between the algorithm and the environment, allowing the algorithm to experience a dataset that changes over time as a result of the interaction with the surrounding environment. The focus of this work will be primarily on supervised learning, which is the typical approach in deep learning. Reinforcement learning will also be considered, primarily considering its integration with deep learning tools, which leads to the recently introduced paradigm of \textbf{deep reinforcement learning} \cite{Li2017_DRL,Arulkumaran2017_DRL}.

Before continuing, it is important to remark that, while the setting described above bears some resemblance to the general problem of classical decision/estimation theory, a fundamental difference exists. Classical decision/estimation theory assumes that the probability distributions of the output vector given the input $p(\by|\bx)$ and that of the input vector $p(\bx)$ are known. Instead, machine learning does not need this assumption and is able to operate based only on some realizations of the underlying distributions, even though the distributions themselves are not known.

\subsection{Overfitting and Underfitting}
Any machine learning algorithm experiences a training set ${\cal S}_{TR}$ that contains some input features $\bx_{1},\ldots,\bx_{N}$. In the supervised scenario, each input feature is also accompanied by the corresponding desired output. While this information is essential to configure the learning scheme, the key problem of any machine learning algorithm is to perform well on \emph{previously unseen} inputs. This means that the algorithm needs to be able to grasp from ${\cal S}_{TR}$ a general rule to produce a suitable output $\by$ also when $\tilde{\bx}\notin{\cal X}$. This is referred to as the algorithm \emph{generalization capability}. During the training phase, the information in the training set is used to set the algorithm parameters in order to minimize any desired performance metric. As it will be detailed in the sequel, this amounts to solving an optimization problem. Machine learning however, is fundamentally different from optimization theory: its ultimate goal is to make the algorithm able to generalize well to new data inputs. In order to evaluate its generalization capability, after the algorithm has been designed as a result of the training phase, its performance is  tested over a new set of different inputs ${\cal S}_{T}$, called the \textbf{test set}. For any given error measure, the error evaluated over the test set is called \emph{generalization error} or \emph{test error}. Similarly, the error evaluated over the training set is called the \emph{training error}. Clearly, in order for the algorithm to generalize well, the data samples in the training set ${\cal S}_{TR}$ and in the test set ${\cal S}_{T}$ need to be drawn from the same distribution, called \emph{data generating distribution}, even though they should be drawn independently of each other. Clearly, the expected generalization error will be larger than the expected training error, and the gap between the two is called the \emph{generalization gap}. Thus, minimizing the training error can be regarded as a necessary but not sufficient condition to obtain also a low generalization error. A machine learning algorithm is said to be: 
\begin{itemize}
\item \textbf{Underfitting} if it is not able to make the error over the training set small.
\item \textbf{Overfitting} if it is not able to make the gap between the training and test error small.
\end{itemize}
The factor that controls whether overfitting or underfitting occurs is the \textbf{capacity} of the algorithm, i.e. the ability of the algorithm to properly fit the training set. Intuitively, the capacity of the algorithm is related to the degrees of freedom or parameters that can be chosen when designing the algorithm. If the algorithm does not have enough free parameters, it will not have enough degrees of freedom to capture the structure of the training set and the algorithm will underfit. Instead, the overfitting scenario is subtler. One may think that increasing the number of free parameters will always lead to better performance, and that an upper limit is represented only by the computational complexity that we can sustain. This is, however, not the case. If the algorithm has too many degrees of freedom, it will learn the structure of the training set too well, memorizing specific properties that are peculiar only to the training set, but that do not hold in general. As a result, there is an optimal capacity that a machine learning algorithm should have to minimize the generalization gap.   
\begin{figure}[!h]
\centering
\includegraphics[width=0.45\textwidth]{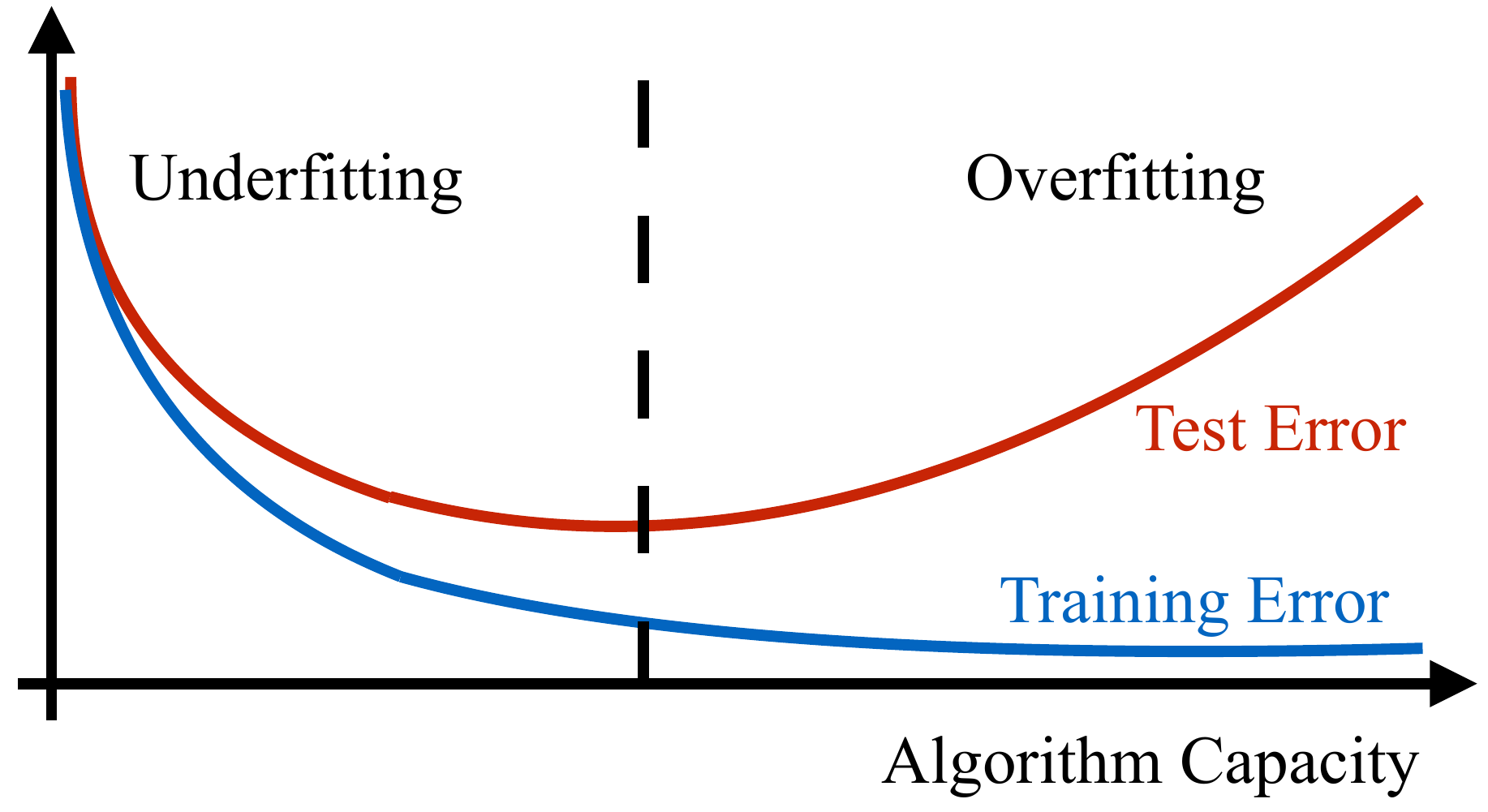}
\caption{Typical behaviors of the training and test errors.} 
\label{Fig:GeneralizationGap}
\end{figure}

As shown in Fig. \ref{Fig:GeneralizationGap}, the training error decreases with the algorithm capacity, asymptotically reaching its minimum value. Instead, the test error has a U-shaped behavior, following the training error up to a capacity value, and then increasing, thereby originating the generalization gap. Fundamental results from statistical learning theory have established that the generalization gap is bounded from above, with the upper bound increasing for larger model capacity, and decreasing for larger training sets \cite{Vapnik1971,Blumer1989,Vapnik1982,Vapnik1995}. On the other hand, a lower-bound to both the training and test error is given by the well-known Bayes error, i.e. the error obtained by an oracle with access to the true underlying distribution sampling from which the training and test set are obtained. 

\begin{figure}[!h]
\centering
\includegraphics[width=0.5\textwidth]{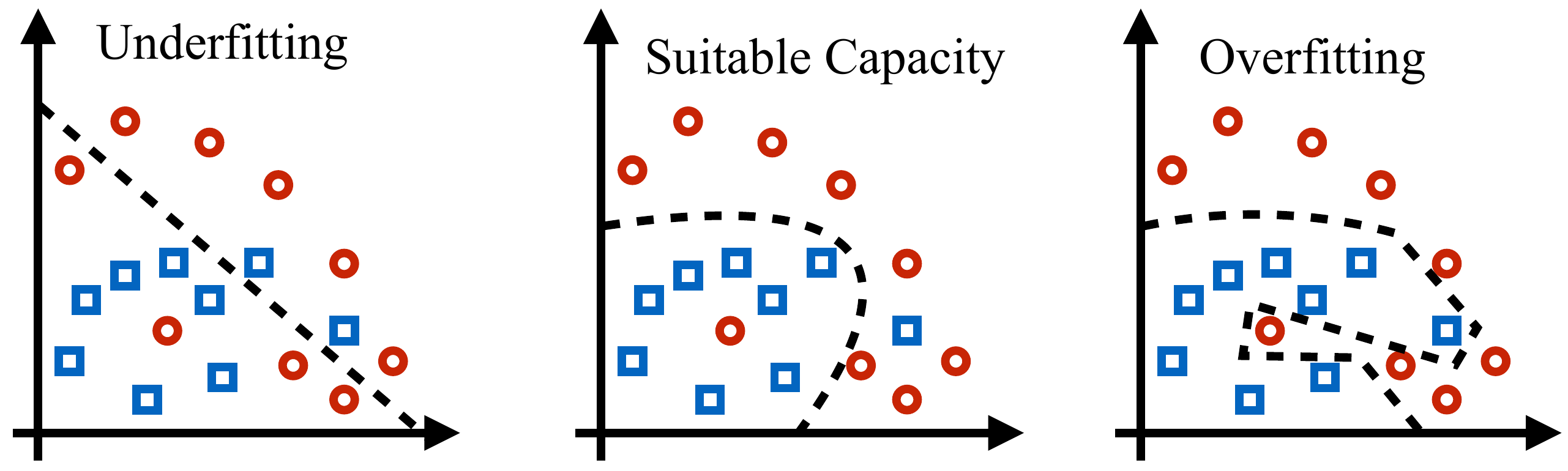}
\caption{Three possible decision boundaries for a classification problem. The left and right figures show  the underfitting and overfitting scenarios. The middle figure shows classifier with the proper capacity.} 
\label{Fig:DecBoundary}
\end{figure}

Another way to interpret the phenomenon of overfitting is to observe that any finite training set will also contain atypical realizations of the underlying distribution, that should be overlooked or given little importance when adjusting the algorithm parameters. However, if too many parameters to optimize are available, the algorithm will try to perfectly fit the complete training set, thus originating the overfitting phenomenon. This concept is illustrated in the example shown in Fig. \ref{Fig:DecBoundary}, where it is assumed that a machine learning classifier must output a decision boundary to separate objects belonging to two different classes. It can be seen how a linear decision boundary is not able to properly separate the samples in the training set, thus causing underfitting. On the other hand, having enough degrees of freedom, one can design a complex boundary to perfectly separate the samples in the training set, even those samples that happen to be surrounded by samples of the other class. However, this leads to including in both decision regions areas that are likely to contain samples from the wrong class, thus causing overfitting. Instead, the curved, but more regular, decision region in the middle better captures the structure of the underlying distribution. 

It is interesting to observe that choosing the decision boundary in the middle illustration of Fig. \ref{Fig:DecBoundary} is in agreement with the Occam's razor principle, stating that among different and equally motivated explanations of a phenomenon, one should choose the simplest one. Of course one should also be careful not to oversimplify the model, so as not to underfit. 

As mentioned above, one of the fundamental features that distinguishes machine learning theory from classical decision theory is the fact that the distribution underlying the task to perform is not known. This could lead to the belief that machine learning algorithms are universal, in the sense that the attainable performance depends only on how the parameters of the algorithm are set and on the size of the training set, but not on the properties of the underlying distribution, and, thus, not on the task to perform. Unfortunately, this belief is disproved by a fundamental result of machine learning, known as the \emph{no free lunch theorem}, which states that the test error of any machine learning algorithm is the same when averaged over all possible underlying distributions. This means that there exists no machine learning algorithm that outperforms any other algorithm at every possible task. Instead, different algorithms will achieve different performance when tackling different tasks, i.e. when the underlying distribution varies. 

\subsection{Hyperparameters and Validation Set}\label{Sec:ValidationSet}
Besides the parameters that are to be optimized by the training procedure, machine learning algorithms also have hyperparameters, i.e. parameters that are not directly set during the training phase, either because they are difficult to optimize, or because they should not be learnt from the training set. The latter case corresponds to the optimization of the parameters that directly affect the capacity of the model. In fact, if a parameter that affects the model capacity is tuned based only on the training set, the result will be that it will be chosen in order to minimize the training error as much as possible. However, we have seen how this would lead to a poor generalization error, due to overfitting.

To be more specific, anticipating some notions about \glspl{ann} to be discussed in the next section, an \gls{ann} is composed of several nodes whose input-output relationship is defined by some weights and bias terms, which are the parameters to be tuned during the training phase. On the other hand, the total number of nodes in the network and the way in which the nodes are interconnected are hyperparameters that are considered fixed while the training algorithm is executed. Besides the difficulty to optimize these discrete parameters, a critical problem is that the number of nodes in an \gls{ann} is directly related to the capacity of the network, since more nodes imply more degrees of freedom. Therefore, if we optimized the number of nodes based only on the training set, the optimum would be to use as many nodes as physically possible, thus causing overfitting. 

On the other hand, it is also not possible to use the test set to tune the hyperparameters, because all choices pertaining to the algorithm design must be independent of the data set that is used to assess the  performance of the algorithm. Otherwise, the estimation of the generalization error will be biased. This implies that we need a third data set for hyperparameter tuning, the \textbf{validation set}. The validation set is typically obtained by partitioning the training data into the training set and the validation set. The training procedure fixes some values of the hyperparameters and optimizes the  network parameters based only on the training set. Afterwards, an \emph{estimate} of the generalization error obtained with the considered hyperparameter configuration is obtained through the validation set. This procedure is repeated for different hyperparameter configurations to identify the best model to use. After both the parameters and hyperparameters have been set, the true generalization error is computed  by using the test set. The main steps of the whole procedure are summarized in Algorithm \ref{Alg:TrainValid}.

\begin{algorithm}
\caption{Hyperparameter and parameters tuning}\label{Alg:TrainValid}
\begin{algorithmic}
\While{\text{Error on validation set not satisfactory}} 
\State \texttt{Choose a set of hyperparameters;} 
\State \texttt{Given the chosen hyperparameters run the learning procedure for parameter optimization using the training set;}
\State \texttt{Evaluate the error on the validation set;}
\EndWhile
\end{algorithmic}
\end{algorithm}

While Algorithm \ref{Alg:TrainValid} provides one with a systematic procedure for training a machine learning algorithm, it does not address how to update the hyperparameter configuration in each loop. In general, there is no simple, algorithmic way to do this, and indeed hyperparameter tuning is more an art than a science. In particular, manual hyperparameter tuning is specific to the task to carry out and some guidelines will be discussed for application to deep learning in Section \ref{Sec:Hyperparameters}. Nevertheless, three systematic approaches for automated hyperparameter selection, which are general enough for many machine learning techniques, can be identified as follows:
\begin{itemize}
\item If the complexity of running the training procedure for a given hyperparameter configuration allows it, the hyperparameters can be learnt by means of a grid search.
\item As a variation of the grid search, a random search has been shown to provide good performance, while at the same time significantly reducing the overall complexity \cite{Bergstra2012}. 
\item A nested learning procedure can be used, in which a second machine learning algorithm is wrapped around the algorithm to be trained, with the task of learning the best hyperparameters for the inner algorithm.
\end{itemize}

\subsection{Beyond classical machine learning}
So far, the general principles at the basis of machine learning have been introduced, and some well-established machine learning algorithms have been mentioned. The rest of this section elaborates on their  inherent limitations, motivating why a different approach is needed, especially when the complexity of the task increases. 

\begin{figure}[!h]
\centering
\includegraphics[width=0.45\textwidth]{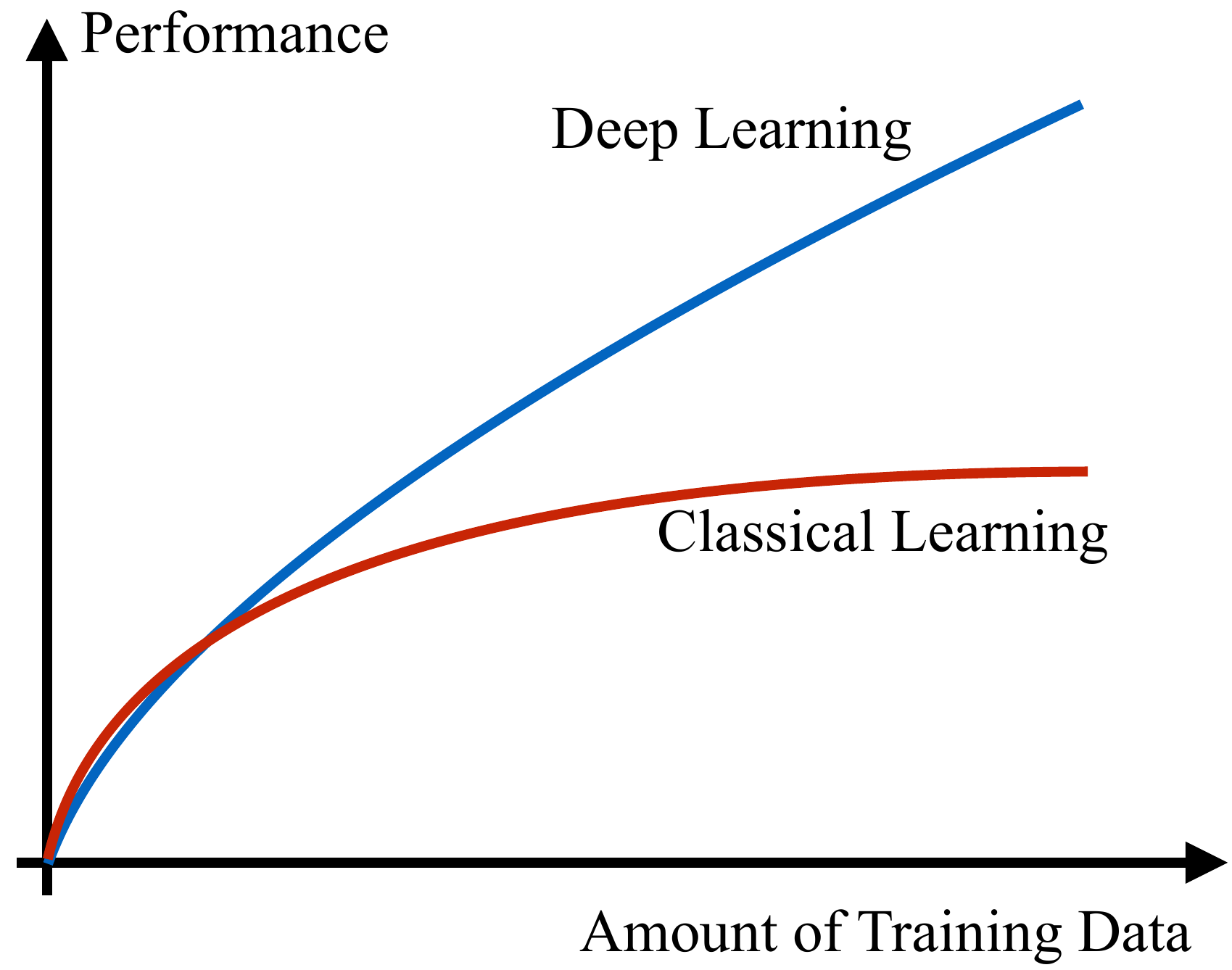}
\caption{Performance of classical and deep learning algorithms as a function of the training set size.}
\label{Fig:Performance}
\end{figure}

The main challenge of machine learning is to learn how to generalize in response to previously unseen inputs. In order to reduce the generalization error, one could train the algorithm over a larger amount of data. In fact, increasing the size of the training set is surely helpful, but there is a limit in terms of computation and storage capacity, to the amount of data that can be processed. Therefore, an essential component of machine learning is the performance of the different algorithms as a function of the size of the training set. Deep learning will be formally introduced in the next section, but Fig. \ref{Fig:Performance} anticipates how deep learning is able to improve the performance at a much faster rate than other machine learning techniques, as the dimension of the training data increases. 

It has to be stressed that, instead, for small-to-medium training set sizes, the relation among deep learning and other machine learning techniques is not well-defined, and in many cases it turns out that classical machine learning algorithms can slightly outperform deep learning. 

How can we explain the behavior in Fig. \ref{Fig:Performance}? The key phenomenon to consider is the so-called \emph{curse of dimensionality}, which refers to the fact that the number of distinct configurations of a set increases exponentially with the number of variables describing each element of the set. Recalling the formal description of a learning algorithm as formulated in the map in \eqref{Eq:MLTechnique}, we emphasize that the dimensionality here does not directly refer to the size of the training set, but instead to the number of features $n$ describing each element $\bx$ in the set of possible inputs ${\cal X}$. Nevertheless, it is clear that as $n$ increases, we need more training samples to successfully learn the structure of ${\cal X}$, thus devising a map ${\cal F}$ that is able to achieve a low generalization error. Conventional machine learning algorithms cope with the curse of dimensionality by using one of the following two approaches:
\begin{itemize}
\item Assuming prior beliefs about the structure that a good function ${\cal F}$ should have, such as the smoothness prior, i.e. assuming that the function ${\cal F}$ does not change drastically when evaluated at two neighboring points $\bx_{1}$ and $\bx_{2}$. However, in high-dimensional spaces even a very smooth function can vary at a different scale along different dimensions. Moreover, even assuming that all the derivatives of the function are similar in the different directions, the smoothness assumption is reasonable only when the points $\bx_{1}$ and $\bx_{2}$ are sufficiently close to each other. Depending on the magnitude of the derivatives this may require an unfeasible amount of training data. 
\item Incorporating task-specific assumptions to perform manual feature selection, i.e. deciding which components of $\bx$ are relevant to the specific problem at hand and performing a customized processing of these features. However, this process requires the analysis of a realistic mathematical model for the problem at hand, which may not be available. Moreover, the settings used for one task are not general in the sense that they may not apply to other problems.
\end{itemize}
Deep learning adopts quite a different approach. It assumes that the data has been generated by a composition of factors with a hierarchical order and develops a learning method that is able to automatically understand the structure of the underlying distribution, extracting directly from the data the features that are important to devise a good map ${\cal F}$. In other words, deep learning assumes that some correlations exist among the behavior of ${\cal F}$ over different regions of space, as a result of the structure of the underlying distribution of the data. This is clearly a more general assumption than the smoothness prior, which constraints the local behavior of ${\cal F}$ in the neighborhood of each point. This has been shown to enable deep learning to generalize non-locally \cite{Bengio2005}. Moreover, deep learning is able to understand the structure of the underlying distribution, without requiring task-specific assumptions, thus enabling more general-purpose algorithms. These improvements are possible thanks to the use of \glspl{ann}, which constitute the tool used by deep learning to implement the learning process.  

\section{Deep learning by artificial neural networks}\label{Sec:Basics}
As anticipated at the end of the previous section, \glspl{ann} are the enablers of deep learning \cite{Schmidhuber2014,Demuth2014}, thanks to their ability to learn, directly from the observed data, complex input-output relationships and statistical structures. \glspl{ann} are organized hierarchically in layers of elementary processing units, called \emph{neurons}. More in detail, an \gls{ann} is characterized by:
\begin{itemize}
\item An input layer, which forwards the input data to the rest of the network. 
\item One or more hidden layers, which process the input data. 
\item An output layer which applies a final processing to the data before outputting it. 
\item Weights and bias terms that model the strength of the connections among the neurons. 
\end{itemize}
If the network has only one hidden layer, it is referred to as a \emph{shallow network}, whereas if it has more than one hidden layer, it is referred to as a \emph{deep network}, hence the name deep learning. As discussed in Section \ref{Sec:FNN}, deep networks are preferred, since they usually require a lower number of neurons to achieve a given accuracy. It is probably the use of deep architectures in which multiple neurons process the information and propagate the result that has motivated the analogy between \glspl{ann} and natural neural networks, i.e. the human brain, which is also composed of a network of elementary processing units, the neurons, that elaborate information and then propagate the results to other neurons.

A first broad classification of \glspl{ann} is based on how the information flows from the input to the  output. Specifically:
\begin{itemize}
\item \gls{fnn} are neural networks in which each neuron is connected only to the neurons in the following layer and thus the input data can only propagate forward, from the input layer to the output layer, without the possibility of any feedback loop. 
\item \gls{rnn} are neural networks in which feedback loops are allowed, and the output of a neuron can become the input of the same neuron, as well as of other neurons in the same or in a previous layer. 
\end{itemize}
Several neural networks architectures exist within each of the two main categories introduced above. A notable example is that of \glspl{cnn}, described in Section \ref{Sec:CNNs}, which have been extensively used for image processing and pattern recognition \cite{Krizhevsky2012}. In this work, we have decided to adopt the broad classification above, because the differences with other neural networks architectures are somewhat blurry, since different kinds of layers can co-exist in the same neural network. Instead, a more specific classification can be made by considering the types of layers composing the \gls{ann}. The most common types of layers are the following:  
\begin{itemize}
\item \textbf{Fully-connected layer.} It is the typical layer employed in FFNs, which is characterized by the fact that each neuron of the layer receives an input from all neurons of the preceding layer, and is connected to all neurons of the following layer. The input data is first linearly processed, then passed through a non-linearity, and finally propagated to the following layer. 
\item \textbf{Convolutional layer.} It is another kind of layer used in FFNs, and more precisely in \glspl{cnn}. Similarly to a fully-connected layer, it filters the input by a linear operation, namely a  convolution, then applies a non-linearity, and finally forwards the result. However, each neuron needs not be connected to all neurons in the following layer. 
\item \textbf{Pooling layer.} It is a layer usually used in \glspl{cnn} which operates by dividing the input data into blocks, and then selecting either the maximum element of each block, or computing the average of the elements within each block. 
\item \textbf{Recurrent layer} It is the typical layer of RNNs. After performing an affine combination of the input and passing it through a non-linearity, the output is not just propagated forward, but a feedback loop is also present. 
\end{itemize}
More details on the operation of the different kinds of layers are provided in the rest of this section. 

\subsection{Feedforward Neural Networks}\label{Sec:FNN}
The focus of this section is on FFNs with fully-connected layers, which is the quintessential \gls{ann} architecture. Instead, convolutional layers will be discussed in Section \ref{Sec:CNNs}. 

\begin{figure}\vspace{-0mm}
  \begin{center}
  \includegraphics[scale=0.5]{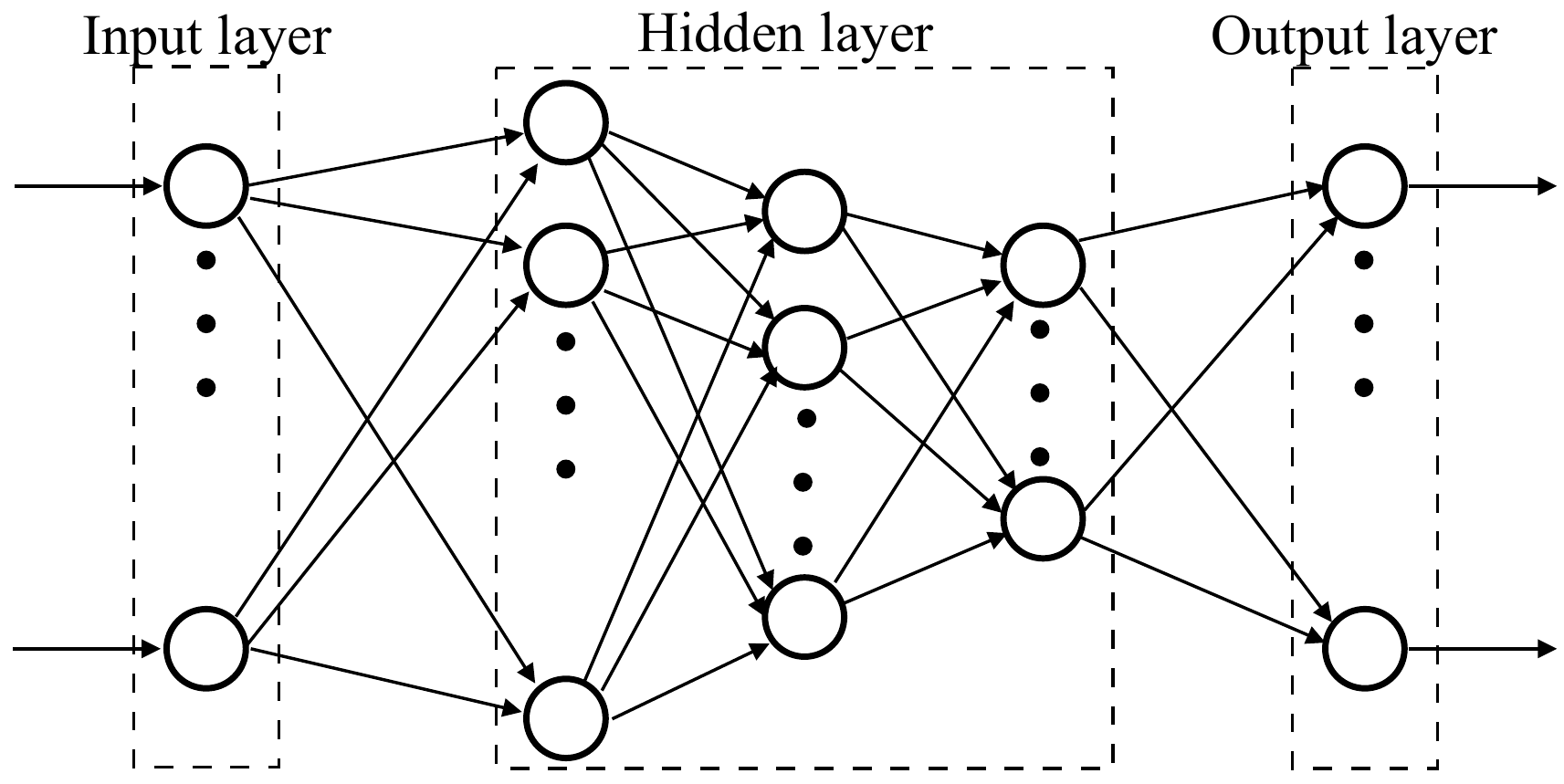}
  \caption{Scheme of a deep \gls{ann} with $L$ hidden layers and $N_{\ell}$ neurons in layer $\ell$, for all $\ell=1,\ldots,L$.}
  \label{Fig:ANN_Scheme}
 \end{center}
\end{figure}

The general structure of a FFN is depicted in Fig. \ref{Fig:ANN_Scheme}. An $N_{0}$-dimensional input vector $\bx_{0}$ is fed to the network through the $N_{0}$ neurons of the input layer.  Afterwards, it passes through $L$ hidden layers, with Layer $\ell$ having $N_{\ell}$ neurons. Finally, the $(N_{L}+1)$-dimensional output is retrieved from the $N_{L}+1$ neurons of the output layer. 
To elaborate, let us denote by $\bx_{\ell-1}$ the input to the $\ell$-th layer of the network. Then, for all $\ell=1,\ldots,L+1$ and $n=1,\ldots,N_{\ell}$, the output $\bx_{\ell}(n)$ of neuron $n$ in layer $\ell$ is obtained as:
\beq\label{Eq:TransferFunction}
\bx_{\ell}(n)=f_{n,\ell}(z_{n,\ell})\;,\;z_{n,\ell}=\bw_{n,\ell}^{T}\bx_{\ell-1}+b_{n,\ell}\;,
\eeq
wherein $\bw_{n,\ell}\in\mathbb{R}^{N_{\ell-1}}$ with $w_{n,\ell}(k)$ being the weight of the link between the $k$-th neuron in layer $\ell-1$ and the $n$-th neuron in layer $\ell$, $b_{n,\ell}\in\mathbb{R}$ is the bias term of neuron $n$ in layer $\ell$, while $f_{n,\ell}$ is the so-called  \textbf{activation function} of neuron $n$ in layer $\ell$. Thus, the processing performed by each neuron can be viewed as a two-step procedure in which first an affine combination of the inputs is computed with weights $\bw_{n,\ell}$ and bias term $b_{n,\ell}$, yielding the intermediate term $z_{n,\ell}$. Then, the final output is obtained by applying the activation function $f_{n,\ell}$ to $z_{n,\ell}$. 

As for the choice of the activation functions, over the years several functions have been considered. The first choice was to use sigmoidal functions
\beq\label{Eq:SigmoidAct}
\sigma(z_{n,\ell})=\frac{1}{1+e^{-z_{n,\ell}}}\;,
\eeq
or hyperbolic tangent functions
\beq\label{Eq:TanhAct}
\text{tanh}(z_{n,\ell})=\frac{e^{z_{n,\ell}}-e^{-z_{n,\ell}}}{e^{z_{n,\ell}}+e^{-z_{n,\ell}}}\;.
\eeq 
The sigmoid function is able to produce feasible probability values, being limited between zero and one, and for this reason nowadays it is typically used as activation function of the output layer for applications that require to estimate a probability. However, its use for the hidden layers is no longer recommended, due to the fact that it saturates for a significant portion of its domain, thus having derivatives very close to zero when the argument is large in modulus. This causes the so-called \emph{vanishing gradient} problem, which slows down the convergence of gradient-based training algorithms. Another way of looking at the problem is to say that sigmoid activation functions are able to learn only when the input is around zero, i.e. in their (approximately) linear region, where the output of the sigmoid function is sensitive to variations of the input. Instead, in other regions of its domain the sigmoid function saturates and the output tends to be approximately constant even in response to significant changes of the input, which does not yield much useful learning information. Similar considerations also apply to the hyperbolic tangent function, which is linked to the sigmoid function by the relation: $\text{tanh}(z_{n,\ell})=2\sigma(2z_{n,\ell})-1$. 

Nowadays, the most widely-used choice for the activation function of the hidden layers is the \gls{relu} function \cite{Jarret2009,Nair2010,Glorot2011}, defined as:
\beq\label{Eq:ReLUAct}
\text{ReLU}(z_{n,\ell})=\max(0,z_{n,\ell})\;.
\eeq
\gls{relu} functions are linear whenever the neuron is active, which makes them easier to optimize. Whenever the neuron produces a non-zero output, the gradient of the activation function is constantly equal to one, and no second-order effects are present. The drawback is that the \gls{relu} function does  not provide any useful learning information when its input is negative. To overcome this issue, some refinements of the \gls{relu} function have introduced a non-zero slope also for negative inputs, considering the function:
\beq\label{Eq:GenReLUAct}
f_{n,\ell}(z_{n,\ell})=\max(0,z_{n,\ell})+c \min(0,z_{n,\ell})\;.
\eeq 
The Leaky \gls{relu} function sets $c=0.01$ as proposed in \cite{Maas2013}; the \emph{absolute value rectification} approach proposed in \cite{Jarret2009} considers $c=-1$, while the parametric \gls{relu} approach proposed in \cite{He2015} treats $c$ as a parameter to be optimized during the training process. 

Another generalization of the ReLU is the exponential linear unit (ELU), which behaves like the ReLU for positive inputs, but outputs 
\beq
f_{n,\ell}(z_{n,\ell})=\alpha (e^{z_{n,\ell}}-1)\;,
\eeq
when the input $x$ is negative, with $\alpha$ a scalar typically set to $1$, \cite{ELU}. 

The properties of the \gls{relu} function and its generalizations seem to lead to the conclusion that the best activation functions are linear functions. In fact, linear activation functions can be used at the output layer to perform specific operations such as computing arithmetic averages. However, their use in the hidden layers is not encouraged, as they might prevent the network from learning non-linear maps. For example, in the extreme case in which all activation functions were linear, the input-output relation of the FNN would reduce to being always linear, when instead one of the strengths of \glspl{ann} lies in their ability to combine multiple non-linearities to emulate virtually any input-output map. This fact was formally established in \cite{Hornik1989}, where it is stated that any deterministic continuous function over a compact set can be approximated arbitrarily well by a single fully-connected layer with enough neurons and sigmoidal activation functions\footnote{The result is proved assuming \emph{squashing activation functions}, which include sigmoid functions as special cases.}. This fundamental result is known as the \textbf{universal approximation theorem} of \glspl{ann} and was later extended to a broader class of activation functions, including the \gls{relu} function and its generalizations \cite{Leshno1993}. Nevertheless, despite its high theoretical importance the universal approximation theorem is not constructive, because:
\begin{itemize}
\item it does not establish the number of neurons that are required in order to obtain the desired level of approximation accuracy.
\item it does not establish whether it is more convenient to use a shallow or deep architecture in order to improve the approximation accuracy or reduce the number of required neurons. 
\item it does not establish how to configure the \gls{ann} in order to obtain the desired approximation accuracy.
\end{itemize}
An answer to the first question was provided in \cite{Barron1993}, which provides bounds for the number of neurons in shallow \glspl{ann} in order to obtain a given approximation accuracy. Unfortunately, the bounds show that, in general, an exponential number of nodes is required. 

As for the second issue, deep architectures seem to require a lower number of neurons, even though a formal proof of this result in a general setting is still an open problem. Nevertheless, some available results prove that certain classes of functions can be represented more efficiently by increasing the network depth, i.e. the number of layers. In \cite{Montufar2014}, for example, it is shown that the number of regions of a piece-wise linear function that can be reliably represented scales exponentially with the number of layers $L$. Moreover, many empirical results have shown that deep architectures provide lower generalization errors than shallow architectures \cite[Sec. 6.4.1]{Bengio2016}. 

Finally, the third issue is perhaps the most problematic. Although the universal approximation theorem ensures that there exists an \gls{fnn} able to learn the desired map, it provides no indication as to how to configure the weights $\bw_{n,\ell}\in\mathbb{R}^{N_{\ell-1}}$ and bias $b_{n,\ell}\in\mathbb{R}$ of each neuron. This shows that configuring the parameters of an \gls{ann} represents the most critical step when employing deep learning. The training process of \glspl{ann} will be addressed in Section \ref{Sec:TrainingANN}.

\subsubsection{Convolutional neural networks}\label{Sec:CNNs}
\glspl{cnn} are FFNs that have established themselves as the main tool for image processing, and, in general, for processing data with a spatial structure. The main ingredient of \glspl{cnn} is the 3D-convolution operation, which amounts to a particular linear processing of the input data. For this reason, \glspl{cnn} can be considered as a sub-category of FFNs.

\begin{figure}
  \begin{center}
  \includegraphics[scale=0.5]{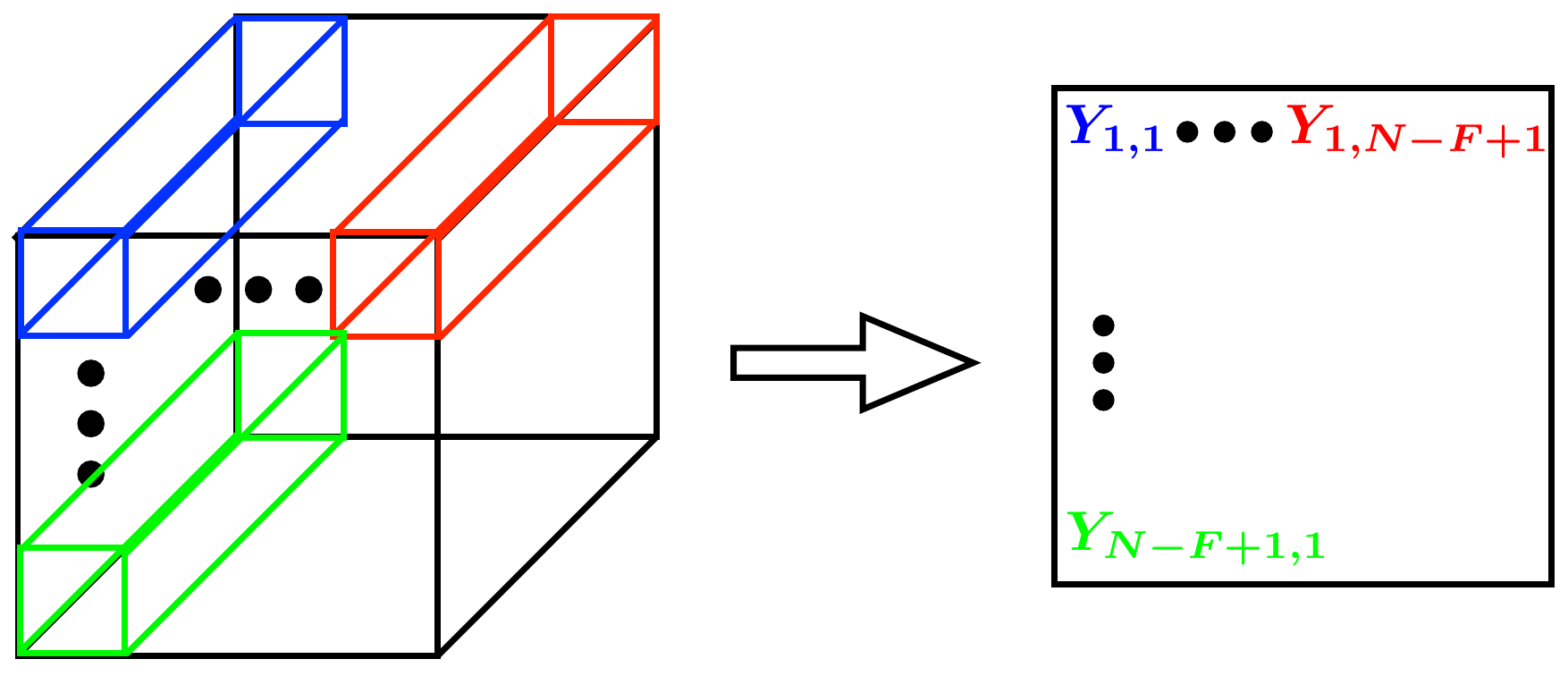}
  \caption{3D-convolution in convolutional neural networks. The input data is arranged in an $N\times N \times N_c$ matrix, which is filtered by a sliding $F\times F\times N_c$ matrix, yielding a $N-F+1\times N-F+1$ output matrix.}
  \label{Fig:3D-Conv}
 \end{center}
\end{figure}

When using a \gls{cnn}, the input data is assumed to be organized in a multi-dimensional matrix $\bX$ with dimensions $N\times  N \times N_{c}$, where the parameter $N_{c}$ is called the number of channels and is typically equal either to $N_{c}=3$ when color images are processed, or to $N_{c}=1$ when black-and-white images are processed. Each node of a convolutional layer is also represented as a multi-dimensional matrix $\bW$ with dimensions $F\times  F \times N_{c}$ (with $F\leq N$) containing the weights of the neuron. The 3D-convolution operation outputs a bi-dimensional matrix $\bY$, with dimensions $N-F+1\times N-F+1$, obtained by sliding the weight matrix over the input matrix, and by  computing each time the cross-correlation between the weight matrix and the corresponding chunk of the input matrix, as depicted in Fig. \ref{Fig:3D-Conv}. 
Mathematically, the $(\ell-m)$-th element of the output matrix $\bY$ is expressed as:
\begin{equation}\label{Eq:3DConv}
\bY_{\ell,m}=\sum_{i=1}^{F}\sum_{j=1}^{F}\sum_{k=1}^{N_{c}}\bW_{i,j,k}\bX_{i+\ell,j+m,k}\;,
\end{equation}
It can be seen that, as already mentioned, each element of the output matrix is obtained through a cross-correlation rather than a convolution, even though the term convolution is universally used in the \gls{ann} jargon to refer to the operation in \eqref{Eq:3DConv}. In the following, we embrace this terminology.  After computing \eqref{Eq:3DConv} for all $\ell$ and $m$, the output of the node is obtained by first summing a scalar bias term $b$ and then applying an activation function to each component of $\bY$, like in a traditional fully-connected layer. Finally, the bi-dimensional output of each node in the layer are stacked together to form a new matrix with dimensions $N-F+1\times N-F+1\times N_{F}$, with $N_{F}$ the number of nodes in the convolutional layer, which is the input of the next layer of the \gls{cnn}.
 
It is interesting to observe that \eqref{Eq:3DConv} can be rewritten as a scalar product similar to a fully-connected layer, upon vectorizing the input and weight matrices. For example, denoting by $\bx$ and $\bw$ the $N^{2}N_{c}\times 1$ and $F^{2}N_{c}\times 1$ vectors obtained by vectorizing $\bX$ and $\bW$, the output element $\bY_{1,1}$ can be obtained as
\beq
\bY_{1,1}=\bx^{T}\widetilde{\bw}\;,
\eeq
wherein $\widetilde{\bw}=[\bw \; \bzero_{(N^{2}-F^{2})N_{c}}]$. All other elements of $\bY$ can be obtained in a similarly way, upon considering suitably zero-padded version of $\bw$. As a result, each node of a convolutional layer is equivalent to $(N-F+1)^{2}$ nodes of a fully-connected layer, in which the weights of many connections are permanently set to zero. This sparsity of the connections is one of the major strengths of \glspl{cnn}, since it enables to process very large data using a relatively small number of parameters, which helps avoid overfitting. On the other hand, the underlying assumption that justifies the use of \glspl{cnn} is the presence of strong spatial correlations in the input. Only if this is fulfilled, as is in image processing, it is possible to apply the same filter to different parts of the input matrix, thus avoiding unnecessary connections among the neurons. 

\begin{figure}
  \begin{center}
  \includegraphics[scale=0.4]{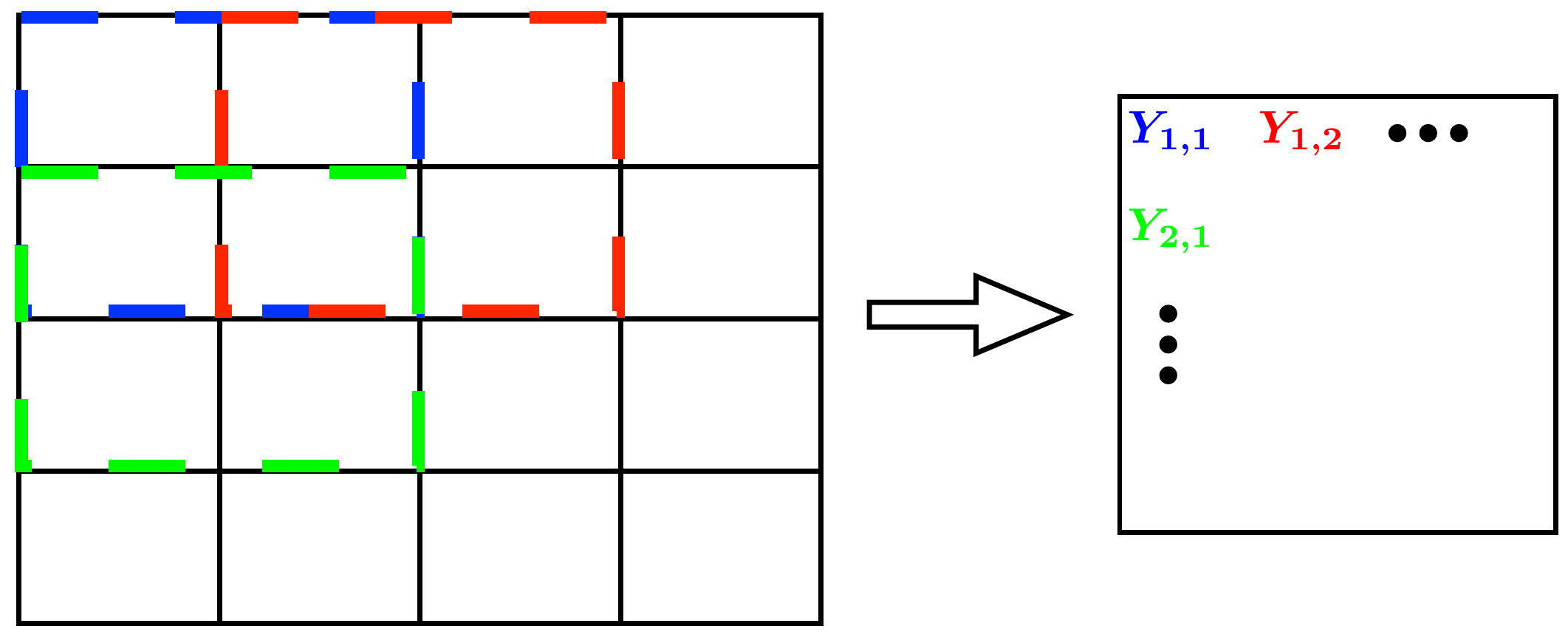}
  \caption{Pooling with $S=1$ of a single channel of a $4\times 4\times N_{c}$ input by a $2\times 2$ filter. From each $2\times 2$ sub-matrix of the input, either the maximum element or the average are computed.}
  \label{Fig:Pooling}
 \end{center}
\end{figure}

The operation defined in \eqref{Eq:3DConv} is the normal convolution employed in \glspl{cnn}. In some cases, it can be slightly modified by applying \textbf{padding} and \textbf{stride}.
\begin{itemize}
\item \textbf{Padding.} When computing \eqref{Eq:3DConv}, the components at the border of the input matrix $\bX$ are used less frequently than the components in the middle. In order to avoid this, it is possible to apply \eqref{Eq:3DConv} to a zero-padded version of $\bX$, in which $P$ rows and columns of zeros are appended to $\bX$. Then, the resulting zero-padded input matrix has dimensions $N+2P\times N+2P$, and the output matrix has dimensions $(N+2P-F+1)\times (N+2P-F+1)$. If $F$ is odd, 
choosing 
\begin{equation}
P=(F-1)/2\;,
\end{equation}
yields an output with the same dimensions as the input. 
\item \textbf{Stride.} The convolution operation in \eqref{Eq:3DConv} slides the weight matrix $\bW$ over the input matrix moving by one position at each step. This can be generalized by sliding the weight matrix by $S$ positions at each step, where $S$ is called the stride parameter. In this case, assuming a padding $P$ is used as well, the output matrix will have dimensions:
\beq
\left\lfloor \frac{N+2P-F}{S}+1 \right\rfloor \times \left\lfloor \frac{N+2P-F}{S}+1 \right\rfloor\;.
\eeq
\end{itemize}
While the convolution operation is the defining feature of \glspl{cnn}, another widely used operation in a \gls{cnn} is the \textbf{Pooling}. Unlike the convolution, which is individually performed by each neuron of a layer before the different bi-dimensional matrices are combined together, the pooling is performed at the layer level and operates separately on each channel of the input matrix $\bX$. Two types of pooling are commonly used:
\begin{itemize}
\item\textbf{Max Pooling.} For each channel of the input matrix $\bX$, say $\bX_{n_{c}}=\bX(:,:,n_{c})$, a max pooling layer with parameter $F$ selects the maximum element out of each $F\times F$ sub-matrix of $\bX_{n_{c}}$.
\item\textbf{Average Pooling.} For each channel of the input matrix $\bX$, say $\bX_{n_{c}}=\bX(:,:,n_{c})$, an average pooling layer with parameter $F$ computes the arithmetic average of each $F\times F$ sub-matrix of $\bX_{n_{c}}$.
\end{itemize}
In both cases, a stride $S$ can also be used, which implies that the sliding window over which the maximum or average are computed moves by $S$ positions each time. An example of pooling with $S=1$ is shown in Fig.  \ref{Fig:Pooling}.

As a final remark before concluding this section, it is worth mentioning that practical FFNs are composed of a mixture of convolutional, pooling, and fully-connected layers, normally performing convolutions and pooling in the first layers, thus decreasing the size of the data, and employing fully-connected layers at the end once the dimension of the data is more manageable.

\subsection{Recurrent neural networks}\label{Sec:RNNs}
If \glspl{cnn} are more suited to processing data exhibiting spatial correlations, RNNs are designed to work on temporal sequences of data with correlated samples. As already anticipated, the main difference compared to FFNs is that the information does not only propagate forward, but loops are allowed. More in detail, each layer of a RNN may receive as input its own activation value. To elaborate, using a similar notation as in Section \ref{Sec:FNN}, the output $\bx_{\ell}^{[t]}(n)$ of neuron $n$ in layer $\ell$ at time $t$ is obtained as:  
\begin{align}\label{Eq:TransferFunctionRNN}
\ba_{\ell}^{[t]}(n)&=f_{n,\ell}(\bw_{n,\ell}^{T}\bx_{\ell-1}^{[t]}+\tilde{\bw}_{n,\ell}^{T}\ba_{\ell}^{[t-1]}+b_{n,\ell})\\
\bx_{n,\ell}^{[t]}&=g_{n,\ell}(\bar{\bw}_{n,\ell}^{T}\ba_{\ell}^{[t]}+\bar{b}_{n,\ell})\;,
\end{align}
wherein $f_{n,\ell}$ and $g_{n,\ell}$ are neuron-dependent activation functions. Thus, each neuron in a recurrent layer combines with different weights not only the current input, but also the intermediate vector $\ba_{\ell}$ that is obtained in the previous step. This introduces a correlation among the different computations that is beneficial to exploit the temporal correlations hidden in the input sequence. Moreover, a recurrent layer has two activation functions, $f$ and $g$. Popular choices here are to use the hyperbolic tangent or the ReLU for $f$ and the sigmoid function for $g$. 

The architecture described above is the general architecture of recurrent layers. Several variants exist that are commonly used in real-world RNNs. In addition, we stress that, typically, a deep RNN has just a few recurrent layers, and it is possible to have hybrid architectures composed of some initial recurrent layers, followed by feed-forward layers. More details on specific RNNs architectures can be found in specialized references on \glspl{ann}, like \cite{Bengio2016}. 

\subsection{Training Neural Networks}\label{Sec:TrainingANN}
For ease of notation, and without loss of generality, this section focuses on FFNs with fully connected layers. Results directly apply to \glspl{cnn} and can be extended to RNNs with minor modifications. Training a neural network is the process that tunes the parameter $\bw_{n,\ell}\in\mathbb{R}^{N_{\ell-1}}$ and $b_{n,\ell}\in\mathbb{R}$ in a supervised learning fashion in order for the \gls{fnn} to learn the desired input-output relation. To elaborate, let us consider a training set composed of $N_{TR}$ input samples with the corresponding desired output, namely 
\beq\label{Eq:TrainSetFNN}
{\cal S}_{TR}=\left\{\left(\bx_{0}^{(1)},\bx_{L+1}^{(1)}\right),\ldots,\left(\bx_{0}^{(N_{TR})},\bx_{L+1}^{(N_{TR})}\right)\right\}\;.
\eeq
For each layer $\ell=1,\ldots,L+1$, let us stack the weight vectors into the $N_{\ell-1}\times N_{\ell}$ matrix $\bW_{\ell}$ and the bias terms into the $N_{\ell}\times 1$ vector $\bb_{\ell}$, respectively defined as $\bW_{\ell}=\left[\bw_{1,\ell},\ldots,\bw_{N_{\ell},\ell}\right]$ and $\bb_{\ell}=\left[b_{1,\ell},\ldots,b_{N_{\ell},\ell}\right]^{T}$.
The \emph{actual} output of the \gls{fnn} when the input is the $nt$-th training sample $\bx_{0}^{(nt)}$  depends on the network weights and bias terms, and is denoted as:
\beq\label{Eq:ActualOutput}
\widehat{\bx}_{L+1}^{(nt)}\left(\left\{\bW_{\ell},\bb_{\ell}\right\}_{\ell=1}^{L}\right)\;,\;\forall\; nt=1,\ldots,N_{TR}\;.
\eeq
The goal of the training algorithm is to optimize the \gls{ann} weights and bias terms in order to minimize the loss incurred between the actual output $\widehat{\bx}_{L+1}^{(nt)}$ in \eqref{Eq:ActualOutput}, and the desired output $\bx_{L+1}^{(nt)}$ defined by the training set in \eqref{Eq:TrainSetFNN}, for all $nt=1,\ldots,N_{TR}$, as quantified by the \textbf{loss function} 
\beq\label{Eq:AverageLoss}
L\!\left(\!\left\{\bW_{\ell},\bb_{\ell}\right\}_{\ell=1}^{L}\!\right)\!=\!\frac{1}{N_{TR}}\!\!\sum_{nt=1}^{N_{TR}}\!\!{\cal L}\left(\bx_{L+1}^{(nt)},\widehat{\bx}_{L+1}^{(nt)}\!\left(\left\{\bW_{\ell},\bb_{\ell}\right\}_{\ell=1}^{L}\!\right)\!\right)\!,
\eeq
wherein ${\cal L}(\bx_{L+1}^{(nt)},\widehat{\bx}_{L+1}^{(nt)})$ is a loss function that models the error  between $\widehat{\bx}_{L+1}^{(nt)}$ and the desired output $\bx_{L+1}^{(nt)}$. A natural and common choice for the loss function is the MSE, namely:
\beq\label{Eq:MSE}
{\cal L}(\bx,\widehat{\bx})=\text{MSE}(\bx,\widehat{\bx})=\sum_{i=1}^{N_{\ell+1}}(\bx(i)-\widehat{\bx}(i))^{2}\;.
\eeq
The MSE has the advantage of being applicable to virtually any scenario, and enables a simple computation of its derivatives. However, in some cases it can slow down the learning algorithm. Instead, faster convergence of the learning algorithm is typically observed by using the cross-entropy loss function, defined as
\beq\label{Eq:CrossE}
{\cal L}(\bx,\widehat{\bx})\!\!=\!H(\bx,\widehat{\bx})\!\!=\!-\!\!\!\sum_{i=1}^{N_{\ell+1}}\!\!\bx(i)\!\log(\widehat{\bx}(i))+(1\!-\bx(i))\!\log(1\!-\widehat{\bx}(i)).
\eeq
However, the applicability of \eqref{Eq:CrossE} is not so wide as that of the MSE function. Indeed, clearly  \eqref{Eq:CrossE} applies only to those cases in which both the desired and actual output data belong to the interval $[0,1]$, and thus can be interpreted as distributions of random variables. A notable case in which this holds true is when sigmoid activation functions are used in the output layer, aiming at estimating a probability distribution. Assuming that both $\bx$ and $\widehat{\bx}$ have entries in $[0,1]$, the cross entropy in \eqref{Eq:CrossE} represents a measure of the divergence between $\bx$ and $\widehat{\bx}$, since the cross entropy of two distributions $p$ and $q$ is equal to the Kullbach-Leibler divergence between $p$ and $q$ plus the entropy of $p$ \cite{cover2006elements}. Applying this result, \eqref{Eq:CrossE} can be rewritten as
\begin{align}
H(\bx,\widehat{\bx})\!&=\!-\!\!\!\sum_{i=1}^{N_{\ell+1}}\!\!\bx(i)\!\log(\widehat{\bx}(i))\!+\!(1\!-\!\bx(i))\!\log(1\!-\!\widehat{\bx}(i))\notag\\
&=-\sum_{i=1}^{N_{\ell+1}}\bx(i)\log\!\left(\frac{\widehat{\bx}(i)}{\bx(i)}\!\right)\!+\!(1\!-\!\bx(i))\log\!\left(\frac{1\!-\!\widehat{\bx}(i)}{1\!-\!\bx(i)}\!\right)\notag\\
&=-\sum_{i=1}^{N_{\ell+1}}\bx(i)\log(\bx(i))+(1-\bx(i))\log(1-\bx(i))\notag\\
&=\sum_{i=1}^{N_{\ell+1}}KL(\bx(i),\widehat{\bx}(i))+H_{b}(\bx(i))\;,
\end{align} 
with $KL(\cdot,\cdot)$ and $H_{b}(\cdot)$ denoting the Kullbach-Leibler divergence and binary entropy, respectively.  Then, since $H_{b}(\bx)$ does not depend on the network parameters, minimizing the cross-entropy in \eqref{Eq:CrossE} is equivalent to minimizing the Kullbach-Leibler divergence between the desired and actual outputs.

In any case, regardless of the loss function that is chosen, the training process mathematically amounts to solving the optimization problem\footnote{In case of RRNs, an additional sum over the time dimension is present to account for the loss over time of each training sample.}
\begin{subequations}\label{Prob:Training}
\begin{align}
&\ds \min\;\frac{1}{N_{TR}}\sum_{nt=1}^{N_{TR}}{\cal L}\left(\bx_{L+1}^{(nt)},\widehat{\bx}_{L+1}^{(nt)}\left(\bW,\bb\right)\right)\label{Prob:aTraining}\\
&\;\textrm{s.t.}\; \bW_{\ell}\in\mathbb{R}^{N_{\ell-1}\times N_{\ell}}\;,\;\forall\;\ell=1,\ldots,L+1\\
&\quad\;\;\;\bb_{\ell}\in\mathbb{R}^{N_{\ell}\times 1}\;,\;\forall\;\ell=1,\ldots,L+1\;,
\end{align}
\end{subequations}
wherein $\bW=\left\{\bW_{\ell}\right\}_{\ell=1}^{L}$, $\bb=\left\{\bb_{\ell}\right\}_{\ell=1}^{L}$. However, as mentioned in previous sections, the goal of deep learning is not so much to minimize the cost function in \eqref{Prob:Training}, i.e. the training error, but rather to ensure a low generalization gap. Tuning the parameters of the network to achieve a low training error is a prerequisite to achieving a low test error, but an equally important task is that of tuning the network hyperparameters, (e.g. the number of layers $L$, the number of neurons per layer $N_{\ell}$, the size of the training set $N_{TR}$), to fit the training data, avoiding both underfitting and overfitting. The coming Section \ref{Sec:Parameters} discusses the design of suitable algorithms to tackle \eqref{Prob:Training} in an efficient and effective way, while Section \ref{Sec:Hyperparameters} provides some guidelines for hyperparameter tuning in \glspl{fnn}. 

\subsubsection{\textbf{Parameter tuning - Tackling \eqref{Prob:Training}}}\label{Sec:Parameters}
Traditionally, in optimization theory, convexity is the critical property that marks the watershed between problems that can be solved with affordable complexity, and problems that require an unfeasible complexity. A convex problem, defined as a problem whose objective and constraint functions are convex in the optimization variables \cite{boyd2004convex,BertsekasNonLinear,TalNemi2001}, enjoys several useful properties, among which the following two have played a critical role in enabling the development of a consolidated theory of convex optimization, and practical algorithms with theoretical optimality guarantees:
\begin{itemize}
\item \textbf{[P.1]:} Every stationary point of a convex function is a global minimum, i.e. the minimization of a convex function can be performed by simply looking for a point where the gradient of the function vanishes. This property establishes that first-order optimality conditions are necessary and sufficient for convex functions. 
\item \textbf{[P.2]:} For any $\varepsilon>0$, the complexity required to find an $\varepsilon$-optimal solution of a generic convex problem with $n$ variables scales, in the worst case, as the fourth power of $n$ and as $\log\left(\frac{1}{\varepsilon}\right)$ \cite[Section 5]{TalNemi2001}. This property establishes that convex problems can be solved with polynomial complexity in the number of variables.
\end{itemize}
Unfortunately, neither of the two properties above holds for Problem \eqref{Prob:Training} because the objective function is not convex with respect to the optimization variables, due to the presence of multiple layers combining several non-linear activation functions. This implies that the cost function of Problem \eqref{Prob:Training} might have stationary points that are either local minima, or local maxima, or saddle points, a circumstance that becomes more and more likely as the dimensionality of the problem increases. In fact, it is quite typical for fairly deep model to have a very large number of points where the gradient vanishes, but that are not global minima. Moreover, the complexity required in order to find the global solution of Problem \eqref{Prob:Training} is not guaranteed to be polynomial, since it scales in general exponentially with the number of variables, which is equal to $\sum_{\ell=1}^{L+1}N_{\ell}(N_{\ell-1}+1)$. As a result, finding the global solution of Problem \eqref{Prob:Training} turns out to be a very challenging task, especially considering that realistic \glspl{ann} have a fairly large number of neurons and layers.

Based on these considerations, it might seem hopeless to perform an effective and efficient training of any reasonably-sized \glspl{fnn}. Fortunately, this is not the case and several efficient algorithms to effectively train \glspl{fnn} exist. To understand why the non-convexity of \eqref{Prob:Training} does not pose a fundamental problem, one must recall that, although the training process amounts to solving an optimization problem, machine learning differs from pure optimization theory, in that the ultimate goal is not so much to minimize the training error, but rather to minimize the generalization error. As discussed in Section \ref{Sec:Basics}, the training error lower bounds the generalization error, but there is no guarantee that a lower training error also results in a lower generalization error. Actually, aiming for a very low training error typically causes overfitting. Therefore, when tackling Problem \eqref{Prob:Training}, it is surely desirable to find a configuration of parameters that yields a low training error, so as to avoid underfitting, but it is also not necessary to pursue the global minimization of the training error, which would most likely lead to overfitting. Any training algorithm will aim at progressively reducing the training error, stopping as soon as the generalization error evaluated over the validation set is below a desired threshold, regardless of the value of the training error. It is not uncommon that a training algorithm stops when the training error is relatively large compared to its global minimum. 

As a result, the presence of stationary points of the cost function of Problem \eqref{Prob:Training} would be a major issue only if the training algorithm were likely to converge to a suboptimal point yielding a too high training error, thus causing underfitting. A definitive formal proof that this does not occur in practice is still an open research problem, but extensive experimental evidence has shown that, for \glspl{ann} with a sufficient amount of neurons, most local minima lead to a satisfactory training error \cite{Saxe2013,Dauphin2014,Goodfellow2015,Choromanska2015}. In addition, especially in higher-dimensional spaces, local minima and local maxima of random functions are much less frequent compared to saddle points \cite{Dauphin2014}. This phenomenon has been proved for some specific shallow \glspl{ann}  \cite{Baldi1989}, while some theoretical arguments as well as experimental evidence that a similar behavior holds also in deep \glspl{ann} is provided in \cite{Saxe2013,Dauphin2014,Choromanska2015}. Therefore, the main issue related to the non-convexity of Problem \eqref{Prob:Training} is not mainly related to local minima, but rather to the presence of saddle-points. In this respect, empirical evidence provided in \cite{Goodfellow2015} shows that first-order methods based on gradient descent are able to escape saddle points. This behavior can be theoretically justified by observing that gradient-based methods are not explicitly designed to find point with zero gradient. Rather, they are designed to reduce the cost function moving in the direction of maximum decrease which is pointed by the gradient. Of course, this implies that the algorithm stops if a point with rigorously zero gradient is reached, but it makes the algorithm capable of moving away from the neighborhood of a saddle point even for relatively small step-sizes. On the other hand, second-order methods like Newton's method do not share this property, having a higher probability of being stuck around saddle points. A training algorithm based on an approximate Newton's method with a regularization strategy is the Levenberg-Marquardt method \cite{Levenberg1944,Marquardt1963}, which yields good performance as long as the negative eigenvalues of the Hessian of the cost function are relatively close to zero. Instead, a recent modification of Newton's method, designed to be more robust to the saddle-point problem in \glspl{fnn}, has been introduced in \cite{Dauphin2014}. Despite enjoying stronger convergence properties in the convex case, at present the use of second-order methods to tackle the non-convex Problem \eqref{Prob:Training} is not so well-established as the use of first-order methods based on gradient descent algorithms. For this reason, the rest of this section is focused on presenting the main first-order training methods for \glspl{fnn}.

\textbf{Backpropagation algorithm.} The first problem that we encounter towards the implementation of a  gradient-based training algorithm for \glspl{fnn} is the complexity related to the computation of the gradient. In large \glspl{ann} with many neurons and large training sets, the direct computation of the derivatives of the training error in \eqref{Prob:aTraining} with respect to all network weights and bias terms would require an unmanageable complexity. Luckily, a fast algorithm to compute the gradient of the training error was developed in \cite{Rumelhart1986}. It makes a clever use of the chain rule from multivariable calculus, and was called backpropagation algorithm, for reasons that will become clear after describing its working operation. 

To begin with, let us observe that the derivative of \eqref{Prob:aTraining} is written as the average of the derivatives of the loss function ${\cal L}(\bx_{L+1},\hat{\bx}_{L+1}(\bW,\bb))$ over the training set. In fact, the backpropagation algorithm provides a way of computing the derivatives of ${\cal L}(\bx_{L+1},\hat{\bx}_{L+1}(\bW,\bb))$. Specifically, given a training input sample $\bx_{0}$, the first step of the backpropagation algorithm is to compute the corresponding actual output $\hat{\bx}_{L+1}(\bW,\bb)$. This step is referred to as \emph{forward propagation} because it propagates the input forward through the network, by computing \eqref{Eq:TransferFunction} for all $n$ and $\ell$. 

After completing the forward propagation, the derivative of the cost function with respect to $z_{n,L+1}$ can be computed as 
\beq\label{Eq:Backprop1}
\frac{\partial \cal {L}}{\partial z_{n,L+1}}=\frac{\partial \cal {L}}{\partial \bx_{L+1}(n)}f_{n,L+1}^{'}(z_{n,L+1})\;,\;\forall\;n=1,\ldots,N_{L+1}
\eeq
The next step consists of computing the derivatives of the loss function with respect to $z_{n,\ell}$, for all $\ell=L,L-1,\ldots,1$, in a recursive way. This is the step that gives the name to the algorithm, since the derivatives are computed backwards, proceeding from the last to the first layer. Specifically, it holds\footnote{Recall that the derivative with respect to $x$ of the function $g(\by(x))$, with $\by(x)=[y_{1}(x),\ldots,y_{I}(x)]$, is given by $\sum_{i=1}^{I}(\nabla_{y} g)^{T}J_{x}\by$, where $J_{x}$ denotes the Jacobian operator with respect to $x$.}
\begin{align}
\frac{\partial {\cal L}}{\partial z_{n,\ell}}&=\sum_{k=1}^{N_{\ell}+1}\frac{\partial {\cal L}}{\partial z_{k,\ell+1}}\frac{\partial z_{k,\ell+1}}{\partial z_{n,\ell}}\notag\\
\label{Eq:Backprop2}
&=\sum_{k=1}^{N_{\ell}+1}\frac{\partial {\cal L}}{\partial z_{k,\ell+1}}w_{k,\ell+1}(n)f^{'}(z_{n,\ell})\;,
\end{align}
which can be easily computed based on the derivatives with respect to $z_{k,\ell+1}$, $k=1,\ldots,N_{\ell+1}$ obtained from Layer $\ell+1$. Finally, based on \eqref{Eq:Backprop2} and recalling \eqref{Eq:TransferFunction}, the derivatives with respect to the weights and bias terms are readily obtained as:
\begin{align}\label{Eq:BackProp3}
\frac{\partial {\cal L}}{\partial w_{n,\ell}(k)}&=\frac{\partial {\cal L}}{\partial z_{n,\ell}}\bx_{\ell-1}(k)\;,\\
\label{Eq:BackProp4}
\frac{\partial {\cal L}}{\partial b_{n,\ell}}&=\frac{\partial {\cal L}}{\partial z_{n,\ell}}\;.
\end{align}
Thus, the backpropagation procedure can be stated as in Algorithm \ref{Alg:BackProp}.
\begin{algorithm}
\begin{algorithmic}\caption{Backpropagation Algorithm.}
\label{Alg:BackProp}
\For{$nt=1\to N_{TR}$}
\State \texttt{Training input $\bx_{0}^{(nt)}$ with desired output $\bx_{L+1}^{(nt)}$};
\State \textbf{Forward Propagation: }\texttt{Compute the actual output $\widehat{\bx}_{L+1}^{(nt)}$ by \eqref{Eq:TransferFunction} for all $\ell=1,\ldots,L+1$};
\State \textbf{Backward Propagation: }\texttt{Compute $\frac{\partial {\cal L}}{\partial z_{n,\ell}}$ by \eqref{Eq:Backprop1} and \eqref{Eq:Backprop2} for all $\ell=L+1,\ldots,1$};
\State \texttt{Compute \eqref{Eq:BackProp3} and \eqref{Eq:BackProp4} for every weight $w_{n,\ell}(k)$  and bias term $b_{n,\ell}$};
\EndFor  
\State $\nabla_{\bW} L(\bW,\bb)=\frac{1}{N_{TR}}\sum_{nt=1}^{N_{TR}}\nabla_{\bW}{\cal L}(\bx_{L+1}^{(nt)},\widehat{\bx}_{L+1}^{(nt)}(\bW,\bb))$;
\State $\nabla_{\bb} L(\bW,\bb)=\frac{1}{N_{TR}}\sum_{nt=1}^{N_{TR}}\nabla_{\bb}{\cal L}(\bx_{L+1}^{(nt)},\widehat{\bx}_{L+1}^{(nt)}(\bW,\bb))$;
\end{algorithmic}
\end{algorithm}

Its strength lies in exploiting the recursive structure of the derivatives to compute, which enables to obtain them by simply computing a forward pass through the network, plus the corresponding backward pass, that has a similar complexity as the forward pass. In contrast to the backpropagation algorithm, the direct computation of the derivatives requires the evaluation of the loss function for each derivative to compute, thus having to perform a number of forward passes equal to the number of weights and bias in the \gls{ann}, which, for large networks, leads to an unfeasible computational complexity.

\textbf{Stochastic Gradient Descent.} While the backpropagation algorithm is computationally more convenient compared to the direct computation of the derivative, its complexity scales with the size of the training set. In order to implement Algorithm \ref{Alg:BackProp}, one must forward-propagate and backward-propagate all $N_{TR}$ samples of the training set. This poses a complexity issue since typically large training sets are used by \glspl{ann}. In more general terms, any algorithm that tried to compute the \emph{true} gradient of the loss function of Problem \eqref{Prob:Training}, i.e.
\begin{align}\label{Eq:DetGrad}
\nabla L\!\left(\bW,\bb\right)=\frac{1}{N_{TR}}\sum_{nt=1}^{N_{TR}}\nabla{\cal L}\left(\bx_{L+1}^{(nt)},\widehat{\bx}_{L+1}^{(nt)}\left(\bW,\bb\right)\right)\;,
\end{align}
would have a complexity proportional to $N_{TR}$. To address this issue, state-of-the-art training algorithms for \glspl{fnn} employ a variant of the gradient descent algorithm known as \gls{sgd} \cite{Bottou1998}. While the standard (or deterministic) implementation of the gradient descent requires computing \eqref{Eq:DetGrad}, the stochastic variant of the gradient descent algorithm computes an estimate of \eqref{Eq:DetGrad} based on a randomly-selected subset of the entire training set, called mini-batch. More precisely, denoting by ${\cal S}_{SGD}$ the set of indexes associated to the selected mini-batch, and by $N_{S}$ the cardinality of ${\cal S}_{SGD}$, an estimate of the gradient is given by:
\begin{align}\label{Eq:DetStoGrad}
&\widehat{\nabla L}\left(\bW,\bb\right)=\frac{1}{N_{S}}\sum_{nt\in{\cal S}_{SGD}}\!\!\!\!\nabla{\cal L}\left(\bx_{L+1}^{(nt)},\widehat{\bx}_{L+1}^{(nt)}\left(\bW,\bb\right)\right)\;.
\end{align}
Each time a gradient descent step is taken, the estimated gradient in \eqref{Eq:DetStoGrad} is evaluated based on a new, randomly selected set ${\cal S}_{SGD}$, and is used in place of the true gradient. The overall procedure is provided in Algorithm \ref{Alg:SGD}.
\begin{algorithm}
\begin{algorithmic}\caption{Stochastic Gradient Descent for \glspl{fnn} training.}
\label{Alg:SGD}
\State \texttt{Set } $\varepsilon>0$, $\bW,\bb$;
\While{\text{Validation Error larger than $\varepsilon$}}
\State \texttt{Sample a random mini-batch ${\cal S}_{SGD}$};
\State \texttt{Compute \eqref{Eq:DetStoGrad} by Algorithm \ref{Alg:BackProp}};
\State $\bW=\bW-\alpha\widehat{\nabla L}(\bW,\bb)$;
\State $\bb=\bb-\alpha\widehat{\nabla L}(\bW,\bb)$;
\EndWhile  
\end{algorithmic}
\end{algorithm}

In Algorithm \ref{Alg:SGD}, $\alpha$ is usually referred to as the \textbf{learning rate} in the machine learning context, and it controls how fast the algorithm reduces the cost function, and thus learns. The learning rate is a key parameter of the \gls{sgd} algorithm and must be carefully selected. While  traditional gradient descent algorithms can use a fixed $\alpha$ and converge as long as $\alpha$ is not too large, the \gls{sgd} uses a variable $\alpha_{k}$ to be used in  iteration $k$, due to the inherent deviation of \eqref{Eq:DetStoGrad} from the true gradient. More formally, a sufficient condition for the convergence of Algorithm \ref{Alg:SGD} is:
\begin{align}
\sum_{k=1}^{\infty}\alpha_{k}=\infty\;,\;\sum_{k=1}^{\infty}\alpha_{k}^{2}<\infty
\end{align}
A common approach is to update $\alpha_{k}$ for the first $t$ iterations according to the formulas:
\begin{equation}
\alpha_{k}=\left(1-\frac{k}{t}\right)\alpha_{0}+\frac{k}{t}\alpha_{t}\;,\;\text{for}\;k\leq t\;,
\end{equation}
while keeping $\alpha$ constant after the $t$-th iteration. Typically, $\alpha_{t}$ should be roughly one hundredth of $\alpha_{0}$, but in practice the parameters $t$, $\alpha_{t}$, and $\alpha_{0}$ are typically chosen by trial and error methods that monitor the error obtained over the validation set for different configurations of parameters. 

\begin{remark}\label{Rem:SGD_1}
The computational complexity of \gls{sgd} depends on the size $N_{S}$ of the mini-batches. If $N_{S}=N_{TR}$ the algorithm reduces to standard gradient descent, also called deterministic or batch gradient descent. Instead, if $N_{S}=1$, the algorithm is referred to as online gradient descent. Typically, \gls{sgd} uses $1<N_{S}<N_{TR}$ and the choice is also dictated by the particular hardware where the algorithm runs, since too low values of $N_{S}$ may underutilize modern multi-core architectures. Also, some architectures, e.g. \glspl{gpu} are more efficient when $N_{S}=2^{n}$, with $n$ an integer number. 
\end{remark}
\begin{remark}\label{Rem:SGD_2}
Since the \gls{sgd} operates based only on an estimate of the true gradient, it typically requires more iterations than its deterministic counterpart to converge. However, each iteration is computationally much faster and the total number of computations required to reach convergence is much lower compared with  the deterministic gradient descent method. In particular, \gls{sgd} has a complexity per update that does not scale with the total size of the training set $N_{TR}$, since it might converge also without having to pass through the entire training set. On the other hand, typically several passes through the training set, called \emph{epochs}, are required to achieve satisfactory training results.
\end{remark}

\textbf{Momentum for Stochastic Gradient Descent.} A drawback of \gls{sgd} is that learning can be sometimes slow due to the fact that only an estimate of the gradient is computed in each iteration. The method of momentum is a general strategy in optimization theory \cite{Polyak1964},  that can be used to accelerate the learning process. The basic idea of the momentum algorithm is to perform the gradient update by an exponentially decaying moving average, as stated in Algorithm \ref{Alg:Momentum}.
\begin{algorithm}
\begin{algorithmic}\caption{Stochastic Gradient Descent with Momentum for \glspl{fnn} training.}
\label{Alg:Momentum}
\State \texttt{Set } $\varepsilon>0$, $\bv\succeq \bzero$, $\bW,\bb$;
\While{\text{Validation Error larger than $\varepsilon$}}
\State \texttt{Sample a random mini-batch ${\cal S}_{SGD}$};
\State \texttt{Compute \eqref{Eq:DetStoGrad} by Algorithm \ref{Alg:BackProp}};
\State $\bv=\delta\bv-\alpha \widehat{\nabla L}(\bW,\bb)$;
\State $\bW=\bW+\bv$;
\State $\bb=\bb+\bv$;
\EndWhile  
\end{algorithmic}
\end{algorithm}

Algorithm \ref{Alg:Momentum} introduces the new parameter $\bv$, which is called \emph{velocity}, in analogy with the fact that it controls the velocity with which the updates move through the parameter space. Due to the presence of the velocity term and to the exponential average of multiple gradient points, 
the magnitude of the step depends on the magnitude of the sequence of gradients, and also on how aligned these gradients are. This tends to smooth out the oscillations of the standard \gls{sgd} algorithm. The velocity $\bv$ represents the cumulative effect of the past gradients, while the term $\delta$ weighs the relative importance of the current gradient with respect to the cumulated gradient. The larger $\delta\in[0,1)$ is with respect to $\alpha$, the more the past gradients affect the direction of the update. If all the gradients of the sequence were equal to $\widehat{\nabla \bar{L}}$, the updates would accelerate in the direction of the common negative gradient until reaching a limit velocity  
\beq
\bv_{\infty}=\frac{\varepsilon\|\widehat{\nabla \bar{L}}\|}{1-\delta}\;.
\eeq
Thus, the parameter $\delta$ determines the relative speed of the updates compared to the \gls{sgd} method without momentum. Common values of $\delta$ are $0.5$, $0.9$, and $0.99$, and it is also desirable to adapt $\delta$ as well as $\alpha$ iteration after iteration, similarly to what is done for the basic \gls{sgd} method.

\textbf{Nesterov Momentum for Stochastic Gradient Descent.} A variant of the momentum for \gls{sgd} appeared in \cite{Sutskever2013}. Following the approach of Nesterov's gradient method \cite{Nesterov2004}, the idea is to compute an estimate of the gradient taking into account the velocity term, as shown in Algorithm \ref{Alg:NesterovMomentum}.

\begin{algorithm}
\begin{algorithmic}\caption{Stochastic Gradient Descent with Nesterov's Momentum for \glspl{fnn} training.}
\label{Alg:NesterovMomentum}
\State \texttt{Set } $\varepsilon>0$, $\bv\succeq \bzero$, $\bW,\bb$;
\While{\text{Validation Error larger than $\varepsilon$}}
\State \texttt{Sample a random mini-batch ${\cal S}_{SGD}$};
\State \texttt{Compute \eqref{Eq:DetStoGrad} evaluated at $\bW+\delta\bv$ and $\bb+\delta\bv$ by Algorithm \ref{Alg:BackProp}};
\State $\bv=\delta\bv-\alpha \widehat{\nabla L}(\bW+\delta\bv,\bb+\delta\bv)$;
\State $\bW=\bW+\bv$;
\State $\bb=\bb+\bv$;
\EndWhile  
\end{algorithmic}
\end{algorithm}

Nesterov's momentum enjoys several convenient properties when applied to convex functions, such as a quadratic convergence rate. However, these advantages are not guaranteed to hold in non-convex scenarios, which is the usual case when training \glspl{fnn}.

\textbf{AdaGrad algorithm.}
The AdaGrad algorithm belongs to the class of gradient-descent algorithms that adapt the learning rate based on the cumulated gradient evaluated over multiple mini-batches. Specifically, the AdaGrad scales the learning rate by a factor that is inversely proportional to the sum of the gradients of all used mini-batches \cite{Duchi2011}. The effect of this strategy is that the parameters with larger partial derivatives of the loss function decrease more rapidly than the parameters with smaller partial derivatives. The AdaGrad algorithm is reported in Algorithm \ref{Alg:AdaGrad}, with the parameter $\delta$ being a small number (typically of the order of $10^{-7}$), which is introduced to avoid a division by zero when updating the parameters.

\begin{algorithm}
\begin{algorithmic}\caption{AdaGrad algorithm for \glspl{fnn} training.}
\label{Alg:AdaGrad}
\State \texttt{Set } $\varepsilon>0$, $\beta>0$, $\br=\bzero$, $\bW,\bb$;
\While{\text{Validation Error larger than $\varepsilon$}}
\State \texttt{Sample a random mini-batch ${\cal S}_{SGD}$};
\State \texttt{Compute \eqref{Eq:DetStoGrad} by Algorithm \ref{Alg:BackProp}};
\State $\br=\br+\widehat{\nabla L}(\bW,\bb)\odot\widehat{\nabla L}(\bW,\bb)$;
\State $\bW=\bW-\frac{\alpha}{\beta+\sqrt{\br}}\widehat{\nabla L}(\bW+\delta\bW,\bb+\delta\bb)$;
\State $\bb=\bb-\frac{\alpha}{\beta+\sqrt{\br}}\widehat{\nabla L}(\bW+\delta\bW,\bb+\delta\bb)$;
\EndWhile  
\end{algorithmic}
\end{algorithm}

\textbf{RMSProp algorithm.} AdaGrad algorithm enjoys several pleasant properties in the convex case. However, when dealing with non-convex problems, it has been empirically observed that summing over all  squared gradients used in the training process can cause a premature and excessive decrease of the  learning rate. As a consequence, the learning rate might have become already too small when the algorithm finally finds a region around a (local) minimum of the loss function. The RMSProp algorithm aims at improving this drawback of AdaGrad, by introducing a moving weighted average of the gradients to reduce the relevance of gradients observed many iterations before. The formal procedure is reported in Algorithm \ref{Alg:RMSProp} and can be readily modified to include the use of Nesterov's momentum to accelerate convergence. 
\begin{algorithm}
\begin{algorithmic}\caption{RMSProp Algorithm for \glspl{fnn} training.}
\label{Alg:RMSProp}
\State \texttt{Set } $\varepsilon>0$, $\beta>0$, $\rho\in(0,1)$, $\br=\bzero$, $\bW,\bb$;
\While{\text{Validation Error larger than $\varepsilon$}}
\State \texttt{Sample a random mini-batch ${\cal S}_{SGD}$};
\State \texttt{Compute \eqref{Eq:DetStoGrad} by Algorithm \ref{Alg:BackProp}};
\State $\br=\rho\br+(1-\rho)\widehat{\nabla L}(\bW,\bb)\odot\widehat{\nabla L}(\bW,\bb)$;
\State $\bW=\bW-\frac{\alpha}{\beta+\sqrt{\br}}\widehat{\nabla L}(\bW+\delta\bW,\bb+\delta\bb)$;
\State $\bb=\bb-\frac{\alpha}{\beta+\sqrt{\br}}\widehat{\nabla L}(\bW+\delta\bW,\bb+\delta\bb)$;
\EndWhile  
\end{algorithmic}
\end{algorithm}

\textbf{Adam algorithm.} The Adam algorithm was introduced in \cite{Kingma2015_ADAM}, and is based on the application of momentum to the RMSProp method. However, the momentum technique is used with a different flavor from the conventional momentum approach. Specifically, the Adam algorithm employs both the first and second moment of the gradient estimated in each mini-batch. Moreover, Adam applies a correction term to both first and second moments, scaling them by a factor approaching one as the algorithm progresses. The procedure is formally stated in Algorithm \ref{Alg:Adam}.
\begin{algorithm}
\begin{algorithmic}\caption{Adam Algorithm for \glspl{fnn} training.}
\label{Alg:Adam}
\State \texttt{Set } $\varepsilon>0$, $\beta>0$, $\rho_{1},\rho_{2}\in(0,1)$, $\bs=\bzero$, $\br=\bzero$, $t=0$, $\bW,\bb$;
\While{\text{Validation Error larger than $\varepsilon$}}
\State \texttt{Sample a random mini-batch ${\cal S}_{SGD}$};
\State \texttt{Compute \eqref{Eq:DetStoGrad} by Algorithm \ref{Alg:BackProp}};
\State $t=t+1$;
\State $\bs=\rho_{1}\bs+(1-\rho_1)\widehat{\nabla L}(\bW,\bb)$;
\State $\br=\rho_{2}\br+(1-\rho_2)\widehat{\nabla L}(\bW,\bb)\odot\widehat{\nabla L}(\bW,\bb)$;
\State $\widehat{\bs}=\frac{\bs}{1-\rho_{1}^{t}}$;
\State $\widehat{\br}=\frac{\br}{1-\rho_{2}^{t}}$;
\State $\bW=\bW-\frac{\alpha\widehat{\bs}}{\beta+\sqrt{\widehat{\br}}}\widehat{\nabla L}(\bW+\delta\bW,\bb+\delta\bb)$;
\State $\bb=\bb-\frac{\alpha\widehat{\bs}}{\beta+\sqrt{\widehat{\br}}}\widehat{\nabla L}(\bW+\delta\bW,\bb+\delta\bb)$;
\EndWhile  
\end{algorithmic}
\end{algorithm}

As far as Adam algorithm is concerned, the suggested value for $\beta$ is $10^{-8}$, whereas the two weighting parameters $\rho_{1}$ and $\rho_{2}$ are suggested to be initialized to $0.9$ and $0.999$. Although Adam is usually quite robust to the choice of the hyperparameters, sometimes the default values need to be adjusted to obtain good convergence properties.

\textbf{Parameters initialization.} A critical issue of any training algorithm is the initialization of the parameters, and in particular of the weights\footnote{The initialization of the bias terms $\bb$ has been found to have a more limited impact on the final performance.} $\bW$. Given the non-convexity of the problem, the training algorithm will converge to some suboptimal point, and thus a suitable initialization point can make the difference between converging to an efficient or inefficient suboptimal point. Unfortunately, the design of efficient initialization strategies for \glspl{ann} is a little understood topic. Consolidated approaches from pure optimization theory should be applied with caution, since they focus on obtaining a low loss function, i.e. a low training error, but there is no guarantee that this will also result in  a low generalization error. 

At present, two general rules are widely used for the initialization of the \gls{ann} parameters:
\begin{itemize}
\item Two hidden nodes connected to the same input and with the same activation function should have different initial parameters. This is needed to avoid any redundancy, since otherwise any deterministic algorithm would update the parameters of these two nodes in the same way.
\item All matrices $\bW_{\ell}$ should be initialized to full-rank matrices, since otherwise some  patterns might be lost in the parameters null-space. 
\end{itemize}
These two guidelines motivate a random initialization of the parameters. Accordingly, initialization values are typically chosen as independent random variables, following either the Gaussian or uniform distribution, but a critical issue is how to choose the parameters of these distributions. These choices affect the initial scale of the parameters, which can have a significant impact on the generalization error. Larger initial weights are able to suppress redundancy more effectively, but might cause vanishing gradients due to the saturation of sigmoidal activation functions, as well as other numerical problems. In \cite{Glorot2010} it is proposed to initialize the weights of Layer $\ell$ with values drawn from a uniform distribution in $[-\frac{-6}{N_{\ell}+N_{\ell-1}}, \frac{-6}{N_{\ell}+N_{\ell-1}}]$. Instead, 
\cite{Saxe2013} recommends initializing the weights to random orthogonal matrices, that are scaled by a specific gain factor depending on the particular non-linearity used in each layer. In \cite{Sussillo2015}, it is shown that, by properly choosing the gain factor, the orthogonality assumption of the weight matrices can be relaxed. In \cite{Martens2010}, a sparse initialization strategy is proposed in which each unit is initialized to have a pre-defined number of non-zero weights. In contrast to these methods, we show, in Section \ref{Sec:Applications}, that the weights and biases can be initialized by using prior knowledge about the system, which can be obtained from (even inaccurate) analytical models.

\textbf{Regularization.} When training an \gls{fnn} it should always be kept in mind that the ultimate goal is to minimize the test error, rather than the training error. To this end, an essential technique is to perturb the training process so as to reduce the capacity of the \gls{ann}, thus avoiding overfitting. Any strategy aimed at reducing the test error at the expense of the training error is a regularization strategy. Empirical results have shown that applying regularization strategies to \glspl{ann} with high capacity is a more effective strategy compared with directly tuning the number of neurons and layers.  Over the years, several regularization methods have been proposed, and the most widely used ones are discussed in the following.

\emph{a) $L^{p}$ regularization}. A major regularization approach is to add a perturbation term proportional to the $p$-th power of the $L^{p}$ norm of the weights, namely modifying  \eqref{Prob:Training} into
\beq\label{Eq:RegCost}
L_{r}(\bW,\bb)=L(\bW,\bb)+\phi \|\bW\|_{p}^{p}\;,
\eeq
wherein $\phi\in[0,\infty)$ is a hyperparameter that weighs the relative contribution of the norm penalty term relative to the standard cost function. It should be stressed that the regularization term depends only on the weights and not also on the bias terms. This is because the weights have a more significant impact on the test error, as they directly link the input and output of a node, whereas the bias terms only directly affect the output. Thus, regularizing the weights is expected to be more important than regularizing the bias terms, which would only add to the complexity of the training process without bringing much improvement. This intuition has been experimentally confirmed in many research works over the years and motivates the current practice in neural networks to perform only weights regularization.

Among the different norms that can be considered in \eqref{Eq:RegCost}, the most widely used is the $L^{2}$ norm. This type of regularization is also called \emph{weight decay} because it can be seen to   reduce the magnitude of the weights, especially for larger $\phi$. This results in limiting the impact of many network connections on the final output, thereby reducing the network capacity. Moreover, reducing the magnitude of the weights causes sigmoidal or hyperbolic tangent activation functions to operate in their  linear regions, thus retaining the advantages of a linear model. 

Another widely used regularization norm is the $L^{1}$ norm. In comparison to $L^{2}$ regularization, $L^{1}$ regularization tends to produce a more sparse weight matrix $\bW$, in which many connections in every layer are effectively turned off. Besides reducing the network capacity, this also reduces the memory required to store the model. 

\emph{b) Early stopping.} Perhaps the simplest form of regularization is represented by the early stopping technique. All training algorithms are designed to minimize the training error in \eqref{Prob:Training} iteration after iteration. However, recalling also Fig. \ref{Fig:GeneralizationGap}, the validation error initially decreases together with the training error, but at some point tends to increase again. Thus, the idea of early stopping is to stop the training phase when the validation error reaches its minimum value. In practice, the network parameters are saved after each gradient update and when the validation error has not improved for a pre-specified number of iterations, the training algorithm stops and the parameters corresponding to the lowest observed validation error are returned. It is observed in \cite{Bishop1995a} and \cite{Sjoberg1995} that limiting the number of training iterations $t$ reduces the volume of parameter space reachable from the initial parameters, thereby reducing the capacity of the \gls{ann} and acting as a regularizer.

\emph{c) Dropout.} The idea of dropout is to introduce a perturbation by randomly changing the topology of the neural network every time a new data sample is used \cite{Srivastava2014}. Specifically,
for each data sample, each neuron in the \gls{ann} has a probability $p$ of being included in the network and if it is not included the corresponding weights are not updated in that particular iteration of the algorithm. Dropout is an effective regularizer due to two main reasons:
\begin{itemize}
\item By randomly removing a subset of connections each time, dropout is actively weakening the coupling among neighboring neurons. This reduces the possibility of performing too complex operations, which could  cause overfitting. 
\item Each time a subset of neurons is randomly disconnected, a different reduced network is being trained. As a result, using dropout effectively trains a large number of different, random \glspl{ann}, and then averages the results, which tends to reduce the net effect of overfitting. 
\end{itemize}

\textbf{Batch Normalization.} One issue when working with gradient-based methods, is the different scale that the features in the input vector, as well as the activation values of each layer, might have. In the presence of vectors with components that have very different magnitude with one another, numerical problems can arise and gradient descent can be slow. In order to avoid this issue, \cite{Ioffe2015_Batch} has proposed to normalize the input data and/or the activation values of each layer in the network. 

Formally speaking, let us consider the training data points $\bx_{0}^{(1)},\ldots,\bx_{0}^{(N_{TR})}$. Then, batch normalization modifies the operation performed by the input layer, which will not simply forward the input vector, but will apply the transformation:
\beq
\tilde{\bx}_{0}^{(nt)}=\frac{\bx_{0}^{(nt)}-\bmu_{0}}{\bpsi+\bsigma_{0}}\;,\;\forall\;nt=1,\ldots,N_{TR}\;,
\eeq
wherein the division is meant component-wise, $\bpsi$ is a vector with positive components of the order of $10^{-8}$, whose purpose is to avoid dividing by zero, while $\bmu_{0}$ and $\bsigma_{0}$ are mean and standard deviation vectors defined as
\begin{align}\label{Eq:NormMu}
\bmu_{0}&=\frac{1}{N_{TR}}\sum_{nt=1}^{N_{TR}}\bx_{0}^{(nt)}\\
\label{Eq:NormVar}
\bsigma_{0}&=\sqrt{\frac{1}{N_{TR}}\sum_{nt=1}^{N_{TR}}(\bx_{0}^{(nt)}-\bmu_{0})\odot (\bx_{0}^{(nt)}-\bmu_{0})}\;,
\end{align}
where the square root operation is meant component-wise. 

Denoting by $\bz_{\ell}^{(nt)}$ the $N_{\ell}$-dimensional vector of activation values of layer $\ell$ when $\bx_{0}^{(nt)}$ is the input of the network, a similar normalization technique can be applied to the vectors $\{\bz_{\ell}^{(1)},\ldots,\bz_{\ell}^{(N_{S})}\}$ in each mini-batch, thus changing the arguments of the activation functions of the $\ell$-th layer to be:
\beq\label{Eq:NormActivation}
\tilde{\bz}_{\ell}=\frac{\bz_{\ell}^{(nt)}-\bmu_{\ell}}{\bpsi+\bsigma_{\ell}}\;,\;\forall\;nt=1,\ldots,N_{S}\;,
\eeq
with $\bmu_{\ell}$ and $\bsigma_{\ell}$ having similar definitions as in \eqref{Eq:NormMu} and \eqref{Eq:NormVar}. In addition, when applied to a hidden layer, it is common to further modify the input to the activation functions in \eqref{Eq:NormActivation} as:
\beq\label{Eq:NormActivation2}
\tilde{\bz}_{\ell}=\bgamma_{\ell}\odot\tilde{\bz}_{\ell}+\bbeta_{\ell}\;,\;\forall\;nt=1,\ldots,N_{S}\;,
\eeq
with $\bgamma_{\ell}$ and $\bbeta_{\ell}$ being $N_{\ell}$-dimensional parameters to be learnt during the training phase. The operation in \eqref{Eq:NormActivation2} is aimed at preserving the representational power of the \gls{ann}, which would be significantly diminished by constraining each layer to have zero-mean and unit-variance activation inputs. This approach might seem counterintuitive, since it seems to defeat the purpose of applying the normalization step in \eqref{Eq:NormActivation} in the first place. The advantage of using \eqref{Eq:NormActivation2} lies in the fact that $\bgamma_{\ell}$ and $\bbeta_{\ell}$ are parameters to be learnt based on the normalized values in $\tilde{\bz}_{\ell}$, which are more conveniently handled by gradient descent algorithms. Moreover, while batch normalization increases the number of parameters to optimize during the training phase, applying \eqref{Eq:NormActivation2} makes the bias terms in each node useless. In other words, when using batch normalization, it should be set $\bb_{\ell}=\bzero$ for any normalized layer, since the role of $\bb_{\ell}$ is played by $\bbeta_{\ell}$. As a consequence, the only new parameters to be trained are the vectors $\bgamma_{\ell}$ for the layers where normalization is applied.

It is also important to mention that batch normalization has a regularization effect, too, due to at least two main reasons:
\begin{itemize}
\item Since $\bmu_{\ell}$ and $\bsigma_{\ell}$ are computed on each mini-batch, they will be slightly different for each mini-batch. This introduces a slight perturbation that has a regularizing effect on the overall \gls{ann}, similarly to the dropout technique. 
\item The fact that batch normalization reduces the variability of the input data to each layer weakens the coupling among different layers, which results in a similar effect as the dropout technique.
\end{itemize}

So far, batch normalization has been described as a technique to aid the training process. However, since it modifies the structure and operation of the \gls{ann}, it also affects the network use at test time. In other words, if an \gls{ann} is trained using batch normalization, at test time  \eqref{Eq:NormActivation2} needs to be computed in each layer, by employing the trained parameters $\bgamma$ and $\bbeta$. However, the issue of this approach is that at test time the dataset at our disposal may not be sufficiently large to compute reliable estimates of mean an variance for each activation input. This problem is typically solved by computing an exponentially-weighted average that accounts for the means and variances computed during the training phase on each mini-batch, in addition to the new data sample at test time. 

\subsubsection{\textbf{Hyperparameter tuning - Fitting the data}}\label{Sec:Hyperparameters}
So far, many techniques have been presented to tune the parameters of an \gls{fnn} in order to achieve a low generalization error. However, the performance of all algorithms that have been presented depends on several hyperparameters, which are not directly tuned during the training phase. Examples of hyper-parameters are the number of layers and neurons per layer, the size of the training set and of each mini-batch, the learning rate, the regularization coefficient, etc. Moreover, other choices that have a significant impact on the overall performance are related to the training algorithm that is used, to the initialization point that is adopted, to the regularization strategy to use, whether or not to use batch normalization, etc. 

As discussed in Section \ref{Sec:ValidationSet}, hyperparameter tuning can be performed either manually or in an automated way. The three automated methods introduced in Section \ref{Sec:ValidationSet}, i.e. grid-search, random search, hyperparameter optimization, are general enough for application not just to deep learning, but to machine learning in general. However, grid search and hyperparameter optimization are rarely used in the context of deep learning. The former is deemed practical only when three or fewer hyperparameters need to be tuned. In this case, a logarithmic search scale is used to span a wider range of values. The latter is problematic due to the lack of an expression of the loss function with respect to some hyperparameters, as well as because any hyperparameter optimization algorithm in turn has its own hyperparameters to set, even though they are typically less problematic to tune. Instead, random search is considered to be a more feasible solution, and has been shown to reduce the validation error to acceptable values much faster than grid search  \cite{Bergstra2012}. 

Along with these automated methods, manual hyperparameter setting represents an effective way to achieve the desired performance at an affordable complexity. Nevertheless, compared to automated approaches, the manual tuning of the hyperparameters requires a higher degree of experience, and is typically carried out by monitoring both training and validation error during the training phase, thereby determining  whether the network is underfitting or overfitting, and modifying the hyperparameters to adjust the network capacity accordingly. To this end, in general a trial and error procedure is required, since it is very challenging to know in advance the optimal configuration of hyperparameters for the specific problem at hand. Nonetheless, some general guidelines can be identified, recalling that the capacity of an \gls{ann} depends on three main factors: 1) the ability of the network to represent the problem at hand; 2) the ability of the learning algorithm to successfully minimize the loss function during the training phase; 3) the degree to which the training procedure regularizes the model, thus avoiding overfitting. 

\begin{figure}
  \begin{center}
  \includegraphics[scale=0.4]{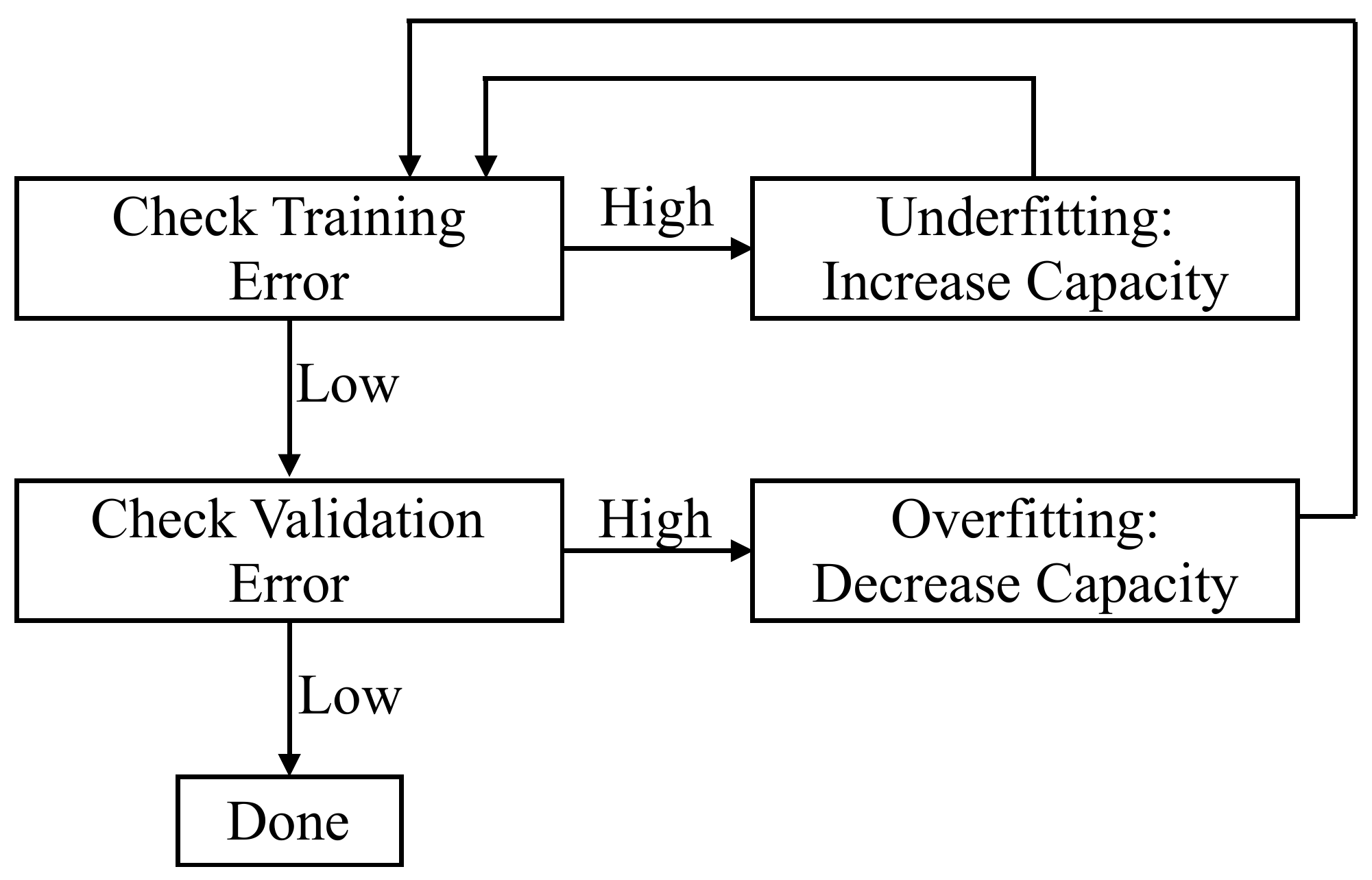}
  \caption{Scheme for manual hyperparameter setting in \glspl{ann}.}
  \label{Fig:HP_Tuning}
 \end{center}
\end{figure}
As shown in Fig. \ref{Fig:HP_Tuning}, when configuring an \gls{ann}, the first issue to take care of is to make sure that the network does not underfit.  If the performance on the training set is not good enough, it means that the \gls{ann} can not fit the available training data and thus it is usually useless to gather more data. In this case, a good approach is to improve the optimization algorithm and the most important hyperparameter to this end is the learning rate. Unfortunately, each task has its own optimal learning rate, and trial and error is the de facto approach to find a learning rate that yields a low enough training error for the task at hand. 

Apart from the learning rate, other strategies to increase the network capacity are to tune the other hyperparameters of the algorithm in use or to consider more sophisticated optimization algorithms. Widely-used choices are \gls{sgd} with momentum, RMSProp, or Adam, possibly coupled with Nesterov's momentum. Moreover, batch normalization can be included if the training error does not decrease as desired. If these strategies are not effective, the problem could be in the size of each mini-batch, which might be too small to provide a reliable estimate of the gradient. Finally, another conceptually simple way to increase the network capacity is to use more neurons and layers. This is a powerful approach to avoid underfitting, but comes at the expense of a larger complexity and its applicability depends on the available computational resources. If none of these strategies work, the problem might just be in the quality of the training data, which might be too noisy and/or might not include the most appropriate features to represent the problem at hand. In this case, it may be needed to collect different data and to use a different training set.

Once a low enough training error is obtained, the validation error needs to be checked. If it is unsatisfactory, then it is likely that overfitting is the issue. In this case, the most effective strategy is to just gather more data. However, gathering more data can be costly and requires higher storage and processing capabilities. A simpler way of reducing the network capacity is to employ a regularization technique. It is advisable to use early stopping as the first approach, while other regularizing techniques could be included during the training phase. Finally, a third approach consists of manually reducing the model size, limiting the number of neurons and layers. If these approaches do not work even after a careful tuning of their hyperparameters, then gathering more data remains the only possible approach to avoid overfitting. 


Finally, it is worth emphasizing once again that the validation error is an estimate of the test error and the  discussion above assumes that such an estimate is reliable. If the test error is high but the validation error is low, then the most effective approach is to increase the size of the validation set. However, if increasing the size of the validation set does not help, then either the validation procedure is not appropriate, or the problem might lie in a more fundamental issue. Typically, the loss function used for training and validation might not be appropriate for the task at hand, or the \gls{ann} model is not properly designed to learn the target objective, or there is a mismatch between validation data and real testing conditions.

\subsection{Deep Reinforcement Learning}
This section presents the framework of deep reinforcement learning, which merges deep learning with reinforcement learning \cite{Arulkumaran2017_DRL,Li2017_DRL}. The framework of reinforcement learning is not directly related to deep learning, but rather it is a different machine learning approach that implements the learning procedure in an adaptive way, namely by interacting with the environment by taking actions and receiving feedback on the result of the actions that have been taken. Nevertheless, recently it has been observed that deep learning can be used to improve and facilitate the implementation of reinforcement learning techniques, which has motivated the cross-fertilization between these two machine learning frameworks, leading to the development of the framework of deep reinforcement learning. The first part of this section provides a short introduction to reinforcement learning, whose purpose is to define basic terminology and provide a brief mathematical description of the typical scenarios where reinforcement learning is employed. For a dedicated and comprehensive treatment of the reinforcement learning framework, we refer the reader to \cite{SuttonBookRL}. 

Reinforcement learning applies to scenarios that can be mathematically described by a \gls{mdp}. An   \gls{mdp} is defined by the following quantities:
\begin{itemize}
\item ${\cal S}$, the set of possible states.
\item ${\cal A}$, the set of possible actions that an agent can take.
\item ${\cal P}$, the set of transition probabilities, with $P(s_{t},s_{t+1},a_{t})$ the probability of moving from state $s_{t}$ to state $s_{t+1}$ by taking action $a_{t}$.
\item ${\cal R}$, the set of rewards, with ${\cal R}(s_{t},a_{t})=\mathbb{E}\left[R_{t+1}|s_{t},a_{t}\right]$, and $R_{t+1}$ the reward obtained at step $t+1$.
\item $\gamma\in[0,1]$, a discount factor adjusting the weight of more recent actions.
\end{itemize}
Based on this notation, it is possible to define the long-term reward as 
\beq
G_{t}=\sum_{k=0}^{+\infty}\gamma^{k}R_{t+k+1}\;,
\eeq
and a (stationary) policy as the probability of taking action $a$ at time $t$, when being in state $s$, namely:
\beq
\pi(s,a)=P(A_{t}=a| S_{t}=s)\;,
\eeq
where the word stationary refers to the fact that the probability of taking action $a$ when in state $s$ does not depend on time. 

A key concept when analyzing an \gls{mdp} is that of \textbf{action-value function}, measuring the value, in terms of expected reward, of being in state $s$ and taking action $a$, following policy $\pi$, namely:
\beq\label{Eq:Qfunction}
Q_{\pi}(s,a)=\mathbb{E}_{\pi}\left[G_{t}|S_{t}=s,A_{t}=a\right]
\eeq
The action-value function can be also rewritten as the sum of the reward at step $t+1$, plus the long-term reward from $t+1$ to $\infty$, namely:
\begin{align}
&Q_{\pi}(s,a)=\\
&\mathbb{E}_{\pi}\left[R_{t+1}+\sum_{k=1}^{+\infty}\gamma^{k} R_{t+k+1}|S_{t}=s,A_{t}=a\right]=\notag\\
&\mathbb{E}_{\pi}\left[R_{t+1}+\gamma\sum_{k=0}^{+\infty}R_{t+k+2}|S_{t}=s,A_{t}=a\right]\;.
\end{align}

Reinforcement learning provides several approaches to determine the optimal sequence of actions to be taken in order to maximize the long-term reward. These approaches can be broadly classified in three main categories, namely, 
\begin{itemize}
\item \textbf{Value-based approaches}, which aim at estimating the action-value function.
\item \textbf{Policy-based approaches}, which aim at estimating the policy function.
\item \textbf{Actor-critic approaches}, which exploit an estimate of both the action-value and the policy function. 
\end{itemize}
Thus, regardless of the particular technique that is chosen, reinforcement learning requires full knowledge about the environment in order to estimate the action-value or the policy functions, which is not realistic in several applications. Moreover, in some cases, the complexity of the estimation rapidly increases with the cardinality of the action-state space, which makes reinforcement problems unfeasible by standard methods when the number of possible states and actions grow too large.

In this context, thanks to their universal function approximation ability, \glspl{ann} provide an efficient way to estimate the action-value and/or the policy functions, thereby enabling the practical solution of complex reinforcement learning problems in the realistic scenario in which the statistics and parameters of the environment are not fully known. 

\subsubsection{\textbf{Deep Q-Network. Estimating the action-value function}}
The goal of the Q-learning method is to compute the optimal action-value function, defined as
\beq\label{Eq:OptQ}
Q^{*}(s,a)=\max_{\pi}\;Q_{\pi}(s,a)\;.
\eeq
Solving \eqref{Eq:OptQ} for each pair $(s,a)$ provides a full characterization of the \gls{mdp} problem, and allows determining the best policy to follow for each possible state and action. To this end, several methods are available, depending on the information available on the \gls{mdp}. An optimality condition for Problem \eqref{Eq:OptQ} is the so-called Bellman's optimality equation, which however requires full knowledge of the \gls{mdp} model and parameters to be solved. 

However, in practical scenarios, assuming complete knowledge of the \gls{mdp} model is often unrealistic. Typically, only the response from the environment is observable, but no information is available as to the statistics regulating the \gls{mdp} process, such as the transition probabilities, which makes it impossible to compute the value of the $Q$ function for any pair $(s,a)$. In these cases, a possible approach is to  obtain the values of the $Q$ function from experience, i.e. by initiating the process from each possible $(s,a)$ pair, and then following different policies, observing the rewards returned by the environment at each step. However, this approach has the clear drawback of requiring a high computational complexity, especially when the number of possible $(s,a)$ pairs is large. A similar drawback is suffered by all other alternative methods aimed at building a table collecting the possible values $Q(s,a)$, for all possible $s\in{\cal S}$ and $a\in{\cal A}$.

In scenarios with a very large (possibly even infinite) number of $(s,a)$ pairs, the state-of-the-art  approach is that of $Q$-learning. As the name implies, this approach is based on learning the values of the $Q$ function. More specifically, $Q$-learning algorithms assume a functional form for the function $Q(s,a)$, namely:
\beq\label{Eq:ModelQ}
Q(s,a)\approx \widehat{Q}(s,a,\bw)\;,
\eeq
with $\widehat{Q}$ a known function, and $\bw$ a set of parameters to be determined by any machine learning method, with the goal of improving the accuracy of the approximation. More specifically, $Q$-learning methods assume that some points of the $Q$ function, say $\{Q(s_{i},a_{i})\}_{i=1}^{N_{T}}$, have been already determined, for example by trying some actions and observing the response of the environment. Then, the parameters in the vector $\bw$ are determined so as to minimize the mean squared error between the samples $\{Q(s_{i},a_{i})\}_{i=1}^{N_{T}}$ and the model \eqref{Eq:ModelQ}. 

Traditional $Q$-learning approaches typically employ a linear model for $\widehat{Q}$, but more recently it has been proposed to adopt an \gls{ann} with weights $\bw$, that takes as input a pair $(s,a)$ and outputs the corresponding value $Q(s,a)$. The parameters $\bw$ are trained by using the samples $\{Q(s_{i},a_{i})\}_{i=1}^{N_{T}}$ as the training set. This implementation of Q-learning is referred to as  the \textbf{Deep Q-Network} approach \cite{Li2017_DRL,Arulkumaran2017_DRL}, which can be considered an algorithm belonging to the family of $Q$-learning methods, with the peculiarity that the approximate function $\widehat{Q}(s,a,\bw)$ is specified through an \gls{ann}. Thus, compared with  other $Q$-learning methods, deep reinforcement learning has the significant advantage of not specifying a-priori the functional form of $\widehat{Q}$, leaving to the \gls{ann} the task of determining the best functional form to use. Since \glspl{ann} are universal function approximators, they will be able to approximate the true function $Q(s,a)$ within any desired tolerance, provided a proper training phase is performed. 

\subsubsection{\textbf{Deep Policy Iteration. Estimating the policy function}}
While the deep Q-network method aims at learning the action-value function, policy iteration methods aim at determining directly the policy function $\pi(a,s)$. To this end, the policy function is parametrized as
\beq\label{Eq:ModelPi}
\pi(s,a)=\widehat{\pi}(s,a,\thetab)\;,
\eeq
with $\thetab$ a vector of parameters to be learnt. Standard policy iteration methods assume a fixed functional form $\widehat{\pi}(\cdot)$, and design $\thetab$ in order to maximize the average reward function, defined as 
\beq\label{Eq:AvRewardDRL}
J(\thetab)=\sum_{s\in{\cal S}}d_{\widehat{\pi}_{\thetab}}(s)\sum_{a\in{\cal A}}\widehat{\pi}(s,a,\thetab){\cal R}(s,a)\;,
\eeq
wherein $d_{\widehat{\pi}_{\thetab}}$ denotes the stationary distribution of $\widehat{\pi}(s,a,\thetab)$. 

The maximization of $J(\thetab)$ with respect to $\thetab$ is carried out by means of the gradient ascent method, wherein an expression of the gradient of \eqref{Eq:AvRewardDRL} is provided by the \emph{policy gradient theorem}, which proves that:
\beq\label{Eq:GradientPolicy}
\nabla_{\thetab}J(\thetab)=\sum_{s\in{\cal S}}d_{\widehat{\pi}_{\thetab}}(s)\sum_{a\in{\cal A}}Q_{\widehat{\pi}}(s,a)\widehat{\pi}(s,a)\nabla_{\thetab}\log(\widehat{\pi}(s,a))\;.
\eeq
In order to implement the gradient ascent algorithm, a standard approach is the so-called Monte-Carlo policy gradient, also known as the REINFORCE method \cite{Williams1992}, which employs stochastic gradient ascent wherein the instantaneous return observed from the environment provides an unbiased sample of the unknown function $Q_{\widehat{\pi}}(s,a)$.

Similarly to the Deep Q-Network case, instead of assuming a fixed functional form for $\pi(s,a)$,  
an \gls{ann} can be trained to output an estimate of the values $\pi(s,a)$. Specifically, it is possible to use an \gls{ann} that takes as input a state $s$, outputs $\pi(s,a)$ for any action $a\in{\cal A}$, and is trained by samples collected according to the target policy. In other words, the training set is built adaptively: given an input state, a realization of the output distribution $\pi(s,a)$ is sampled and used as training label. Next, the sampled action is performed and the reward obtained from the environment is used to weigh the training loss function in order to refine the training. Also, the action that is taken brings the agent into a new state and the whole procedure is iterated.

\subsubsection{\textbf{Deep Actor-Critic. Estimating the action-value and policy functions}}
Instead of employing the instantaneous returns as an estimate for the action-value function $Q_{\widehat{\pi}}(s,a)$, deep actor-critic approaches improve purely policy-based methods by merging them with a Deep Q-Network that provides an estimate of $Q_{\widehat{\pi}}(s,a)$. Thus, in order to maximize \eqref{Eq:GradientPolicy}, actor-critic approaches assume both the models in \eqref{Eq:ModelQ} and \eqref{Eq:ModelPi}, using a first \gls{ann}, called the critic \gls{ann}, to estimate the value $Q_{\widehat{\pi}}(s,a,\bw)$, and a second \gls{ann}, called the actor \gls{ann}, to estimate the policies $\widehat{\pi}(s,a,\thetab)$.

Actor-critic methods typically perform better than purely policy-based methods and during the last years several improvements have been proposed. A notable example is the use of a so-called advantage function to reduce the estimation variance by subtracting it from the value function \cite{Schulman2018}. Namely, the method exploits the fact that:
\begin{align}\label{Eq:GradientPolicyAdv}
&\nabla_{\thetab}J(\thetab)=\sum_{s\in{\cal S}}d_{\widehat{\pi}_{\thetab}}(s)\sum_{a\in{\cal A}}Q_{\widehat{\pi}}(s,a)\widehat{\pi}(s,a)\nabla_{\thetab}\log(\widehat{\pi}(s,a))\notag\\
&=\sum_{s\in{\cal S}}d_{\widehat{\pi}_{\thetab}}(s)\sum_{a\in{\cal A}}[Q_{\widehat{\pi}}(s,a)-B(s)]\widehat{\pi}(s,a)\nabla_{\thetab}\log(\widehat{\pi}(s,a))\;,
\end{align}
since 
\beq
\sum_{a\in{\cal A}}\widehat{\pi}(s,a)\nabla_{\thetab}\log(\widehat{\pi}(s,a))=\nabla\left(\sum_{a\in{\cal A}}\widehat{\pi}(s,a)\right)=0\;,
\eeq
wherein $A_{\widehat{\pi}}(s,a)=Q_{\widehat{\pi}}(s,a)-B(s)$ is the advantage function. 

Other improvements of the actor-critic approach have been proposed in \cite{Mnih2016},  \cite{Lillicrap2016}, and \cite{Lowe2018}. In \cite{Mnih2016} the so-called \emph{asynchronous advantage actor-critic} (A3C) approach is introduced, in which multiple actors and critics are deployed. The critics learn the action-value function while the actors are trained in parallel, being synchronized with each other with global parameters from time to time. A deterministic version of the A3C method, called \emph{synchronous advantage actor-critic} (A2C) is also proposed, in which all critics are synchronized with the global parameters at the same time, hence the name \enquote{synchronous}. In \cite{Lillicrap2016}, a deterministic version of the deep actor critic approach, the \emph{deep deterministic policy gradient} (DDPG) is presented, in which the policy is no longer modeled as a distribution over actions, but rather as a deterministic function $a=\pi(s)$. The authors of \cite{Lillicrap2016} merge deep learning with the DPG approach, first introduced in \cite{Silver2014}. Finally, in \cite{Lowe2018} the DDPG approach is extended to multi-agent environments, i.e. to scenarios in which multiple decision-makers coordinate among themselves to complete tasks based only on local information.

\subsection{Deep unfolding}
As discussed, one of the issues of \glspl{ann} is to determine the number of neurons and layers to use. However, in some cases it is possible to match the iterations of iterative algorithms to the layers of an \gls{ann} by a technique called deep unfolding \cite{Hershey14}. This provides a systematic approach to determine the hyperparameters of an \gls{ann} that implement a given number of iterations of a recursive algorithm. 

To elaborate, the idea of deep unfolding applies to all algorithms that take as input a vector $\bx=[x_{1},\ldots,x_{N}]$ and produce as output a vector $\by=[y_{1},\ldots,y_{M}]$ expressed by
\beq\label{Eq:OutputDU}
y_{i}=g_{i}(\bx,\phib,\thetab)\;,\;\forall\;i=1,\ldots,M\;,
\eeq
wherein $\thetab$ is a vector containing all the parameters of the algorithm, while $\phib=[\phi_{1},\ldots,\phi_{N}]$ is iteratively updated according to the formula
\beq\label{Eq:HiddenDU}
\phi_{i}^{(k)}=f_{i}(\bx,\phib^{(k-1)},\thetab)\;,
\eeq
with $k$ the iteration index and $\phib^{(0)}$ the initial value. This formalism applies to detection tasks \cite{He2018}, as well as to the computation of posterior probabilities by the belief propagation method, or to inference techniques aimed at estimating a distribution by minimizing its divergence from an approximate distribution \cite{Hershey14}. 

\begin{figure}
  \begin{center}
  \includegraphics[scale=0.5]{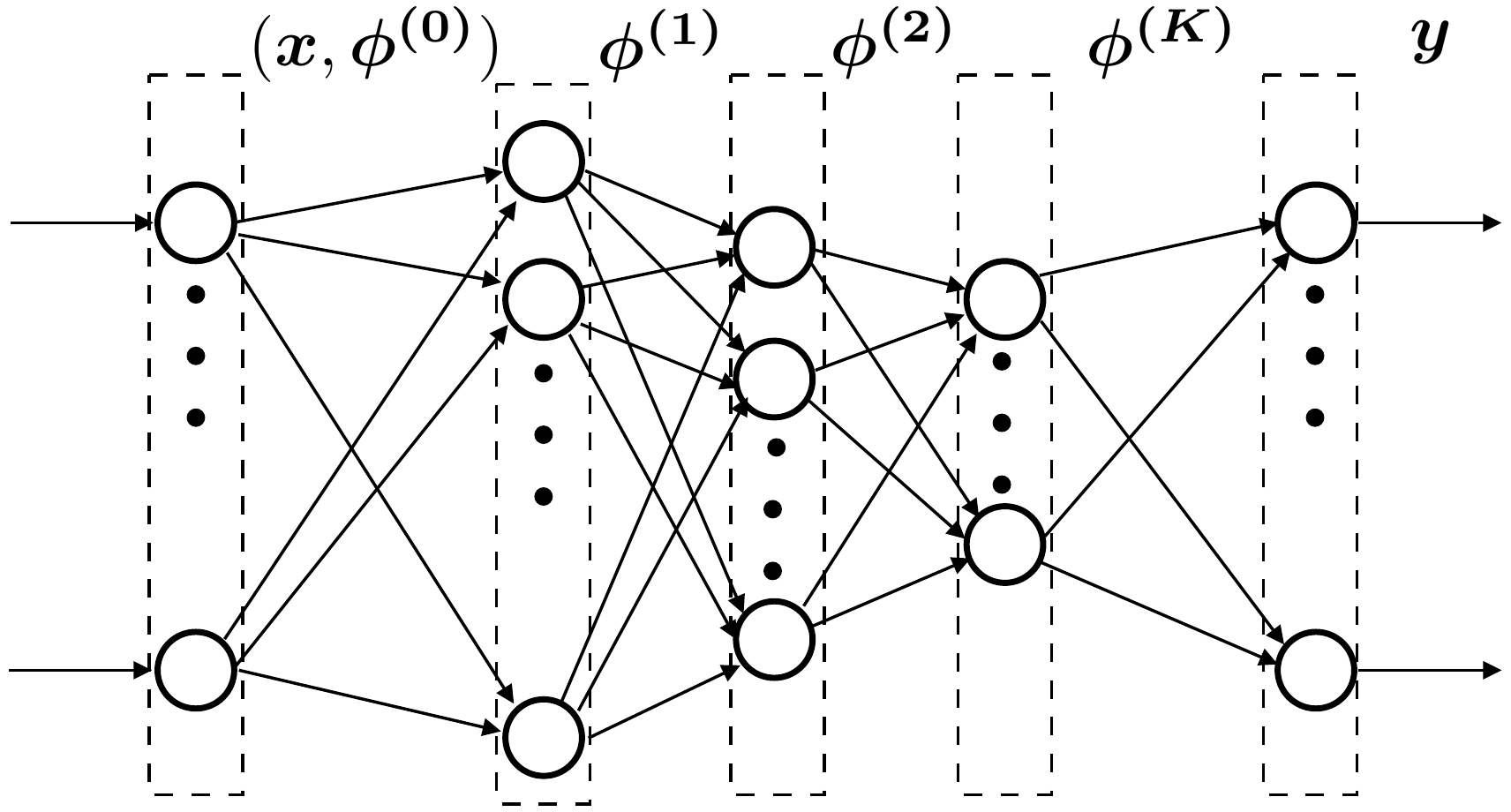}
  \caption{$K+1$ iterations of an iterative algorithm are unfolded onto the $K$ hidden layers and onto the output layer of an \gls{ann}. The output of each layer is equal to the output produced by one iteration of the algorithm, and the output of the last layer is equal to the output of the last iteration of the algorithm.}
  \label{Fig:DeepUnfolding}
 \end{center}
\end{figure}

The main idea of deep unfolding lies in the observation that \eqref{Eq:OutputDU} can be regarded as the input-output relationship of an \gls{ann}, with \eqref{Eq:HiddenDU} being the input-output relationship of Layer $k$, and $\thetab$ representing the parameters of the \gls{ann}, i.e. all weights and bias of each layer. Then, the iterative algorithm can be \emph{unfolded} by mapping each iteration onto one layer of the \gls{ann}, which takes as inputs $\bx$ and $\phib_{0}$, compute $\phib^{(k)}$ at the output of the $k$-th hidden layer, and finally produce $\by$ as output, as displayed in Fig. \ref{Fig:DeepUnfolding}.

Two main points are to be highlighted:
\begin{itemize}
\item In deep unfolding, in contrast to typical \glspl{ann}, the number of nodes and layers is determined by the particular algorithm that is unfolded. Specifically, the number of layers is fixed by the number of iterations of the algorithm, while the number of nodes in each layer is fixed by the sizes of the vectors $\bx$, $\phib$, and $\by$.  
\item The advantage of unfolding an algorithm onto an \gls{ann} rather than implementing it directly, 
lies in the fact that the parameters $\thetab$ of the algorithm are determined by an \gls{ann}, instead of being set by more conventional methods. Moreover, once the parameters are determined, the \gls{ann} can be directly used as an alternative and efficient implementation of the iterative algorithm to compute $\by$ based on the chosen parameters $\thetab$.
\end{itemize}
In the context of jointly exploiting model-based and AI-based methods, deep unfolding, in combination with deep transfer learning described in the next section, offers the possibility of initializing a model-based ANN by unfolding the model onto the layers of the \gls{ann}, and then refining it by using empirical data. This approach has the advantage of not requiring the tuning of the number of layers and neurons, as they are obtained by directly unfolding the model on the \gls{ann} architecture. 

\subsection{Deep Transfer learning}
Deep transfer learning is another recent framework that combines deep learning with another machine learning framework, namely transfer learning. In the broadest sense, transfer learning studies how to transfer the knowledge that is used in a given context to execute a given task, into a different, but related context, to execute  another task. Formally speaking, four fundamental components can be identified in a transfer learning problem:
\begin{itemize}
\item A source task, ${\cal T}_S$, i.e. the original task for whose execution the knowledge to be transferred was developed.
\item A source domain, ${\cal D}_{S}$, i.e. the context in which the task ${\cal T}_S$ was executed. 
\item A target task, ${\cal T}_{T}$, i.e. the new task to be executed thanks to the knowledge transfer. 
\item A target domain, ${\cal D}_{T}$, i.e. the new context in which the task ${\cal T}_T$ must be executed. 
\end{itemize}
Clearly, such a problem formulation is very general, and need not be related to any deep learning problem. However, transfer learning can be successfully used to facilitate the implementation of deep learning algorithms, especially by reducing the amount of data to be acquired for training and validation purposes. Indeed, the availability of large quantities of data is a prerequisite for deep learning to outperform other machine learning methods, but in the context of wireless communication networks the  acquisition of large amount of data can be too expensive and/or not practical. In these cases, transfer learning can be used by transferring knowledge from other related scenarios in which data acquisition has been already performed. For example, datasets for similar communication systems can be used, and/or datasets generated according to (possibly inaccurate) mathematical models can be used. Concrete examples about the latter approach are analyzed in the next section.

Despite being a relatively recent approach, many techniques for deep transfer learning have already appeared in the literature and it is difficult to provide a general taxonomy. Here, following the taxonomy by the  recent tutorial \cite{TanTL2018}, we categorize transfer learning techniques into four main classes. 
\subsubsection{\textbf{Instance-based transfer learning}}
This approach assumes to have data from both the source domain ${\cal D}_{S}$ and target domain ${\cal D}_{T}$. Then, the idea is to exploit both datasets to carry out the target task ${\cal T}_{T}$, by assigning a different weight to each instance of the source and target data. Otherwise stated, data from the source domain is used to augment the data from the target domain, but it must be weighted differently to ensure that instances that are specific to the source domain are given less or no importance during the training process. After this re-weighting step, the augmented data set is used as training set for the target task by any traditional training algorithm, with the re-weighting factors acting as hyperparameters to be adjusted during the validation process. 

In principle, this method does not require having labeled data, in the sense that, once the new dataset has been built, it can be used in conjunction with any machine learning method. However, as far as training a neural network is concerned, it is required that the training set be labelled in order to implement available training algorithms. Recently, instance-based transfer learning has proved effective when employed in  conjunction with the AdaBoost training algorithm, addressing both classification and regression problems \cite{Yao2010,Pardoe2010}. 

\subsubsection{\textbf{Mapping-based transfer learning}}
Mapping-based transfer learning redefines the training cost function in order to account for the presence of data from both the source and target domains. Specifically, the cost function used during the training phase is defined as:
\beq\label{Eq:CostMTL}
{\cal L}(\bW,\bb)={\cal L}_{S}(\bW,\bb)+\lambda{\cal L}_{T}(\bW,\bb)+R^{2}(\bW,\bb)\;,
\eeq
wherein ${\cal L}_{S}$ is the cost function for the source task, taking as input training samples from the source domain, ${\cal L}_{T}$ is the cost function for the target task, taking as input training samples from the target domain, $\lambda$ is a non-negative term weighting the relative importance of the two cost functions, and $R$ is a regularization function that accounts for the differences between source and target domains. More in detail, the regularizer $R$ is typically chosen as the \emph{maximum mean discrepancy} function between the source and target domains, with respect to a generic representation $\phi(\cdot)$, namely \cite{TzengTL}
\beq
\text{MMD}=\left\|\frac{1}{|{\cal X}_{S}|}\sum_{x\in{\cal X}_{S}}\phi(x)-\frac{1}{|{\cal X}_{T}|}\sum_{x\in{\cal X}_{T}}\phi(x)\right\|\;,
\eeq
wherein ${\cal X}_{S}$ and ${\cal X}_{T}$ denote the source and target available datasets. Thus, this approach requires having labelled data from both the source and target domains.
Based on \eqref{Eq:CostMTL}, any standard training algorithm can be executed, exploiting all available labeled data. 

Recent studies on mapping-based transfer learning have focused on analyzing the performance when other regularizers are used. In \cite{LongICML} it is proposed to use a multiple kernel variant of the MMD (MK-MMD), while in \cite{Long2016} it is proposed to use the joint maximum mean discrepancy as regularizer. Finally, we mention \cite{Arjovsky2017}, where Wasserstein's distance is used as regularizer and is shown to achieve better performance than the MDD in some cases. 

\subsubsection{\textbf{Network-based transfer learning}}
Network-based deep transfer learning implements the transfer of knowledge by first training an \gls{ann} to execute the source task ${\cal T}_S$ in the source domain ${\cal D}_S$, and then reusing and/or refining the obtained network configuration to execute the target task ${\cal T}_T$ in the target domain ${\cal D}_T$. This general concept can be applied in several different ways. For example, it is possible to identify a part of the \gls{ann} that extracts general features that describe both the source and target tasks. Then, after training the \gls{ann} in the source domain, the part of the \gls{ann} that applies to both source and target tasks need not be trained again. This approach is taken in  \cite{Huang2013}, where a language processing application is considered, and it is proposed to divide the \gls{ann} in two parts. The former extracts language-independent features, which can be reused for all languages, while the latter is language-specific and needs to be trained for each new language. 

Nevertheless, a more common approach is to perform a two-step training. At first, the \gls{ann} is  trained to execute the source task, yielding a tentative configuration of the network parameters. Next, a second training phase is performed in the target domain, which uses the configuration of the weights and bias from the first phase as the initialization point for the training algorithm. This approach is very useful in all situations in which a lot of training data is available in the source domain, whereas only a few labeled training samples are available (or are difficult/expensive to obtain) in the target domain. As described in Section \ref{Sec:Applications}, this is the typical scenario in wireless communications, and indeed Section \ref{Sec:Applications} will present several case-studies wherein this particular transfer learning method proves extremely useful. Techniques inspired to network-based transfer learning have been recently proposed for resource allocation in wireless communications in \cite{ZapCOMMAG2018,Shen2018}.

\subsubsection{\textbf{Adversarial-based transfer learning}}
The main idea of adversarial transfer learning is to identify the common features between source and target tasks through the use of an another deep neural network, called generative adversarial network (GAN) \cite{GoodfellowGANs}. The first step of the approach is to divide the \gls{ann} that implements the source task into two segments, one that extracts the salient features of the source domain, and one that exploits these features to carry out the source task. Then, the output of the first segment of the \gls{ann} is also fed to another \gls{ann}, the GAN, which has the task of discriminating whether the input comes from the source domain or from the target domain. 
The two \glspl{ann} are trained together as if they were a single \gls{ann}, even though they have competing goals: the adversarial \gls{ann} aims at minimizing the error in the discrimination between target and source inputs, while the main \gls{ann} aims at minimizing the error on the source task, while at the same time aiming at \emph{maximizing} the error that the adversarial \gls{ann} makes in discriminating between data coming from the source or target domain. If the adversarial \gls{ann} is not able to distinguish between source and target domains, then the first segment of the main \gls{ann} has determined a representation of the source domain that is virtually indistinguishable from the target domain, and thus the main \gls{ann} can be used to execute both the source and target tasks. The contrasting goals during the training process are modeled by defining the overall training cost function as:
\beq\label{Eq:CostGAN}
{\cal L}(\bW,\bb,\bV,\bc)={\cal L}_{m}(\bW,\bb)-\lambda {\cal L}_{a}(\bW,\bb,\bV,\bc)\;,
\eeq
wherein ${\cal L}_{m}$ is the error on the source task, ${\cal L}_{a}$ is the error in discriminating between source and target inputs, $\lambda$ is a factor weighting the relative importance of these  two errors, $\bW$ and $\bb$ are the weights and bias terms of the main network, while $\bV$ and $\bc$ are the weights and bias of the adversarial \gls{ann}, and the overall cost function needs to be minimized with respect to $\bW,\bb$, and maximized with respect to $\bV,\bc$. By minimizing \eqref{Eq:CostGAN} with respect to $\bW,\bb$, the primary \gls{ann} minimizes ${\cal L}_{m}$ while at the same time maximizing ${\cal L}_{a}$. Instead, by maximizing \eqref{Eq:CostGAN} with respect to $\bV$ and $\bc$ the adversarial network is minimizing ${\cal L}_{a}$. As a result, unlike typical training procedures that aim at minimizing the training cost function, the goal here is to determine a saddle point of \eqref{Eq:CostGAN}, which can be accomplished by several saddle-point algorithms based again on stochastic gradient descent techniques, as in regular training procedures \cite{GaninAdversarial,CaoTL}. 
It is to be stressed that, in order to find a saddle point of \eqref{Eq:CostGAN}, it is not required to know the desired output for each training sample. Indeed, each training sample must simply carry a label discriminating whether the sample comes from the source or target domain, but the desired output is required only if the sample comes from the source domain. This means that adversarial training can be used for \gls{ann} training even when the available target data is not labeled. 

\section{Applications to wireless communications}\label{Sec:Applications}
After presenting the main concepts and tools of the deep learning framework, this section describes practical applications to the design of wireless communication systems. First, a literature survey is performed, reviewing available contributions about the application of deep learning to wireless communication systems, and then several novel applications are presented.

\subsection{State-of-the Art Review}\label{Sec:SOA}
The application of deep learning to the design of the physical layer of wireless communication networks has started attracting research attention only very recently, mostly in the last couple of years. For this reason, fewer contributions have appeared than in other areas of wireless communications. Nevertheless, two main research directions can be identified:
\begin{itemize}
\item deep learning to operate the physical layer, simplifying the execution of tasks such as data detection, decoding, channel estimation, localization, etc. 
\item deep learning to manage the physical layer, simplifying radio resource allocation tasks. 
\end{itemize}

\subsubsection{Operation of the physical layer}
The first area of application of deep learning at the physical layer of wireless networks has been the use of \gls{ann} to simplify the implementation of detection and/or estimation operations such as information decoding, channel estimation, localization, etc. \cite{OShea17,Kim2018,Samuel2018,Xue2018,Jin2018,Felix2018,Dorner2018,Oshea2018,OShea2018a,Ye2018a,Aoudia2018,Raj2018,Farsad2017,Qian2018,Neumann17,Vieira2017,Navabi2018,Ding2018,Decurninge2018,Schibisch2018,Koller2018,Ma2018,Xu2018,Wang2017,Qin2018,Javid2018,Ye2018}.

In \cite{OShea17}, the authors use deep FFNs to emulate the transmitter and the receiver of point-to-point communication systems, while assuming the communication channel is known. The end-to-end system is modeled as a deep \gls{ann} composed of the cascade of an \gls{ann} implementing the data transmission process, one layer implementing the known channel (whose parameters are fixed and not trainable), and another \gls{ann} implementing the reception process. The overall network receives as input the information signal and provides as output the corresponding symbol estimate. This architecture is referred to as an \emph{auto-encoder}, since the goal of the network is to reproduce the input data at the output. It is shown that, without having any information about the implementation of the transmitter/receiver chains, the auto-encoder is able to outperform traditional approaches that design the system based on (approximate) mathematical models of the transmitter/receiver chains. The work in \cite{OShea17} paved the way for many subsequent studies that exploited \glspl{ann} at the physical layer of wireless devices. In \cite{Kim2018} it is proposed to use an auto-encoder to jointly minimize the system bit error rate and peak to average power ratio, and again an improvement over traditional methods is obtained. Deep learning is used for data detection in MIMO systems in \cite{Samuel2018,Xue2018}, in decode-and-forward relay channels \cite{Jin2018}, and for equalization and synchronization in OFDM systems in \cite{Felix2018}. 

In all of these works, perfect knowledge about the communication channel is assumed. Several subsequent works have tried to relax this assumption. In \cite{Dorner2018} a two-stage approach is taken. At first, a synthetic channel model is used to provide a first training of the \gls{ann}. Next, this initial training is refined at the receiver based on the true channel characteristics. GANs are used in \cite{Oshea2018,OShea2018a,Ye2018a}, by exploiting a surrogate channel for training purposes. A combination of supervised training and reinforcement learning is used in \cite{Aoudia2018} to remove the need of channel knowledge. In \cite{Raj2018}, the auto-encoder approach is further extended to the case in which no channel state information is available by exploiting a stochastic perturbation approach. A similar scenario is considered in \cite{Farsad2017}, where the auto-encoder approach is used for data detection without any channel knowledge, considering molecular communications as a main application scenario. The use of fully connected \glspl{ann} for molecular communications is also investigated in \cite{Qian2018}. 

In \cite{Neumann17} it is shown that a deep neural network can reliably learn the MMSE channel estimator, while in \cite{Vieira2017} convolutional neural networks are successfully used to implement a fingerprinting-based scheme for user localization. Channel estimation through neural networks is successfully demonstrated in \cite{Navabi2018} and also in \cite{Ding2018}, where an FDD massive MIMO system is considered, and the channels are assumed to be representable by a finite-size dictionary. Experiments showing the performance of deep learning methods for users localization in outdoor environments are provided in \cite{Decurninge2018}, showing that even simple \glspl{ann} architectures can achieve satisfactory performance. In \cite{Schibisch2018} it is shown that deep learning can be successfully used to implement error correction tasks, while \cite{Koller2018} shows that machine learning is able to provide reliable channel estimation from compressed measurements. Channel estimation in rapidly time-varying environments is discussed in \cite{Ma2018}, and it is shown that deep architectures are able to cope with this more challenging setup, while \cite{Xu2018} proposes a deep learning approach for joint equalization and decoding in wireless networks. Surveys on the use of \glspl{ann} to implement encoding/decoding operation as well as channel estimation tasks with limited side information have appeared in \cite{Wang2017,Qin2018}. An information-theoretic study of the mutual information between input and output of a shallow neural network is provided in \cite{Javid2018}. Channel estimation and signal detection are also performed through deep learning in \cite{Ye2018}, showing that similar performance as traditional methods can be achieved, but with a much lower computational complexity. 

\subsubsection{Management of the physical layer}
A second emerging application area is the use of deep learning to perform radio resource allocation at the physical layer, with minimum complexity and/or side-information requirements \cite{Li2017,Calabrese2017,Fang18,Chen2017b,Sun2017,ZapSPAWC2018,ZapASILOMAR2018a,ZapASILOMAR2018b,Nasir2018,Liang2018,Kerret2018,ZapCOMMAG2018}. 

The works \cite{Li2017} and \cite{Calabrese2017} put forward the idea of using \glspl{ann} for network resource management, providing an overview of potential applications of \gls{ai} for network resource management in future 5G wireless networks, and discussing supervised, unsupervised, and reinforcement learning. In \cite{Sun2017}, a fully connected FNN is used for sum-rate maximization in interference-limited networks, by learning the input-output map of each iteration of the iterative weighted MMSE power control algorithm \cite{LuoWMMSE}. The proposed approach is able to mimic the performance of the weighted MMSE resource allocation algorithm, while at the same time significantly reducing the computational complexity. In \cite{BhoEE2019,ZapCOMMAG2018}, the problem of energy efficiency maximization in wireless interference networks by a fully connected FNN is tackled. Unlike \cite{Sun2017}, in \cite{BhoEE2019,ZapCOMMAG2018} the FNN is directly trained based on the optimal energy-efficient power allocation, which can be computed offline using the novel global optimization procedure also proposed in \cite{BhoEE2019}. The results indicate that the optimal performance can be approached with limited online complexity, thus enabling an online implementation. A similar approach is proposed in \cite{ZapASILOMAR2018a,ZapASILOMAR2018b} for power control and user-cell association in massive MIMO multi-cell systems. Instead, a different approach is taken in \cite{Liang2018}, where a fully connected \gls{ann} is trained to solve the sum-rate maximization problem subject to maximum power and minimum rate constraints. In order to reduce the complexity of building the training set, the authors propose to train the \gls{ann} using directly the system sum-rate as training cost function. The results show a gain compared with previous low-complexity optimization methods. 

In \cite{Chen2017b} a cloud-RAN system with caching capabilities is considered. Echo-state neural networks, an instance of \glspl{rnn}, are used to enable base stations to predict the content request distribution and mobility pattern of each user, thus determining the best content to cache. It is shown that the use of deep learning increases the network sum effective capacity of around 30\% compared with baseline approaches based on random caching. In \cite{Fang18}, deep reinforcement learning is used to develop a power control algorithm for a cognitive radio system in which a primary and secondary user share the spectrum. It is shown that both users can meet their QoS requirements despite the fact that the secondary user has no information about the primary user's transmit power. The use of deep reinforcement learning is also considered in \cite{Nasir2018}, where it is used to develop a power control algorithm for weighted sum-rate maximization in interference channels subject to maximum power constraints. The proposed algorithm exhibits fast convergence and satisfactory performance. A decentralized robust precoding scheme in a network MIMO system is developed in \cite{Kerret2018} by \glspl{ann}. In \cite{SharmaDRL2019}, online power allocation policies for a large and distributed system with energy-harvesting nodes are developed by merging deep reinforcement learning and mean field games. It is shown that the proposed method outperforms all other available online policies and suffers a limited gap compared to the use of non-causal offline policies. 

\subsection{Learning to optimize}
The rest of this section describes several applications, primarily focusing on the most recent area of \gls{ann}-based physical layer resource allocation. In this context, a promising approach is to develop methodologies to embed prior available (expert) knowledge about the problem to solve into deep learning, rather than using only empirical data. The motivation for this approach lies in the consideration that purely data-driven approaches may become too complex for large-scale applications, due to the large amount of required data, and to the related processing complexity. Expert-knowledge-aided deep learning is an emerging topic even in fields of science where data-driven deep learning techniques are a consolidated reality. In \cite{Inoue2017}, image processing for object position detection in robotics applications is considered, and it is observed that augmenting a small training set of real images with a large dataset of synthetic images significantly improves the estimation accuracy with respect to processing only the small dataset of real images. Similar results have been obtained in \cite{Kim_AcousticDNN} with reference to speech recognition applications. 

In the context of wireless communications, leveraging data-driven techniques based on deep learning, with expert knowledge coming from (even approximate) theoretical models holds an even greater potential. Indeed, despite their possible inaccuracy or cumbersomeness, theoretical wireless models provide important prior information compared to what is available in other fields of science. In our opinion, this clear advantage of wireless communications should not be wasted. More specifically, when performing resource allocation, depending on the system complexity, one is faced with one of the four cases shown in Tab. \ref{Fig:TabCases}:

\begin{table}[!h]
  \includegraphics[scale=0.5]{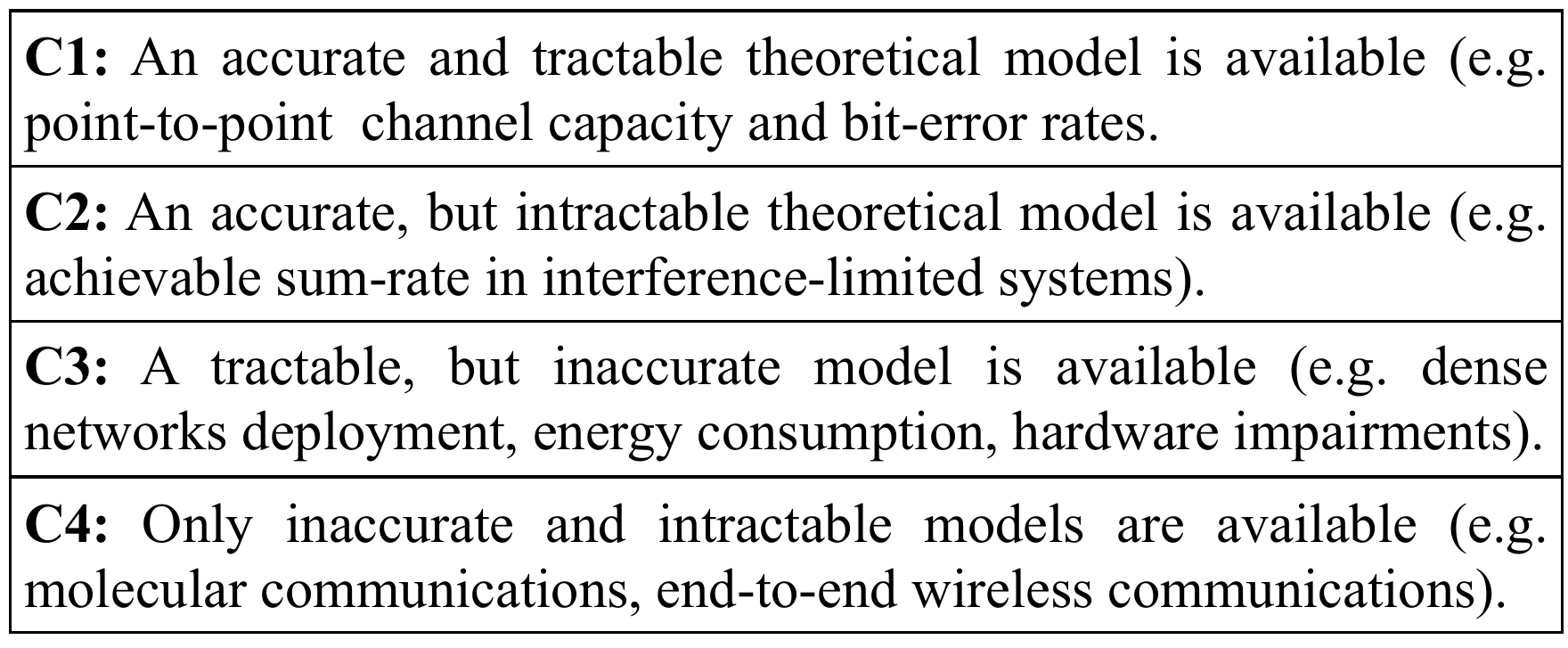} 
  \caption{Scenarios for resource management in wireless networks}
  \label{Fig:TabCases}
\end{table}

While, it is clear that C.1 and C.4 should be handled by traditional model-based approaches, and fully data-driven techniques, respectively, the most appropriate way of tackling C.2 and C.3 is an open issue. Indeed, C.2 and C.3 offer the possibility of a cross-fertilization between model-aided and data-driven approaches, due to the fact that a model is available, even though it is inaccurate or cumbersome to optimize. Moreover, C.2 and C.3 are the typical situations in wireless communications, where models and optimization algorithms are usually available, despite being the result of some approximations and simplifications. 

\begin{figure}[!h]
  \includegraphics[scale=0.35]{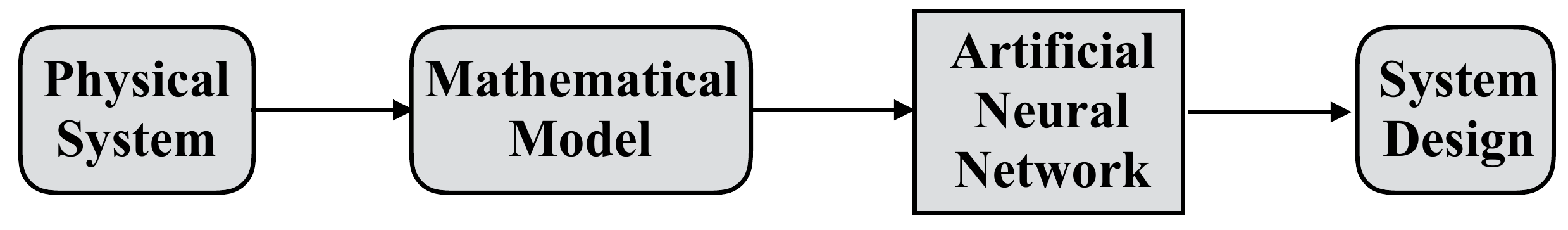} 
  \caption{\textbf{Optimizing a model.} An \gls{ann} is trained based on data generated from the theoretical models. No measurement campaign is needed.}
  \label{Fig:CaseC2}
\end{figure}

\begin{figure}[!h]
  \includegraphics[scale=0.28]{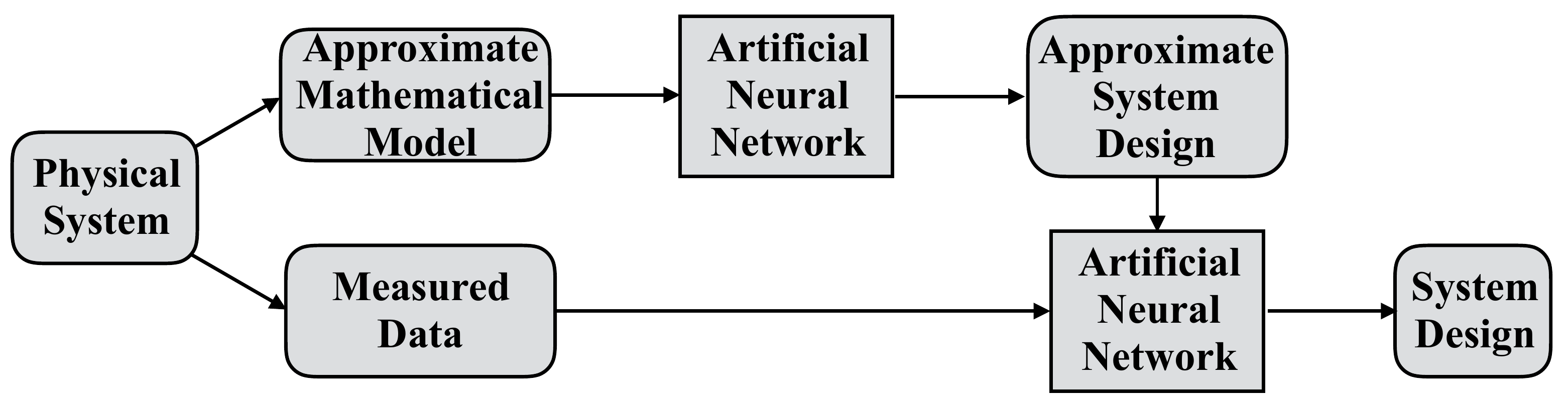} 
  \caption{\textbf{Refining a model.} An \gls{ann} is first trained based on data generated from the approximate theoretical models, and then refined based on a small dataset of measured data.}
  \label{Fig:CaseC3}
\end{figure}

In order to tackle C.2 and C.3, we propose the following two methodological approaches:
\begin{itemize}
\item \textbf{Optimizing a model.} In Case C.2, an analytical expression of the performance metric to optimize is available. Then, an \gls{ann} can be trained to learn the map between the system parameters and the corresponding optimal resource allocation, following the technique anticipated in Section \ref{Sec:DLforRM}. This approach is depicted in Fig. \ref{Fig:CaseC2}.
\item \textbf{Refining a model.} In Case C.3, a two-step approach can be exploited. In the first step, an \gls{ann} is trained based on synthetic data generated from the approximate model. Next, a second training phase based on true, measured data can be used to refine the \gls{ann} configuration. This approach is depicted in Fig. \ref{Fig:CaseC3}.
\end{itemize}
As it will become clear from the applications illustrated in the sequel, the main advantages of the proposed approaches are:
\begin{itemize}
\item The significant complexity reduction compared to purely model-based methods, thus enabling real-time resource allocation with near-optimal performance. 
\item The significant reduction of the amount of empirical data compared to purely data-driven methods, thus   dispensing with expensive and unpractical measurement campaigns. 
\end{itemize}
With the exception of one case-study related to the auto-encoder approach, all applications described in the following address resource allocation problems by using one of the two methodologies described above. 

\subsubsection{Physical layer design: Optimizing the receiver of a molecular communication system}\label{Sec:Molecular}
In this section, we consider the typical case study of optimizing the receiver of a communication system. As an example, we focus our attention on a molecular communication system, where chemical signals instead of electromagnetic signals are used to convey information \cite{Farsad2016}. The motivation of this choice is the complexity of modeling molecular communication systems, and the possibility of leveraging data-driven methods in this context \cite{farsad2018sliding}. A similar approach can be used to design and optimize the receivers of different communication systems. The objective is to prove that, by assuming that the system model is accurate, model-based and data-driven methods yield the same optimal receiver designs if they are both appropriately designed.

As a practical case study, we consider a molecular communication system where diffusion is employed for allowing information particles to propagate from a transmitter to a receiver. Due to the intrinsic characteristics of diffusion, the resulting transmission channel is usually affected by a non-negligible Inter-Symbol Interference (ISI), which, if not taken into account for system optimization, may severely degrade the system performance. For this reason, we focus our attention on optimizing the receiver operation in the presence of ISI. In particular, we consider a threshold-based demodulator and denote by $\tau$ the demodulation threshold. Let $\bar{s}_i$ be the estimate of symbol ${s}_i$ at time-slot $i$, a threshold-based demodulator operates as follows:
\begin{eqnarray}
   \bar{s}_i = \left\{ \begin{array}{cl}
                0, & r_i\leq \tau   \\
                1, & r_i > \tau   \\
                \end{array}\right.
\end{eqnarray}
where $r_i$ is the number of molecular received at time-slot $i$.

Under the typical operating conditions discussed in detail in \cite{MDR_Molecular} for a binary modulation scheme, the error probability as a function of $\tau$ can be formulated as follows:
\begin{eqnarray} \label{Eq:MolError}
   P_e(\tau) = \frac{1}{2^{L}}\sum\limits_{\bm{s}_{i-1}}P_e(\bm{s}_{i-1},\tau)
\end{eqnarray}
where:
\begin{equation}
   \begin{split}
   P_e(\bm{s}_{i-1},\tau) &= \frac{1}{2} [ Q(\lambda_0 T  + \sum\limits_{j=1}^L s_{i-j}C_j,\lceil\tau\rceil)\\
   & + 1 - Q(\lambda_0 T  + \sum\limits_{j=1}^L s_{i-j}C_j+C_0,\lceil\tau\rceil) ]
   \end{split}
\end{equation}
and $Q(\lambda,n)= \sum\nolimits_{k=n}^\infty \frac{e^{-\lambda}\lambda^k}{k!} $ is the incomplete Gamma function, $L$ is the memory of the chemical channel, i.e., the length of the ISI, $ \lambda_0 $ is the background noise power per unit time, $T$ is the duration of the time-slot, and $C_j$ is the average number of received information particles at the $j$th time-slot.

\begin{figure}[!t]
\centering\includegraphics[width=\columnwidth]{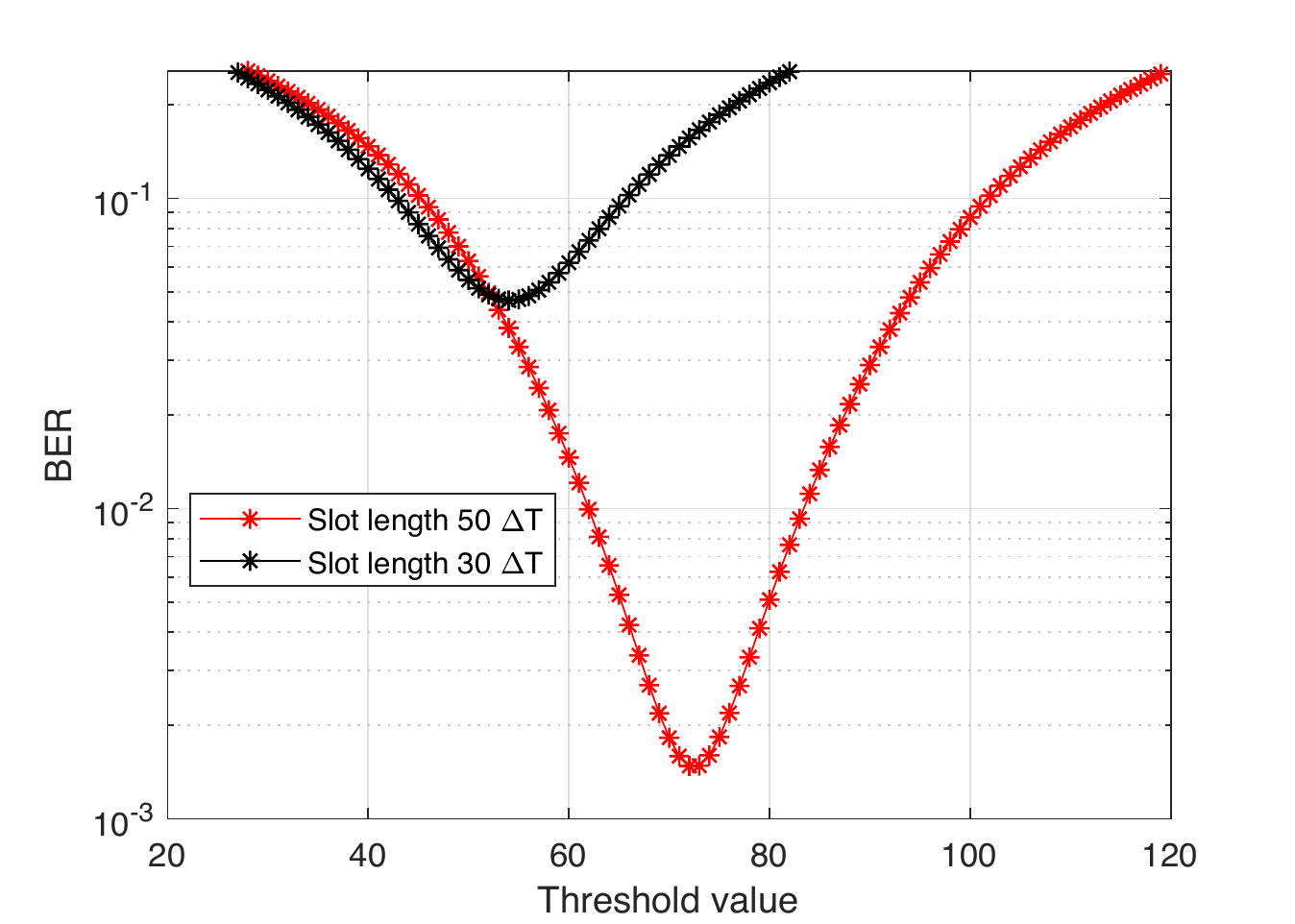}
\caption{Error probability as a function of $\tau$ (the signal-to-noise-ratio is equal to 30 dB) for two different durations of the time-slot (amount of ISI).}
\label{fig:diff_tau}
\end{figure}
In order to obtain appropriate performance and, thus, reduce the error probability, the detection threshold, $\tau$, needs to be appropriately chosen and optimized. In Fig. \ref{fig:diff_tau}, we depict the error probability as a function of $\tau$ for a typical system setup. We observe that an optimal value of $\tau$ exists that minimizes the error probability and that depends on the time slot duration $T$, i.e., the amount of ISI for a given channel.

In mathematical terms, the optimal threshold that minimizes the error probability can be formulated as follows:
\begin{eqnarray}
  \tau^*=\mathop{\arg\min}_{\tau}\quad P_e(\tau)
  \label{equ:optimal_threshold}
\end{eqnarray}

Due to the analytical complexity of \eqref{Eq:MolError}, it is not possible to compute $\tau^*$ explicitly, but it can be obtained numerically at an affordable complexity.

An alternative approach is to employ a data-driven approach that does not rely on any model but uses only empirical data, e.g., a large set of values for $r_j$. More precisely, we consider an \gls{ann} whose aim is to demodulate the transmitted data by minimizing the error probability. An \gls{ann}-based demodulator is a system whose input is the number of received information particles, $ {r}_i $ at the $i$th time-slot, and the outputs are the probabilities that the transmitted bit is 0 or 1, i.e., $ {P}_i(s_i=0|r_i) $ and $ {P}_i(s_i=1|r_i) $, respectively. Since, $ {P}_i(s_i=1|r_i)+{P}_i(s_i=0|r_i)=1 $, only one of the two probabilities is needed. We use the notation ${P}_i = {P}_i(s_i=1|r_i) $. Based on the outputs, the ANN demodulate the received bits as follows:
\begin{eqnarray}
   \bar{s}_i = \left\{ \begin{array}{cl}
                0, & P_i\leq 0.5   \\
                1, & P_i > 0.5   \\
                \end{array}\right.
\end{eqnarray}
where the threshold 0.5 accounts for the fact that the bits are equiprobable.

In order to train the \gls{ann}, we consider a supervised learning approach, i.e., we compute the parameters (e.g., the bias factors and the weights) of the \gls{ann} by using a known sequence of transmitted bits. In particular, we use the Bayesian regularization back propagation technique, which updates the weights and biases by using the Levenberg-Marquardt optimization algorithm. The set of parameters to train and operate the \gls{ann} are as follows: The number of layers is 10, the learning rate is 0.01, the training epoch is 200, the number of validation bits is 100000, and the replication time is 50. In particular, the training is performed in a batch mode, and the replication time denotes the number of batches each of which is 1000-bit long.

\begin{figure}[!t]
\centering\includegraphics[width=\columnwidth]{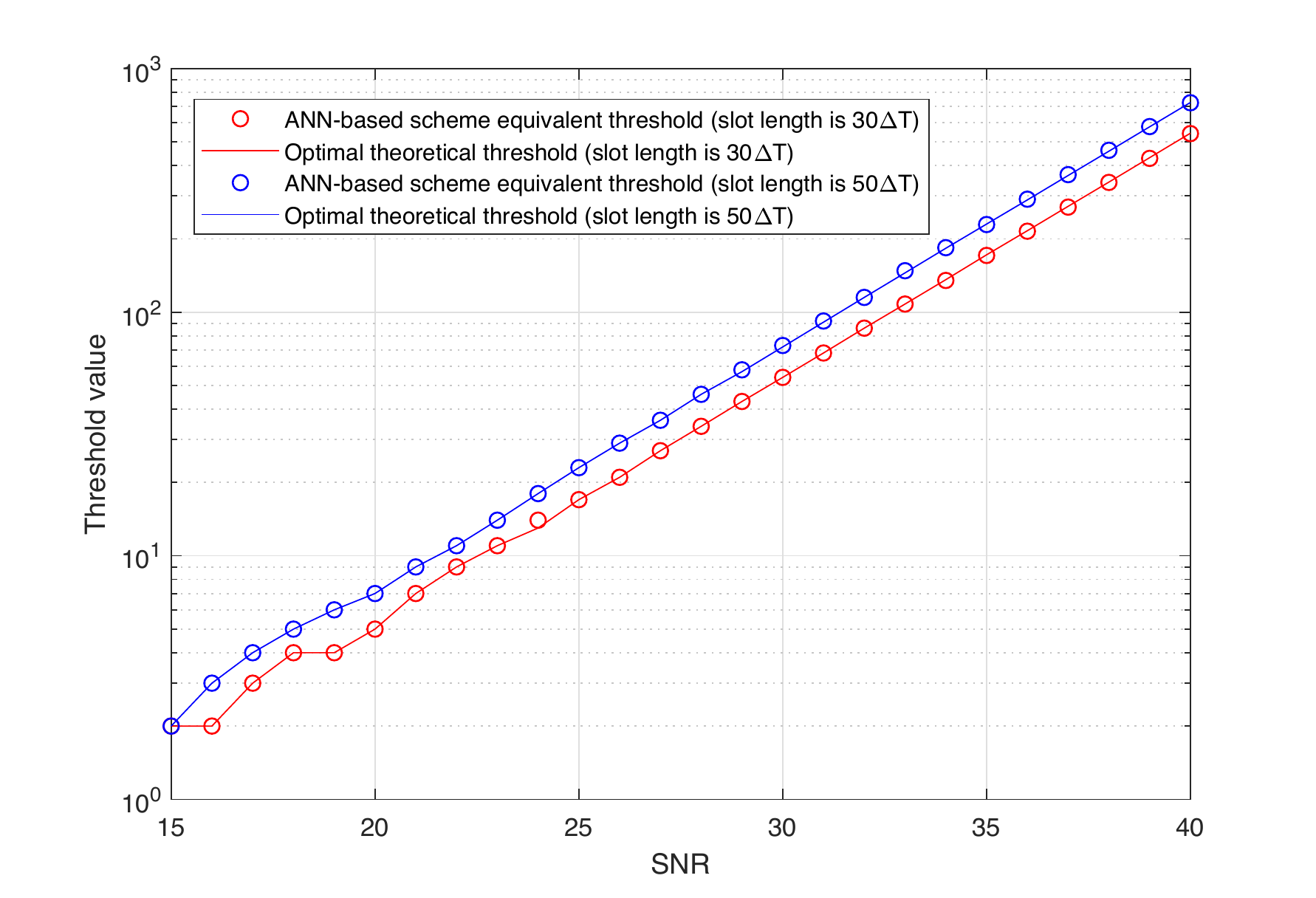}
\caption{Demodulation thresholds: Model-based vs. data-driven for two different durations of the time-slot (amount of ISI).}
\label{Optimal_threshold_vs_ANN_threshold}
\end{figure}
In Fig. \ref{Optimal_threshold_vs_ANN_threshold}, we compare the optimal threshold computed numerically from \eqref{equ:optimal_threshold} as a function of the signal-to-noise-ratio, and the demodulation threshold that is learnt by the \gls{ann}-based demodulator. In the latter case, the threshold is obtained, after completing the training of the \gls{ann}, and identifying the input, i.e., the number of information particles, for which the output probability is equal to 0.5. We observe that the \gls{ann}-based implementation is capable of learning the demodulation threshold in a very accurate manner.

\begin{figure}[!t]
\centering\includegraphics[width=\columnwidth]{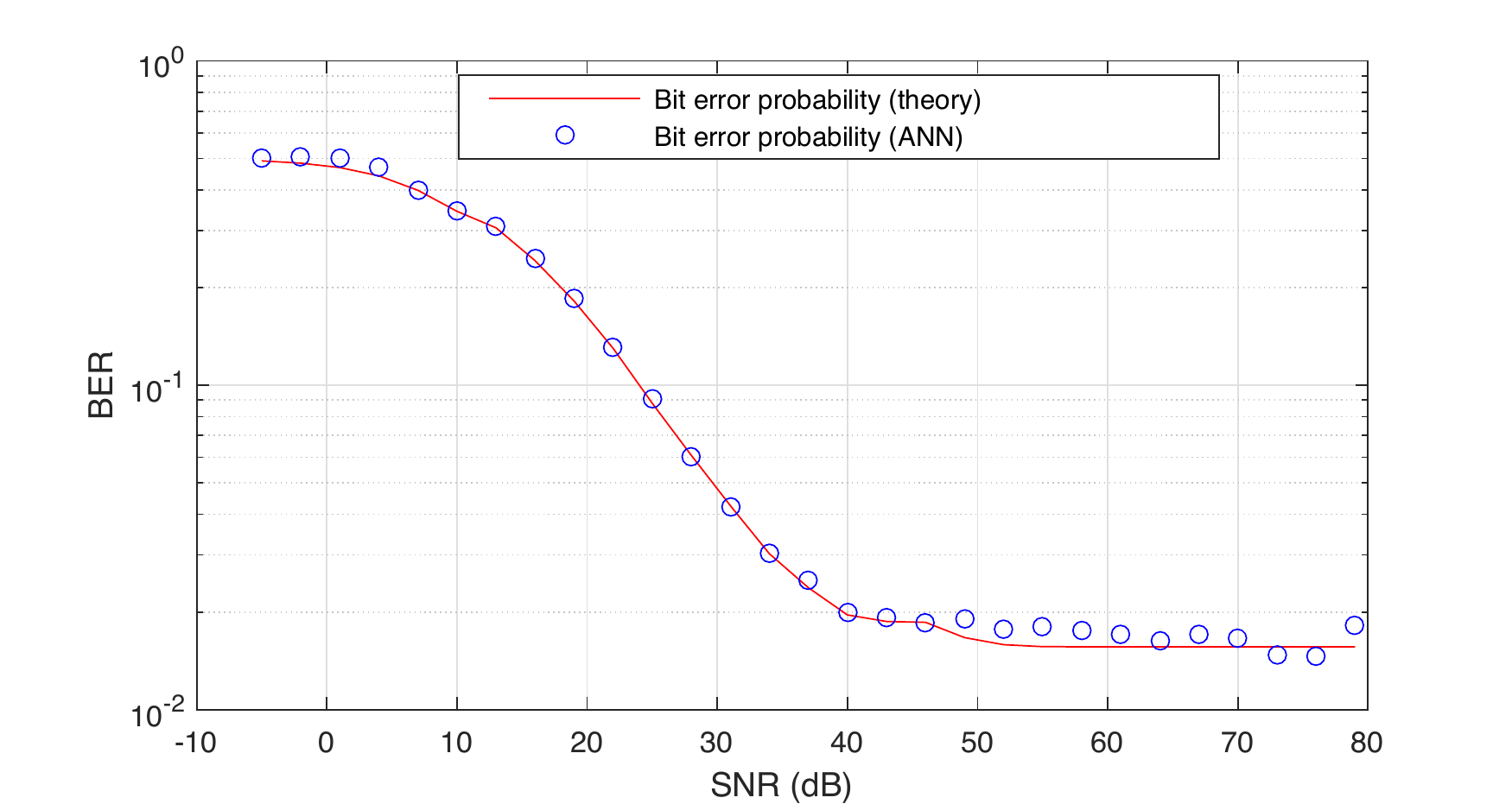}
\caption{Bit error probability of the \gls{ann}-based demodulator vs. the analytical framework - $T = 30 \Delta T $.}
\label{MolCom__BER_ErrorFloor}
\end{figure}

\begin{figure}[!t]
\centering\includegraphics[width=9cm]{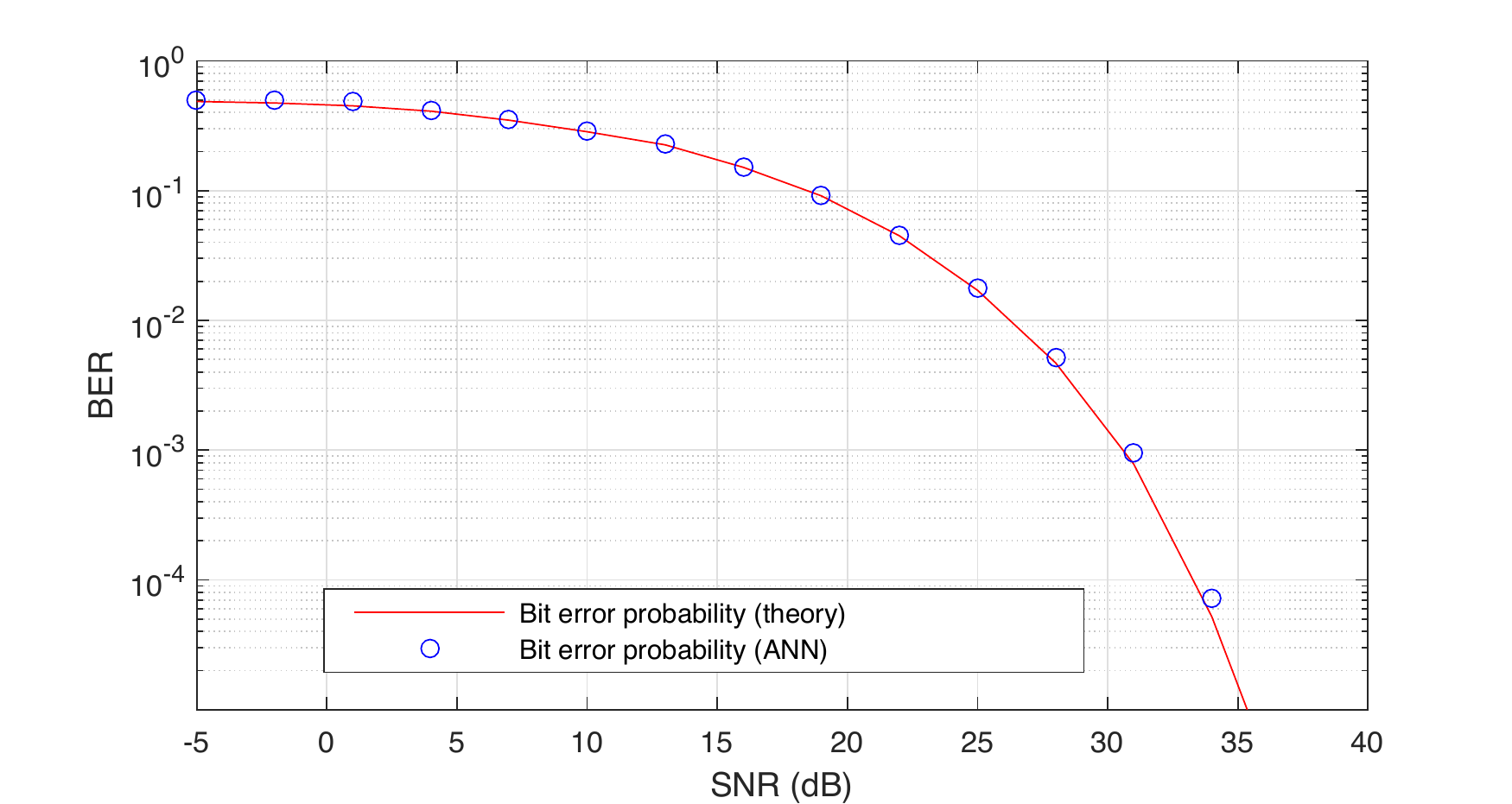}
\caption{Bit error probability of the \gls{ann}-based demodulator vs. the analytical framework - $T = 50 \Delta T $.}
\label{MolCom__BER_NoErrorFloor}
\end{figure}

In Fig. \ref{MolCom__BER_ErrorFloor} and Fig. \ref{MolCom__BER_NoErrorFloor}, we compare the bit error probability of the \gls{ann}-based demodulator against the bit error probability in \eqref{Eq:MolError} by considering a short symbol time (small ISI) and a long symbol time (large ISI), respectively. As for the analytical model, the optimal threshold is estimated from \eqref{equ:optimal_threshold} for each value of the signal-to-noise-ratio. We note a very good agreement even with only 10 layers.

In summary, this section shows that an optimal receiver design can be obtained by relying solely on data-driven methods and that the resulting \gls{ann} can be used for system optimization, e.g., to optimize the demodulation threshold.

\subsubsection{Optimizing a model: power control in wireless networks}\label{Sec:WSEE}
This application focuses on the maximization of the bit-per-Joule energy efficiency in interference-limited networks. The importance of the energy efficiency as a key performance metric in communication systems has emerged recently, motivated by the need to  provide 1000x higher data rates compared to present systems, while at the same time halving the energy consumption. Already 5G wireless networks are requested to increase the bit-per-Joule energy efficiency  by a factor 2000 compared to previous wireless networks \cite{5GNGMN,GEJSAC16}. 

Traditional approaches for energy efficiency maximization in wireless networks are based on the theory of fractional programming, the branch of optimization theory that focuses on the optimization of fractional functions. A tutorial on fractional programming methods for energy efficiency maximization in wireless networks is available in \cite{ZapNow15}. Therein, it is observed that achieving the global maximum of the energy efficiency metric requires exponential complexity whenever the communication system is interference-limited. Here, we will show how the global maximum of the energy efficiency can be approached with limited complexity by using \glspl{ann}.

To elaborate, let us consider an interference-limited network in which $K$ single-antenna transmitters communicate with $M$ receivers, each equipped with $N$ antennas. Denote by $\bh_{k,m}$ the $N\times 1$ channel from transmitter $k$ to receiver $m$, by $p_{k}$ the transmit power of transmitter $k$, by $\bc_{k}$ the $N\times 1$ receive vector used by the receiver associated to transmitter $k$, and by $\sigma_{m}^{2}$ the received noise power at receiver $m$. Then, the \gls{sinr} enjoyed by transmitter $k$ at its associated receiver $m_{k}$ is expressed as:
\beq\label{Eq:SINR}
\gamma_{k}=\frac{p_{k}|\bc_{k}^{H}\bh_{k,m_{k}}|^{2}}{\sigma^{2}+\sum_{j\neq k}p_{j}|\bc_{k}^{H}\bh_{j,m_{k}}|^{2}}=\frac{p_{k}d_{k,k}}{\sigma^{2}+\sum_{j\neq k}p_{j}d_{k,j}}\;,
\eeq
with $d_{k,j}=|\bc_{k}^{H}\bh_{j,m_{k}}|^{2}$, for all $k$ and $j$. 

Based on \eqref{Eq:SINR}, the network weighted sum energy efficiency (WSEE) is given by 
\begin{align}\label{Eq:WSEE}
\text{WSEE}&=\sum_{k=1}^{K}w_{k}\frac{B\log_{2}(1+\gamma_{k})}{P_{c,k}+\mu_{k}p_{k}}\quad[\text{bit/Joule}]\;,
\end{align}
wherein $B$ is the communication bandwidth, $P_{c,k}$ is the hardware static power consumed to operate the $k$-th communication link, $\mu_{k}$ the inverse of the power amplifier efficiency of transmitter $k$, and $w_{k}$ is a non-negative weight modeling the priority given to the energy efficiency of user $k$. It is important to stress that $P_{c,k}$ depends on system parameters such as the number of antennas and the efficiency of the system hardware components, but it is assumed not to depend on the transmit powers, and therefore the specific model expressing $P_{c,k}$ as a function of the system hardware components is inessential as far as maximizing \eqref{Eq:WSEE} as a function of the transmit powers is concerned.

Thus, the power control problem is stated as the maximization of the \gls{wsee} subject to power constraints, namely
\begin{subequations}\label{Eq:WSEEMax}
\begin{align}
&\ds\max_{\{p_{k}\}_{k=1}^{K}}\;\text{WSEE}(p_{1},\ldots,p_{K})\label{Eq:aWSEEMax}\\
&\;\text{s.t.}\;P_{min,k}\leq p_{k}\leq P_{max,k}\;,\forall\; k=1,\ldots,K\label{Eq:bWSEEMax}
\end{align}
\end{subequations}
with $P_{max,k}$ and $P_{min,k}$ being the maximum feasible and minimum acceptable transmit powers  for user $k$. The challenge in tackling \eqref{Eq:WSEEMax} lies both in the fact that the numerators of \eqref{Eq:aWSEEMax} are not concave functions of $\bp=\{p_{k}\}_{k=1}^{K}$ due to the presence of  multi-user interference, and to the sum-of-ratios functional form, which is regarded as the hardest fractional problem to tackle. Therefore, showing that an \gls{ann} can be used to solve \eqref{Eq:WSEEMax} makes a very strong case towards the development of \gls{ann}-based solutions of generic energy-efficient resource allocation problems. To solve \eqref{Eq:WSEEMax}, global optimization methods are required to find the optimal power allocation, while more practical approaches guarantee only first-order optimality with a polynomial complexity. Moreover, Problem \eqref{Eq:WSEEMax} needs to be solved anew whenever the channel realizations $\{\bh_{\ell,m_{k}}\}_{k,\ell}$ change. This represents a critical drawback, especially considering that the resource allocation process must be completed well before the end of the channel coherence time in order for the optimized power vector to be practically useful. This observation makes it difficult to employ even polynomial-complexity algorithms to perform resource allocation in real-time, i.e. following the small-scale variations of the channel coefficients. 

In oder to address this issue and enable real-time resource allocation, it is possible to resort to deep \glspl{ann} paired with the use of energy efficiency models and traditional optimization approaches. Specifically, this case study is an instance of C.2 of Table \ref{Fig:TabCases}, since a model is available and has allowed us to formulate Problem \eqref{Eq:WSEEMax}. However, the model is too complex (for practical implementations) to be optimized by directly using traditional optimization methods. The idea is, therefore, to exploit the model by using it to train an \gls{ann} in order to learn the map between the system parameters, and the corresponding optimal power allocation. To elaborate, let us observe that Problem \eqref{Eq:WSEEMax} can be regarded as an unknown function mapping from the coefficients $\{d_{k,\ell}\}_{k,\ell}$ and the maximum/minimum transmit powers $P_{max}$ and $P_{min}$, to the optimal power allocation vector $\bp^{*}$, namely
\beq\label{Eq:Fpower}
{\cal F}:\bd=\{d_{k,\ell},P_{min,k},P_{max,k}\}_{k,\ell}\in\mathbb{R}^{K(M+2)}\to \bp^{*}\in\mathbb{R}^{K}
\eeq
Since \glspl{ann} are universal function approximators, it is possible to train an \gls{ann} so that its input-output relationship reproduces the unknown map \eqref{Eq:Fpower}. This leads to considering an \gls{ann} with $K(M+2)$ input nodes and $K$ output nodes, to be trained so that it outputs the optimal $K\times 1$ power vector $\bp^{*}$ corresponding to a given $K(M+2)\times 1$ input of system parameters $\bd$. This enables to update the resource allocation without having to solve any optimization problem every time that the system parameters change, but by simply feeding the new vector $\bd$ to the \gls{ann}, and obtaining the corresponding power allocation as the output of the \gls{ann}. 

It is important to emphasize that this entails a negligible computational complexity compared to using sophisticated numerical optimization algorithms. Indeed, once all the parameters and hyperparameters of the \gls{ann} are fixed, the \gls{ann} provides a closed-form expression of its input-output relationship, whose complexity amounts to computing $\sum_{\ell=1}^{L+1}N_{\ell-1}N_{\ell}$ real multiplications\footnote{The complexity related to additions is negligible compared to that related to multiplications} and evaluating $\sum_{\ell=1}^{L+1}N_{\ell}$ activation functions, with $N_{\ell}$ denoting the number of neurons in Layer $\ell$ in accordance with the notation of Section \ref{Sec:FNN}. 

Instead, a higher complexity is required to generate a suitable training set, because this requires to consider many different system parameters realizations $\{\bd_{nt}\}_{nt=1}^{N_{T}}$, and to compute the corresponding desired power allocation vector $\{\bp_{nt}^{*}\}_{nt=1}^{N_{T}}$ by solving \eqref{Eq:WSEEMax} $N_{T}$ times. At a first sight, this might seem to result in a complexity overhead that defeats the purpose of using \glspl{ann} to reduce the computational complexity of resource allocation problems. However, this is not the case for at least two major reasons that make the generation of the training set fundamentally different from solving Problem \eqref{Eq:WSEEMax} in real-time:
\begin{itemize}
\item The training set can be generated and used \emph{offline} to train the \gls{ann}. Thus, a higher complexity can be afforded and real-time constraints do not apply.
\item The training set needs to be updated at a much longer time-scale than that with which the network parameters change.
\end{itemize}
In other words, the training process needs not be executed each time a system parameter changes, and the solution needs not be obtained within the channel coherence time. Thus, the use of traditional optimization theory to generate the training set does not defeat the practicality of the proposed \gls{ann}-based approach. On the contrary, the use of mathematical models to formulate the optimization problem and the use of traditional optimization techniques to build the training set, represent the expert knowledge that is exploited to facilitate the use of \glspl{ann} for real-time power control in wireless networks. In addition, we mention that recently a more efficient branch-and-bound solution to globally solve energy-efficient problems has been proposed in \cite{BhoEE2019}, which further facilitates the global solution of Problem \eqref{Eq:WSEEMax}.


\textbf{Numerical performance analysis.}
Consider the uplink of a wireless interference network with $K=4$ single-antenna \cglspl{ue} placed in a square area with edge $2\,\textrm{km}$ and communicating with 4 access points placed at coordinates $(0.5, 0.5)\,\textrm{km}$, $(0.5, 1.5)\,\textrm{km}$, $(1.5, 0.5)\,\textrm{km}$, $(1.5, 1.5)\,\textrm{km}$, and equipped with $n_{R}=2$ antennas each. The path-loss is modeled following \cite{PathLossModel}, with carrier frequency $1.8\,\textrm{GHz}$ and power decay factor equal to 4.5, while fast fading terms are modeled as realizations of zero-mean, unit-variance circularly symmetric complex Gaussian random variables. Moreover, $P_{c,k} =1\,\textrm{W}$ and $\mu_{k} = 4$ for all $k=1,\ldots,K$, respectively, while the noise power at each receiver is $\sigma^{2}=F{\cal N}_{0}B$, with $F=3\,\textrm{dB}$ the receiver noise figure, $B=180\,\textrm{kHz}$ the communication bandwidth, and ${\cal N}_{0}=-174\textrm{dBm/Hz}$ the noise spectral density. All users have the same maximum transmit powers $P_{max,1}= \ldots= P_{max,K}=P_{\text{max}}$, while $P_{min,k}=0$ for all $k=1,\ldots,K$. 

The proposed \gls{ann}-based solution of Problem \eqref{Eq:WSEEMax} is implemented through a feedforward \gls{ann} with $L+1$ fully-connected layers, with the $L=5$ hidden layers having 128, 64, 32, 16, 8 neurons, respectively. The training set has been generated by solving Problem \eqref{Eq:WSEEMax} for different realizations of the vector $\bd$. When doing this, due to numerical reasons, the parameter  vectors $\bd$ and the optimal output powers in the training set have been expressed in logarithmic units rather than in a linear scale. On the other hand, the use of logarithms may cause numerical problems when the optimal transmit powers are very close to zero. For this reason, logarithmic values approaching $-\infty$ have been clipped at $-M$ for $M > 0$. In our experiments, $M = 20$ worked well.\footnote{Note that, although using a logarithmic scale, the transmit powers are not expressed in dBW, since the logarithmic values are not multiplied by 10. Thus $-M=-20$, corresponds to $-200\,\textrm{dBW}$.} Summarizing, the considered normalized training set is
\begin{equation*}
	{\mathcal S}_{T} = \{ (\log_{10} \bd_n, \max\{-20, \log_{10} \tilde{\bp}_n^{*}\}) \,|\, n = 1, \dots, N_T \},
\end{equation*}
where all functions are applied element-wise to the vectors in the training set. 

The activation functions have been set as follows. The first hidden layer has an ELU activation, the other hidden layers alternate ReLU and ELU activation functions, while the output layer uses a linear activation function. The use of a linear activation in the output layer is motivated by the consideration that it allows the \gls{ann} to produce low training error as a result of a proper configuration of the hidden layers, instead of artificially reducing the output error thanks to the use of cut-off levels in the activation function. In other words, a linear output activation function allows the \gls{ann} to learn whether the present configuration of weights and biases is truly leading to a small output error.

The \cgls{ann} is implemented in Keras~2.2.4 \cite{keras} with TensorFlow~1.12.0\cite{tensorflow} as backend, using Glorot uniform initialization \cite{Glorot2010}, the Adam training algorithm with Nesterov momentum, and the mean squared error as the loss function. The training is obtained by solving Problem \eqref{Eq:WSEEMax} for 102,000 \cgls{iid} realizations of \cglspl{ue}' positions and propagation channels, and different values of $P_{max}$. In each scenario, the \glspl{ue} are associated to the access point towards which they enjoy the strongest effective channel. A validation and a test set of 10,200 and 510,000 samples, respectively, were also generated following a similar procedure.

\begin{figure}[!h]
  \includegraphics[scale=1]{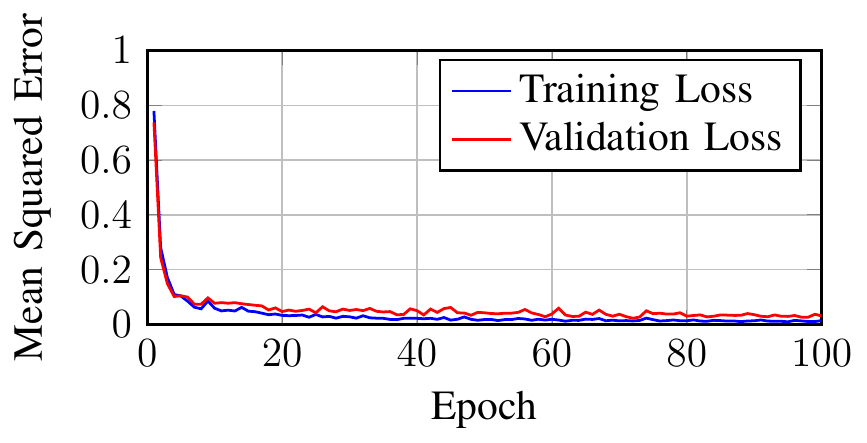} 
  \caption{Training and validation losses versus training epoch number. It is seen that after the training phase, the \gls{ann} neither underfits nor overfits.}
  \label{Fig:LearningCurve}
\end{figure}

Considering training, validation, and test sets, 622,200 data samples were generated, which required solving the NP-hard Problem \eqref{Eq:WSEEMax} 622,200 times. This has been accomplished by the newly proposed branch-and-bound method developed in \cite{BhoEE2019}, which required 8.4 CPU hours to solve all 622,200 instances of the WSEE maximization problem, on Intel Haswell nodes with Xeon E5-2680 v3 CPUs running at $2.50\textrm{GHz}$. This strongly supports the argument that the offline generation of a suitable training set for \gls{ann}-based power control is quite affordable. Finally, all performance results reported in the sequel have been obtained by averaging over 10 realizations of the network obtained by training the \cgls{ann} \emph{on the same training set} with different initialization of the underlying random number generator.\footnote{Note that this is not equivalent to \emph{model ensembling} \cite[Sect.~7.3.3]{Chollet2017} or \emph{bagging} \cite[Sect.~7.1]{Bengio2016}.}
The average training and validation losses for the final \cgls{ann} are shown in Figure \ref{Fig:LearningCurve}. It can be observed that both errors quickly decrease and approach a very small value, thus showing that the adopted \gls{ann} configuration is able to properly fit the training data, without underfitting or overfitting. 

\begin{figure}[!h]
  \includegraphics[scale=1]{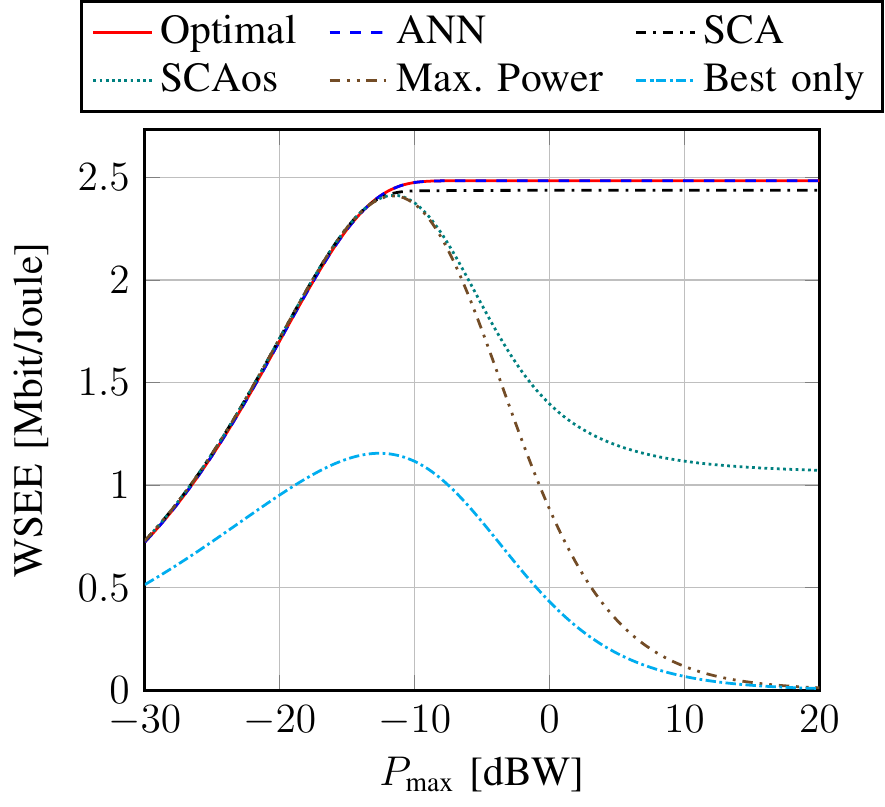} 
  \caption{WSEE performance of the proposed ANN-based method compared to the global optimum and to several state-of-the-art algorithms.}
  \label{Fig:TestPerformance}
\end{figure}

Next, we present the performance of the proposed method over the test set. Specifically, we have compared the proposed \gls{ann}-based method with the following benchmarks:
\begin{itemize}
\item \textbf{SCAos:} A first-order optimal method from \cite{BhoEE2019} that leverages sequential convex approximation methods. For each value of $P_{\text{max}}$, the algorithm initializes the transmit power to $p_{i}=P_{\text{max}}$, for all $k=1,\ldots,K$. 
\item \textbf{SCA:} Again the first-order optimal method based on sequential convex approximation developed in \cite{BhoEE2019}, but with a double-initialization approach. Specifically, at $P_{\text{max}} = -30\,\textrm{dBW}$ maximum power initialization is used. However, for all values of $P_{\text{max}} > -30\,\textrm{dBW}$, the algorithm is run twice, first with the maximum power initialization, and then  initializing the transmit powers with the optimal solution obtained for the previous $P_{\text{max}}$ value. Then, the power allocation achieving the better \gls{wsee} value is retained.
\item \textbf{Max. Power:} All \cglspl{ue} transmit at maximum power, i.e. $p_{k}=P_{\text{max}}$, for all $k=1,\ldots,K$. This strategy is known to perform well in interference networks for low $P_{\text{max}}$ values.
\item \textbf{Best only:} Only one \cgls{ue} is allowed to transmit, specifically that with the best effective channel. This approach is motivated for high $P_{\text{max}}$ values, as a naive way of nulling out multi-user interference.
\end{itemize}
The results are shown in Figure \ref{Fig:TestPerformance} and indicate that the \gls{ann}-based approach outperforms all other practical approaches. The only benchmark that performs comparably with the \gls{ann}-based approach is the SCA algorithm with the more sophisticated initialization rule, which requires to solve the \gls{wsee} maximization problem twice and for the complete range of $P_{\text{max}}$ values. Thus, this SCA approach is quite more complex than the \gls{ann}-based method, but, despite this, it performs slightly worse. In conclusion, we can argue that the \cgls{ann} approach strikes a much better complexity-performance trade-off than state-of-the-art approaches, and thus it enables online power allocation in wireless communication networks.

\subsubsection{Optimizing a model: user-cell association in massive MIMO networks}\label{Sec:User-Cell}
This application has a similar flavor as that in Section \ref{Sec:WSEE}, with the difference that instead of allocating the users' transmit powers, the problem consists of deciding the assignment between transmitters and receivers in an interference network. This means that, while the case-study in Section \ref{Sec:WSEE} tackles a continuous resource allocation problem, and thus can be regarded as a regression problem, here the focus is on a discrete resource allocation problem, which can be viewed as a classification problem. To elaborate, consider a massive MIMO multi-cell network with $K$ single-antenna users and $M$ base stations equipped with $N$ antennas each. Also, assume that each user can be associated to only one access point, and that each access point $m$ can serve at most $a_{m}$ users. In this context, the user-cell association sum-rate maximization problem is cast as:
\begin{subequations}\label{Prob:UserBS}
\begin{align}
&\ds\max_{\rhob}\sum_{k=1}^{K}\sum_{m=1}^{M}\rho_{k,m}d_{k,m}\label{Prob:UserBSa}\\
&\;\textrm{s.t.}\;\sum_{m=1}^{M}\rho_{k,m}\leq 1\;,\;\forall\;k=1,\ldots,K\label{Prob:UserBSb}\\
&\;\;\quad\;\sum_{k=1}^{K}\rho_{k,m}\leq a_{m}\;,\;\forall\;m=1,\ldots,M\label{Prob:UserBSc}\\
&\;\;\quad\; \sum_{m=1}^{M}\rho_{k,m}d_{k,m}\geq R_{min,k}\;,\;\forall\;k=1,\ldots,K\label{Prob:UserBSd}\\
&\;\;\quad\; \rho_{k,m}\in\{0,1\}\;,\;\forall\;m=1,\ldots,M\;,\;\forall\;k=1,\ldots,K\;,\label{Prob:UserBSe}
\end{align}
\end{subequations}
wherein $d_{k,m}=\log_{2}(1+\gamma_{k,m})$ is the spectral efficiency enjoyed by transmitter $k$ if associated to receiver $m$, with $\gamma_{k,m}$ the corresponding \gls{sinr} accounting for typical massive MIMO impairments such as pilot contamination and imperfect channel state information, $\rho_{k,m}$ is a binary variable taking value $1$ when transmitter $k$ is served by receiver $m$, $\rhob=\{\rho_{k,m}\}_{k,m}$, and $B$ is the communication bandwidth. Constraints \eqref{Prob:UserBSb} and \eqref{Prob:UserBSc} ensure that each transmitter can be associated to only one receiver and that each receiver can serve at most $a_{m}$ transmitters, while Constraint \eqref{Prob:UserBSd} guarantees minimum \gls{qos} for each transmitter, and Constraint \eqref{Prob:UserBSe} is due to the integrality of the association variables.

Typical approaches to solve linear programs such as \eqref{Prob:UserBS} resort to branch-and-cut techniques, which require solving a series of continuous relaxations of \eqref{Prob:UserBS}. In some special cases, i.e. when $R_{min,k}$ is integer for all $k$, the constraint matrix of Problem \eqref{Prob:UserBS} can be shown to be totally uni-modular, which enables to solve \eqref{Prob:UserBS} through just one continuos relaxation. Nevertheless, this still requires to employ numerical optimization algorithms, whose complexity might still be quite high, especially in large networks. Moreover, as in the power control example of Section \ref{Sec:WSEE}, the optimal association rule needs to be computed in real-time, thus implying that Problem \eqref{Prob:UserBS} needs to be solved anew each time any of the coefficients $\{d_{k,m}\}_{k,m}$ changes. Moreover, in order to be useful, the solution needs to be obtained well before the coefficients $\{d_{k,m}\}_{k,m}$ change again. 

In order to reduce the complexity of the resource allocation process, we observe that the considered problem is again an instance of C.2 in Table \ref{Fig:TabCases}, since a model is available and has allowed us to formulate Problem \eqref{Prob:UserBS}. Then, following a similar approach as in Section \ref{Sec:WSEE}, the optimization program in \eqref{Prob:UserBS} can be seen as the problem of determining the unknown map:
\beq\label{Eq:F}
\hspace{-0.2cm}{\cal F}\!:\!\bd\!=\!\{d_{k,m},\!R_{min,k},\!a_{m}\}_{k,m}\!\in\!\mathbb{R}^{KM+K+M}\!\!\to\!\! \rhob^{*}\!\!\in\!\!\{0,1\}^{KM}\!\!\!\!\!\!\!\!\!\!,
\eeq
which can be tackled by resorting again to a fully-connected FFNs, taking $(KM+K+M)$-dimensional inputs and producing $KM$-dimensional outputs, with similar implementation and complexity considerations as those in Section \ref{Sec:WSEE}.   

\textbf{Numerical performance analysis.}
Consider the uplink of a massive MIMO system wherein $4$ \glspl{bs} are deployed in a square area with edge $1\,\textrm{km}$ at points with coordinates $(250,250)\,\textrm{m}$, $(250,750)\,\textrm{m}$, $(750,250)\,\textrm{m}$, $(750,750)\,\textrm{m}$, serving $40$ users randomly placed in the coverage area. Each \gls{bs} is equipped with $N_{R}=64$ antennas, while all mobile users have a single antenna. A uniform uplink power $p$ of $20\,\textrm{dBm}$ is considered for all users, while a common receive noise power $\sigma^2$ of $-94\,\textrm{dBm}$ is assumed for all \glspl{bs}. The communication bandwidth is $20\,\textrm{MHz}$ and the propagation channels follow the local scattering model \cite{EmilNowPub17}. 

A training set of $N_{T}=155000$ samples has been generated by considering independent realizations of the users' positions in the service area, and solving the corresponding instance of Problem \eqref{Prob:UserBS}, with $a_{m}=15$ for all $m$. Out of these $N_{T}$ samples, 140000 have used as training set, while the remaining 15000 have been used as validation set for hyperparameter tuning. The considered \gls{ann} architecture is composed of $L=3$ fully connected layers with $128$, $64$, $64$ neurons, respectively, plus an output layer with $KM=40$ neurons. Layers $1$ and $3$ have a ReLU activation function, while Layer $2$ and the output layer have a sigmoidal activation function. The Adam  training algorithm with Nesterov's momentum has been employed for training, using the mean squared error as loss function.

The training and validation MSEs are reported in Tab. \ref{Tab:MSE} versus the training epoch number. The result show that the considered \gls{ann} architecture fits well the training data, without underfitting or overfitting. 
\begin{table}[!h]
\caption{Training and validation errors versus training epoch.}
\centering
\label{Tab:MSE}
\begin{tabular}{|c|c|c|}
\hline
 & Training MSE & Validation MSE \\
\hline 
Epoch 1 & $0.0116$ & 0.0113 \\
\hline
Epoch 5 & $0.0100$ & 0.0116 \\
\hline
Epoch 10 & $0.0093$ & 0.0104 \\
\hline
Epoch 15 & $0.0091$ & 0.0096 \\
\hline
Epoch 20 & $0.0090$ & 0.0091 \\
\hline
Epoch 25 & $0.0089$ & 0.0089 \\
\hline
Epoch 30 & $0.0087$ & 0.0092 \\
\hline
Epoch 35 & $0.0085$ & 0.0087\\
\hline
Epoch 40 & $0.0083$ & 0.0089 \\
\hline
Epoch 45 & $0.0082$ & 0.0087 \\
\hline
Epoch 50 & $0.0081$ & 0.0090\\
\hline
\end{tabular}
\end{table}
 
After training and validation, the performance of the resulting ANN has been evaluated over a test set of $15000$ data samples that have been generated independently from the training and validation samples. For each test sample, denoting by ${{\boldsymbol {\rho}}}_{ANN}=\{\rho_{k,m}\}_{k,m}$ the \gls{ann} output, user $k$ has been associated to \gls{bs} $\bar{m}$ if $\bar{m}=\text{arg}\max_{m}\;\rho_{k,m}$, and then the resulting sum-rate performance has been compared to the optimal solution of Problem \eqref{Prob:UserBS}.

\begin{figure}[!h]
\centering
\includegraphics[scale=0.33]{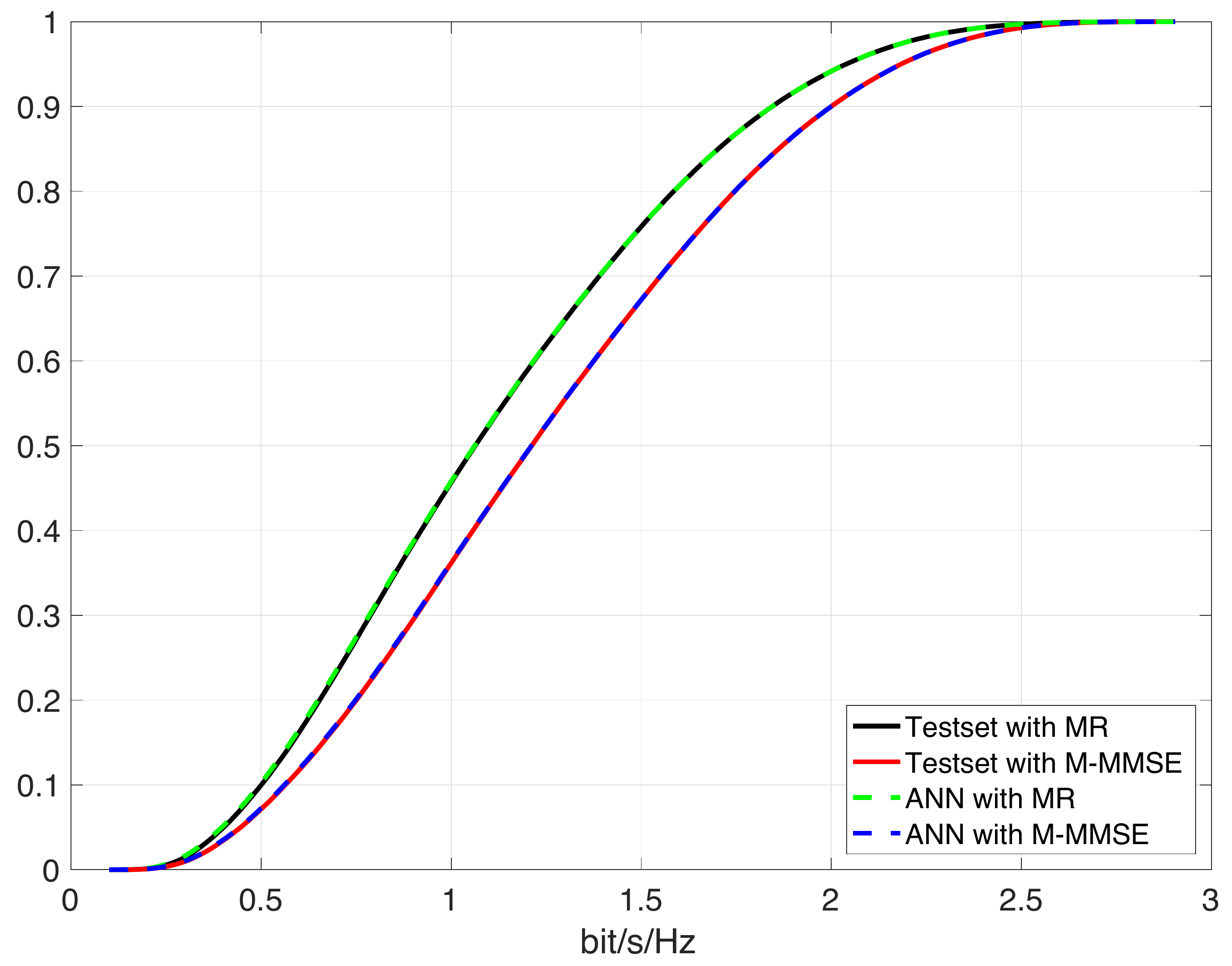}
\caption{CDF of the average users' rate over the testset for the \gls{ann}-based approach and the optimal allocation, with MMSE and MR reception.} \label{Fig:CDF}
\end{figure}

Fig. \ref{Fig:CDF} shows the cumulative distribution function (CDF) of the average users' rate over the test set for the following schemes:
\begin{itemize}
\item \gls{ann}-based association with MMSE reception.
\item Optimal association with MMSE reception.
\item \gls{ann}-based association with MR reception.
\item Optimal association with MR reception.
\end{itemize}
It is seen that in all cases the \gls{ann}-based method performs similarly as the optimal user-cell association, while requiring a much lower computational complexity. Thus, once again, this motivates the use of \gls{ann}-based resource allocation methods. 

\subsubsection{Refining a model by deep transfer learning - Cellular networks beyond the Poisson point process}\label{Sec:NonPPP}
In this section, we consider the case study in which an analytical model exists and is analytically tractable, but it is not considered to be sufficiently accurate for system optimization. We assume, in addition, that more accurate network models are difficult to develop and/or are not suitable for system optimization. As a practical example, we consider the optimization of the Energy Efficiency (EE) \cite{TWC_Paper2018} in non-Poisson cellular networks \cite{124}, which is known to be an intractable optimization problem because of the analytical complexity of the utility function to optimize.

As discussed in Section I-C, we propose to solve this issue by relying on deep transfer learning. Our proposed idea consists of jointly exploiting model-based and data-driven optimization. The approach consists of first optimizing the network using a mismatched, but simpler for optimization, model, and then refining the result with (few) empirical data. Let us assume, as a practical example, that the mismatched (approximated) model is the Poisson model. More precisely, we assume that the only inaccuracy of the system model is the spatial distribution of the cellular base stations, while all the other parameters and modeling assumptions as considered to be accurate. More general system setups can be considered, and another example is studied in the next section. In detail, the approximated model is assumed to be the Poisson point process model, while the \enquote{exact} point process model is assumed to be the square grid model \cite{83}. This is a simple example that is chosen in order to shed light on our proposed approach, and that is also easy to simulate and reproduce.

From \cite{TWC_Paper2018}, we know that the EE in Poisson cellular networks is available in closed-form and is amenable to optimization. Thus, a large dataset of optimal values for the EE as a function of any system parameters can be readily obtained. This dataset is used to train a (mismatched) \gls{ann} with the desired accuracy. The issue, as mentioned, is that the original network model is non-Poisson. We assume, however, that the considered cellular network deployment is equipped with a sensing platform, e.g., by using the meta-surfaces discussed in Section I-C, that can sense and report some contextual data about the network, which is used to obtain a dataset of just a few empirical but optimal values of the EE, which account for the actual non-Poisson spatial model. This dataset is used to tune the \gls{ann} and to correct the mismatch. The intuition behind this proposed approach is that, despite mismatched, the initial \gls{ann} embeds the most important features of the cellular network already, and thus less data is needed compared with the case study in which no pre-training is performed. The objective of this section is to study the amount of empirical samples that the proposed approach based on transfer learning, which jointly combines model and data, requires to achieve similar performance as a pure data-driven method. If the amount of empirical data is not that large, the proposed approach will be successful and will also reduce the amount of overhead, to collect the empirical samples, that is needed for network optimization.

In the rest of this section, we discuss both pure model-based and data-driven approaches, and then combine them together based on transfer learning principles, and, more precisely on network-based transfer learning.

\textbf{Model-based optimization.}
From \cite{TWC_Paper2018}, the EE in Poisson cellular networks can be formulated as follows:
\begin{equation}
\centering
{\rm{EE}}\left( {{\lambda _{{\rm{BS}}}}} \right) = \frac{{{\rm{SE}}\left( {{\lambda _{{\rm{BS}}}}} \right)}}{{{{\rm{P}}_{{\rm{grid}}}}\left( {{\lambda _{{\rm{BS}}}}} \right)}}
\end{equation}
where
\begin{equation} \label{PSE}
\centering
\begin{split}
{\rm{SE}}\left( {{\lambda _{{\rm{BS}}}}} \right) &= {{\rm{B}}_{\rm{W}}}{\log _2}\left( {1 + {\gamma _{\rm{D}}}} \right)\frac{{{\lambda _{{\rm{BS}}}}{\rm{L}}\left( {{{{\lambda _{{\rm{MT}}}}} \mathord{\left/
 {\vphantom {{{\lambda _{{\rm{MT}}}}} {{\lambda _{{\rm{BS}}}}}}} \right.
 \kern-\nulldelimiterspace} {{\lambda _{{\rm{BS}}}}}}} \right)}}{{1 + \Upsilon {\rm{L}}\left( {{{{\lambda _{{\rm{MT}}}}} \mathord{\left/
 {\vphantom {{{\lambda _{{\rm{MT}}}}} {{\lambda _{{\rm{BS}}}}}}} \right.
 \kern-\nulldelimiterspace} {{\lambda _{{\rm{BS}}}}}}} \right)}}\\
 & \times {\rm{Q}}\left( {{\lambda _{{\rm{BS}}}},{{\rm{P}}_{{\rm{tx}}}},{{{\lambda _{{\rm{MT}}}}} \mathord{\left/
 {\vphantom {{{\lambda _{{\rm{MT}}}}} {{\lambda _{{\rm{BS}}}}}}} \right.
 \kern-\nulldelimiterspace} {{\lambda _{{\rm{BS}}}}}}} \right)
\end{split}
\end{equation}
\begin{equation} \label{Pgrid}
\centering
\begin{split}
{{\rm{P}}_{{\rm{grid}}}}\left( {{\lambda _{{\rm{BS}}}}} \right) &= {\lambda _{{\rm{BS}}}}{{\rm{P}}_{{\rm{tx}}}}{\rm{L}}\left( {{{{\lambda _{{\rm{MT}}}}} \mathord{\left/
 {\vphantom {{{\lambda _{{\rm{MT}}}}} {{\lambda _{{\rm{BS}}}}}}} \right.
 \kern-\nulldelimiterspace} {{\lambda _{{\rm{BS}}}}}}} \right)\\
 &+ {\lambda _{{\rm{MT}}}}{{\rm{P}}_{{\rm{circ}}}} + {\lambda _{{\rm{BS}}}}{{\rm{P}}_{{\rm{idle}}}}\left( {1 - {\rm{L}}\left( {{{{\lambda _{{\rm{MT}}}}} \mathord{\left/
 {\vphantom {{{\lambda _{{\rm{MT}}}}} {{\lambda _{{\rm{BS}}}}}}} \right.
 \kern-\nulldelimiterspace} {{\lambda _{{\rm{BS}}}}}}} \right)} \right)
\end{split}
\end{equation}
are the spectral efficiency and the power consumption of the cellular network, respectively.

Equations \eqref{PSE} and \eqref{Pgrid} depend on many parameters, which are all defined in \cite{TWC_Paper2018}. As far as the present paper is concerned, we are interested in four main parameters: ${{\lambda _{{\rm{BS}}}}}$, which is the deployment density of the base stations, ${{{\rm{P}}_{{\rm{tx}}}}}$, which is the transmit power of the base stations, ${{\rm{P}}_{{\rm{circ}}}}$, which is the circuit power consumption of the base stations, and ${{\rm{P}}_{{\rm{idle}}}}$, which is the idle power consumption of the base stations. In this section ${{\rm{P}}_{{\rm{circ}}}}$ and ${{\rm{P}}_{{\rm{idle}}}}$ are assumed to be fixed, and they are further analyzed in the next section. The objective is to identify the optimal deployment density of the base stations, ${{\lambda _{{\rm{BS}}}}}$, given some values of the transmit power ${{{\rm{P}}_{{\rm{tx}}}}}$. In \cite{TWC_Paper2018}, it is proved that this optimization problem has a unique solution, which is formulated as the unique root of a non-linear equation. Therefore, the optimal density of the base stations that maximizes the EE can be computed efficiently, for any given values of the transmit power. By solving this optimization problem, we can easily obtain the optimal pairs $\left( {{{\rm{P}}_{{\rm{tx}}}},\lambda _{{\rm{BS}}}^{\left( {{\rm{opt}}} \right)}} \right)$, where $\lambda _{{\rm{BS}}}^{\left( {{\rm{opt}}} \right)} = \mathop {\arg \max }\nolimits_{{\lambda _{{\rm{BS}}}}} \left\{ {{\rm{EE}}\left( {{\lambda _{{\rm{BS}}}}} \right)} \right\}$. These pairs can then be used to train an \gls{ann}, with ${{{\rm{P}}_{{\rm{tx}}}}}$ as the input, and ${\lambda _{{\rm{BS}}}^{\left( {{\rm{opt}}} \right)}}$ as the output. \\

\textbf{Data-driven optimization.}
Let us assume now that we cannot rely on any analytical models and that the EE needs to be estimated by collecting empirical samples from the cellular network, from which the optimal cellular network deployment needs to be inferred. In particular, the spectral efficiency and the power consumption can be estimated, respectively, as follows:
\begin{equation}
\centering
\begin{split}
{\rm{PSE}}\left(  \bullet  \right) &= \frac{1}{{{\rm{AreaNet}}}}\sum\limits_{{\rm{Cell}}\left( 1 \right) \in {\rm{Net}}} {\sum\limits_{{{\rm{N}}_{{\rm{MT}}}} \in {\rm{Cell}}\left( 1 \right)} {\left\{ {\frac{{}}{{}}} \right.} } \\
& \left. {\frac{{{{\rm{B}}_{\rm{W}}}}}{{{{\rm{N}}_{{\rm{MT}}}}}}{{\log }_2}\left( {1 + {\gamma _{\rm{D}}}} \right){\bf{1}}\left( {{\rm{SIR}} \ge {\gamma _{\rm{D}}},\overline {{\rm{SNR}}}  \ge {\gamma _{\rm{A}}}} \right)} \right\}
\end{split}
\end{equation}
\begin{equation}
\centering
\begin{split}
{{\rm{P}}_{{\rm{grid}}}}\left(  \bullet  \right) &= \frac{1}{{{\rm{AreaNet}}}}\left( {\sum\limits_{{\rm{Cell}}\left( 0 \right) \in {\rm{Net}}} {{{\rm{P}}_{{\rm{idle}}}}} } \right.\\
& \left. { + \sum\limits_{{\rm{Cell}}\left( 1 \right) \in {\rm{Net}}} {\left( {{{\rm{P}}_{{\rm{tx}}}} + {{\rm{P}}_{{\rm{circ}}}}\sum\limits_{{{\rm{N}}_{{\rm{MT}}}} \in {\rm{Cell}}\left( 1 \right)} {{{\rm{N}}_{{\rm{MT}}}}} } \right)} } \right)
\end{split}
\end{equation}

These two formulas can be interpreted as follows. Let us consider the spectral efficiency as an example. Each mobile terminal in the cellular network determines, based on the received signal, whether it is in coverage. This is performed by measuring the average signal-to-noise-ratio during the cell association phase and the signal-to-interference-ratio during data transmission (if the first phase was successful). This condition corresponds to the term ${{\bf{1}}\left( {{\rm{SIR}} \ge {\gamma _{\rm{D}}},\overline {{\rm{SNR}}}  \ge {\gamma _{\rm{A}}}} \right)}$, where ${\bf{1}}\left( \cdot \right)$ is the indicator function. Each mobile terminal, reports whether it is in coverage or not to a network controller (one bit of information). Based on the number of mobile terminals that are in coverage on a given cell (say ${{{\rm{N}}_{{\rm{MT}}}}}$), the corresponding base station equally allocates the available spectrum (say ${{{\rm{B}}_{\rm{W}}}}$) among them, and transmit data with a fixed rate ${\frac{{{{\rm{B}}_{\rm{W}}}}}{{{{\rm{N}}_{{\rm{MT}}}}}}{{\log }_2}\left( {1 + {\gamma _{\rm{D}}}} \right)}$. Based on the information gathered by all the mobile terminals, it is possible to identify the base stations that have at least one mobile terminal in their corresponding cells (say ${{\rm{Cell}}\left( 1 \right)}$) and to compute the number of mobile terminals that lie in each of them for each network realization. The spectral efficiency can then be estimated by summing the rates all of active base stations and by normalizing by the area of the network under analysis. It is worth mentioning that in order to identify, e.g., the optimal deployment density of the base stations, we need to repeat this procedure by considering all possible combinations of base station patterns, given the number of base stations actually deployed. If the optimization variable is the transmit power of the base stations, all possible values of transmit power need to be tested and the value corresponding to the optimal EE needs to be recorded and used to train an \gls{ann}, similar to the approach discussed for model-based optimization. Based on this simple description, we can readily understand that the amount of empirical data that is necessary to train an \gls{ann} may not be negligible, and, in any case, may strongly affect the overhead for network optimization. \\

\textbf{Network-based transfer learning optimization.}
Network-based transfer learning is a solution to overcome the limitations of model-based and data-driven approaches, since it is apparent that both have advantages and limitations. As already mentioned, the idea is to first train and optimize an \gls{ann} by using a model-based approach, and then refine the obtained \gls{ann} by using some empirical data (data-driven approach). Once the first model-based \gls{ann} is obtained, in particular, we consider that its configuration, i.e., the number of layers, neurons, weights, and biases, constitute the initial configuration of the second \gls{ann} that is refined based on empirical data. In our case study, we assume that, during the refinement phase, the number of layers and neurons are not  modified, while the weights and biases are finely-tuned in order to account for the empirical data and to capture those features of the actual network setup that the assumed model, in order to keep its complexity at a low level, is not capable of doing.

\begin{figure}[!t]
\centering
\includegraphics[width=\columnwidth]{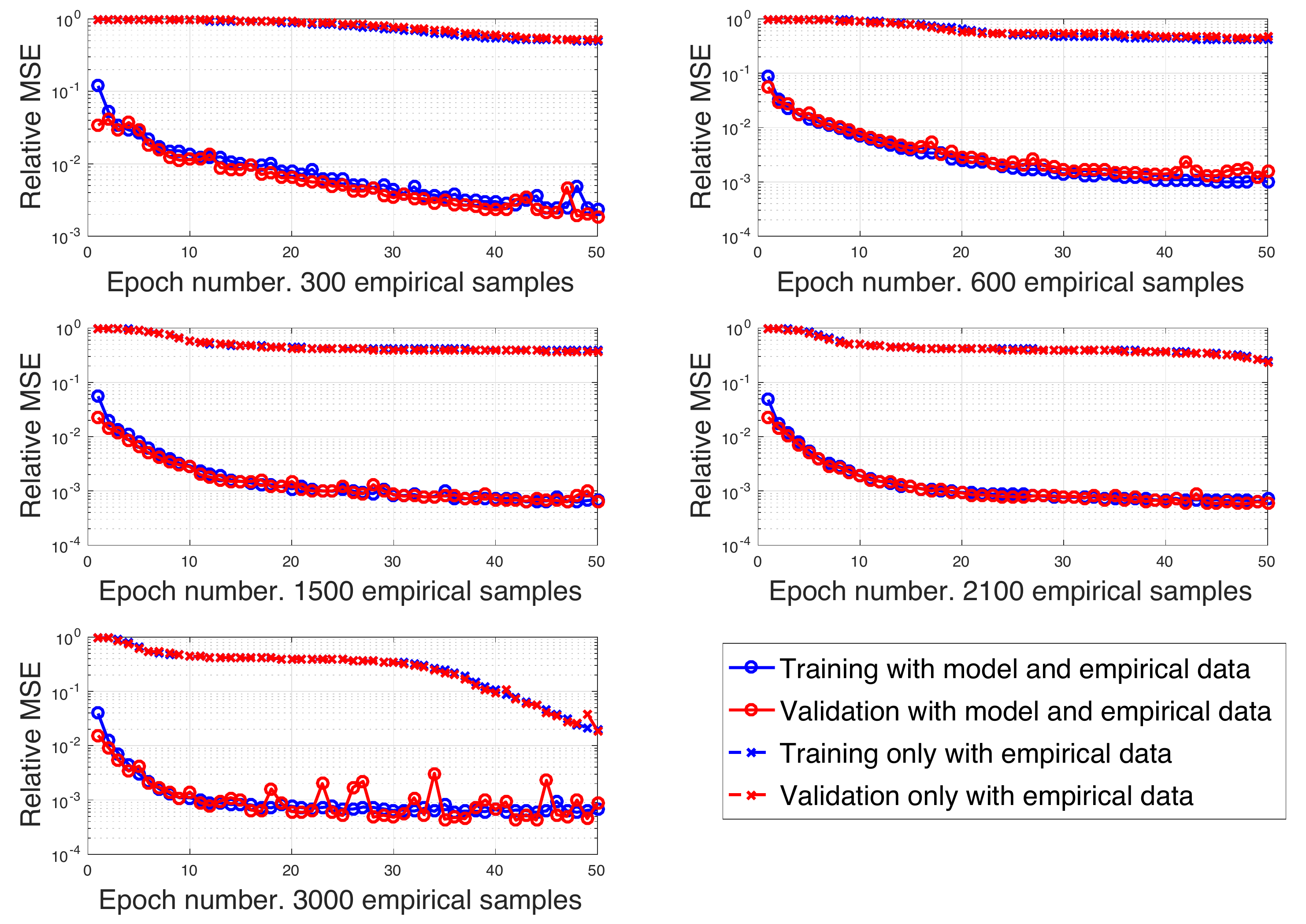}
\caption{Learning and validation relative MSE vs. training epochs for x = 300, 600, 1500, 2100, and 3000 samples. For each case, the performance with and without PPP-based samples is reported. It is seen how the use of PPP-based data significantly improves the performance.}
\label{Fig_nonPPP_Learning}
\end{figure}
\begin{figure}[!t]
\centering
\includegraphics[width=\columnwidth]{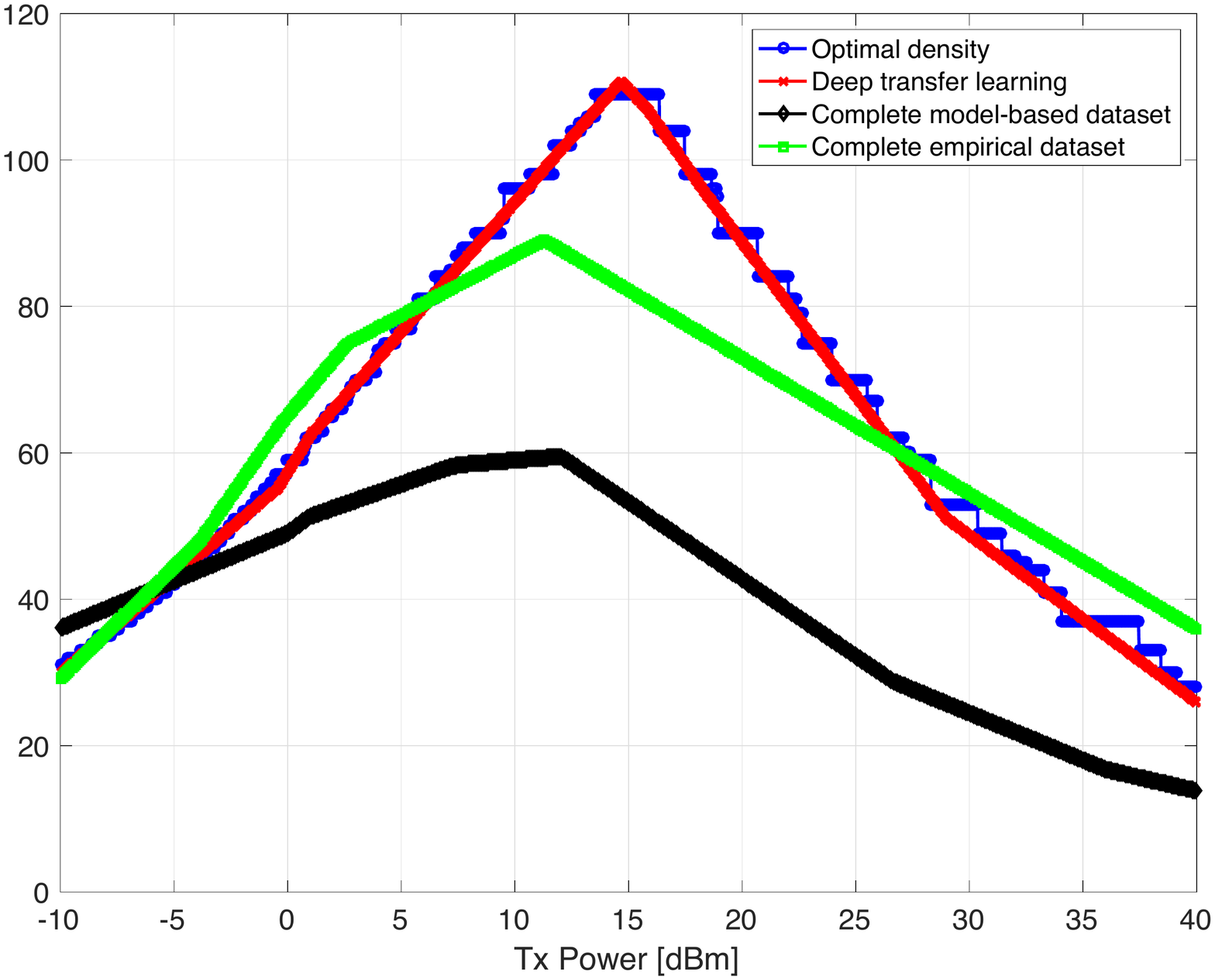}
\caption{Comparison between model-based, data-driven, and deep transfer learning based optimization - Optimal deployment.}
\label{RcellOpt__PPP_vs_nonPPP}
\end{figure}

In Figures \ref{Fig_nonPPP_Learning} and \ref{RcellOpt__PPP_vs_nonPPP}, we illustrate some numerical examples that compare the performance of the three proposed approaches. A feed-forward ANN architecture with fully-connected layers and ReLU activation functions is considered. Specifically, after trying many different \gls{ann} configurations, we found that an \gls{ann} with three hidden layers equipped with 8, 8, and 2 neurons, respectively, yields comparable performance as a much larger \gls{ann} that contains six hidden layers with 64, 32, 16, 8, 4, 2 neurons, respectively. Thus, in all our experiments, we have adopted the 8, 8, 2 \gls{ann} configuration, since it provides the best complexity-performance trade-off among all \gls{ann} architectures we tested. 

Figure \ref{Fig_nonPPP_Learning} shows the training and validation relative MSE versus the number of training epochs for the following approaches:
\begin{itemize}
\item the proposed deep transfer learning technique that employs both model-based and empirical data samples.
\item the baseline approach, where only empirical data samples are used.
\end{itemize} 
As for the first approach, the size of the training set is always set equal to 30,000 samples, out of which x samples follow the true base station distribution (square grid model), and (30,000-x) samples follow the Poisson distribution. As for the second approach, the adopted training set contains only the x empirical samples. Thus, this comparison is fair in terms of number of empirical data samples employed and is aimed at showing the performance that can be obtained by augmenting a small dataset of empirical data with a larger dataset of model-based data. For both approaches, the results for the values x = 300, 600, 1500, 2100, and 3000 are shown, and, for each value of x, it is seen that the proposed deep transfer learning method achieves much lower training and validation errors compared to the baseline approach.

This result is confirmed also in the testing phase. Fig. \ref{RcellOpt__PPP_vs_nonPPP} shows the density of base stations as a function of their transmit power, considering a test set of 8,000 new transmit powers, which were not used during the training phase. Four schemes are compared:
\begin{itemize}
\item the optimal density  computed through exhaustive search
\item the density predicted by means of deep transfer learning, where 3,000 empirical samples are used in the second training step
\item the density obtained without transfer learning and performing the training by using only  3,000 empirical samples
\item the density obtained without transfer learning and performing the training by using only  30,000 model-based samples
\end{itemize}
Notably, we observe that using only the 3,000 empirical samples yields inaccurate estimates of the optimal deployment density of the base stations. Instead, combing model-based data with the same 3,000 samples of empirical data provides one with near-optimal performance. This highlights the relevance of performing the model-based pre-training before employing actual measurements for system optimization, while overcoming their inherent limitations. Moreover, it is interesting to observe that using only the 30,000 model-based samples does not lead to satisfactory performance, thus showing that it is necessary to merge model-based and empirical samples to obtain accurate performance.

In summary, based on the results reported in Figs.  \ref{Fig_nonPPP_Learning} and \ref{RcellOpt__PPP_vs_nonPPP}, we conclude that the proposed approach based on transfer learning constitutes a suitable approach to take the best of both model-based and data-driven methods. 

\subsubsection{Refining a model by deep transfer learning - Cellular networks with inaccurate power consumption models}\label{Sec:PowerMismatch}
In this section, we consider a similar optimization problem as in the previous section. Rather than focusing on the impact of the spatial distribution of the cellular base stations, we focus our attention on the power consumption model of the base stations. More precisely, we assume that the Poisson point process is sufficiently accurate to account for the distribution of the cellular base stations. As far as the power consumption model of the cellular base stations is concerned, on the other hand, we assume a model based on a uniform distribution for ${{{\rm{P}}_{{\rm{circ}}}}}$ and ${{{\rm{P}}_{{\rm{idle}}}}}$, while the empirical model is assumed to be based on the Gaussian distribution. The optimization problem that we are interested in is still concerned with identifying the optimal deployment density of the base stations, but as a function of three variables: ${{{\rm{P}}_{{\rm{tx}}}}}$, ${{{\rm{P}}_{{\rm{circ}}}}}$, and ${{{\rm{P}}_{{\rm{idle}}}}}$. The model-based, the data-driven, and the transfer learning based approach are obtained by using the same approach as the one described in the previous section. As far as the architecture of the \gls{ann} is concerned, on the other hand, we consider a different \gls{ann} architecture, which is made of six layers and four neurons. The adopted \gls{ann} is, therefore, more complicated because three input parameters instead of one are considered in this case study.

\begin{figure}[!t]
\centering
\includegraphics[width=\columnwidth]{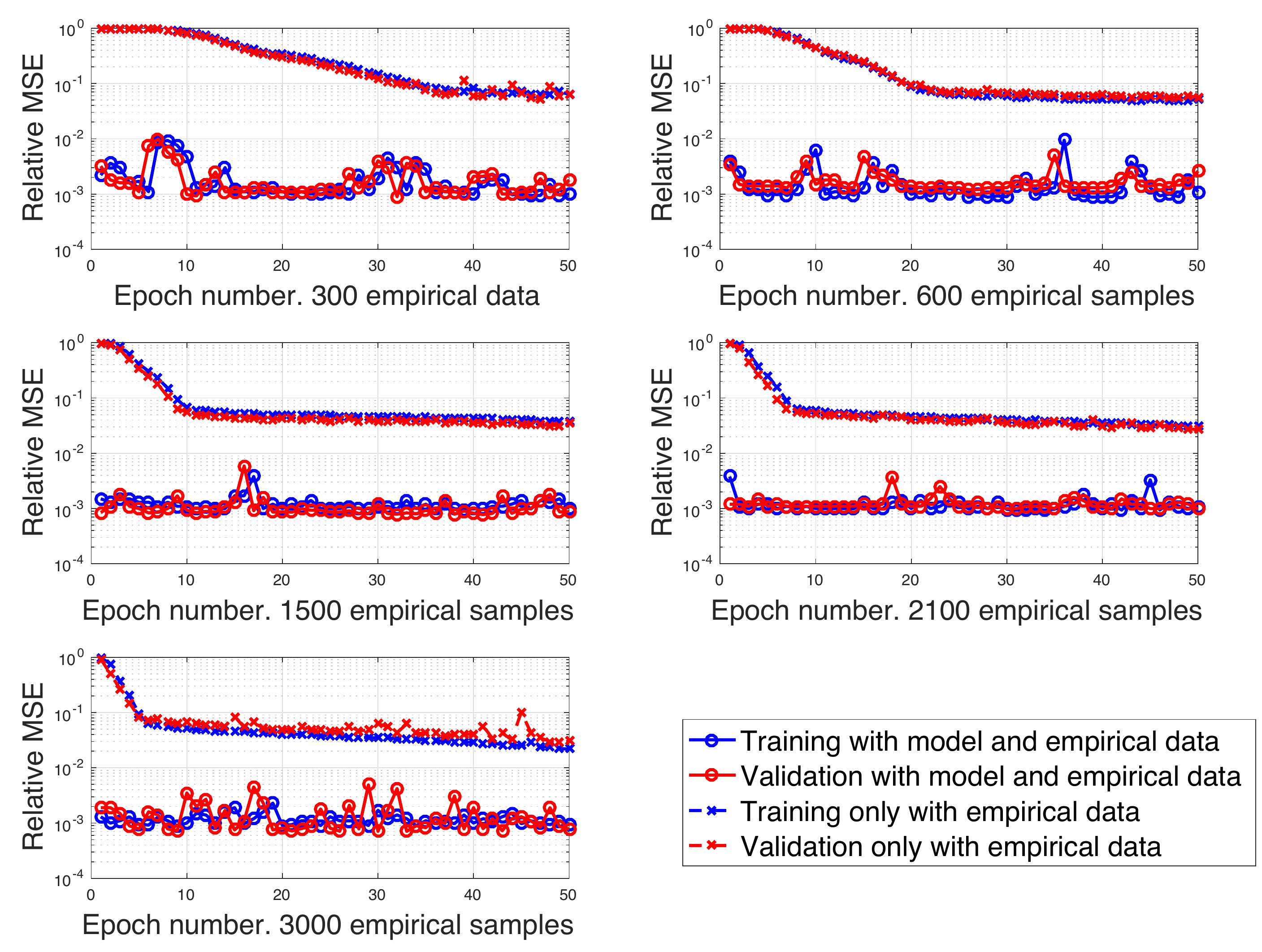}
\caption{Learning and validation relative MSE vs. the training epochs for x=300, 600, 1500, 2100, and 3000. For each case, the performance with and without empirical samples is reported. The use of model-based data significantly improves the performance.}
\label{Fig:Learning_Case2}
\end{figure}


For this scenario, the \gls{ann} configuration with the best complexity-performance trade-off has been found to be one with five hidden layers equipped with 8 neurons each, and ReLU activation functions. Remarkably, the performance granted by this \gls{ann} architecture is slightly worse than that of a much more complex \gls{ann} with 128-64-32-16-8 neurons in the five hidden layers. The training and validation performance of the adopted \gls{ann} are reported in Figure \ref{Fig:Learning_Case2}, and similar considerations as for Figure \ref{Fig_nonPPP_Learning} apply. 
Thus, also in this case the proposed network-based transfer learning approach is a promising alternative to bridge the critical tension between modeling accuracy, optimization complexity, and sensing overhead for network optimization.

\subsubsection{Deep reinforcement learning for power control in energy-harvesting wireless systems}
As a last case-study, we consider the use of deep reinforcement learning, in the context of energy-harvesting communication systems.

Specifically, consider a time-slotted energy-harvesting node transmitting its data over block fading channels to an access point powered by traditional energy sources. Denote by $g_{n}\in {\cal G}$ the fading complex channel gain between the transmitter and the access point in time-slot $n$, by $e_{n}$ the energy harvested during time-slot $n$, which is modeled as a realization of a random variable with unknown distribution, and by $B_{n}$ the energy stored in the transmitter battery at time-slot $n$. The battery is assumed to be perfectly efficient, with maximum capacity $B_{max}$. At the transmitter, only causal information about energy arrivals and communication channels is assumed, i.e. neither the distribution of the energy arrival and channel processes, nor their future realizations are known at each time-slot $n$. Also, denote by $p_{n}\leq P_{\max}$ the transmit power  in the $n$-th time-slot, with $P_{\max}$ the maximum feasible transmit power. 

In this context, the goal is to maximize the system long-term achievable rate, by solving the following problem:
\begin{subequations}
\begin{align}
&\max_{\{p_{n}\}_{n}} \liminf_{N\to \infty}\frac{1}{N}\sum_{n=1}^N\log\left(1+p_{n}\frac{g_{n}}{\sigma^{2}}\right)\\
&\text{s.t.}\;  0\leq  T p_{n}\leq \min\{B_{n},TP_{\max}\}\;, 
\label{eq:optim_prob_const}
\end{align}
\label{eq:optim_prob}
\end{subequations}
wherein $\sigma^{2}$ is the receive noise power, $T$ is the time-slot duration, and the battery state evolves as  
\beq
B_{n+1}=\min\{[B_{n}+e_{n}-Tp_{n}]^+, B_{\max}\}\;.
\label{eq:battery_evol}
\eeq
Constraint \eqref{eq:optim_prob_const} captures the fact that the maximum energy that can be used in time-slot $n$ is limited by the minimum between the amount of energy available in the battery, $B_n$, and the maximum allowed transmit energy $TP_{\max}$. 

Since the information about the random energy arrivals and the channel realizations is only causally available, and the battery evolves in a Markovian fashion, according to \eqref{eq:battery_evol}, Problem \eqref{eq:optim_prob} is a stochastic control problem which could be formulated as a \gls{mdp}, with state space ${\cal S}=\left\{(B,g)\in [0,B_{max}]\times {\cal G}\right\}$, action space ${\cal A}=\left\{p_{n}\in[0,\min\{B_{n},TP_{\max}]\;,\;n=1,\ldots,N\right\}$, and reward at time-slot $n$ given by $R_{n}=\log\left(1+p_{n}\frac{g_{n}}{\sigma^{2}}\right)$. Thus, in principle, upon discretization of the state space, standard \gls{mdp} techniques can be used to solve \eqref{eq:optim_prob}. However, this  poses at least the following three major challenges:
\begin{itemize}
	\item Large feedback overhead, since global information about the battery and channel states of each network node is needed for the operation of the policy.
	\item The solution of the \gls{mdp} requires statistical information about the energy-harvesting process and the wireless channel, which is often difficult to obtain. 
	\item In order to obtain a good solution, a fine discretization step needs to be employed, which results in very large state and action spaces, thus further increasing the problem complexity.
\end{itemize}
These reasons motivate the use of deep reinforcement learning to tackle Problem \eqref{eq:optim_prob}.

\textbf{Numerical performance analysis.} 
Consider an energy-harvesting system in which the transmitter harvests energy according to a non-negative truncated Gaussian distribution\footnote{The energy-harvesting distribution is not assumed known at the design stage.} with mean $m$ and variance $v$. The harvested energy is stored in a battery with capacity $B_{\max}=0.2\,\textrm{J}$ and the maximum feasible transmit power in each time slot is $P_{\max}=0.15\,\textrm{W}$. 

The Deep Q-Network method is implemented by an \gls{ann} with 10 hidden layers equipped with 60, 60, 58, 58,  56, 56, 54, 54, 52, 52 neurons, respectively. The input layer contains $3$ neurons, the output layer contains 150 neurons, which implies that a discretization of the feasible transmit power levels with step $10^{-3}$ has been considered. All hidden layers have ReLU activation functions, while the output layer employs linear activations, motivated by similar considerations as in previous case-studies. The Q-learning algorithm adopts a forgetting factor of $\gamma=0.99$ and the performance of the three following algorithms has been compared:
\begin{itemize}
\item The deep reinforcement learning method that employs the deep Q-Network described above.
\item The solution of the \gls{mdp}. This approach yields, in principle, the optimal online policy, but on the other hand requires a complexity that increases proportionally with the number of considered power levels. For the problem at hand, the complexity of the \gls{mdp} approach becomes unfeasible when the same discretization step of $10^{-3}$ as in the deep reinforcement learning case is used. Therefore, a discretization step of $10^{-2}$ has been used for the \gls{mdp} approach.
\item An offline policy that assumes non-causal knowledge of the channels and energy-harvesting realizations. Clearly, this approach is not practically implementable, and is considered only as a performance upper-bound of any online method.
\end{itemize}

Table~\ref{table_P2P_mean10} shows the performance of the three schemes above, with mean $m=10$ and different values of the variance $v$. The results indicate that the deep Q-Network method is able to achieve  performance very close to that of the offline policy that exploits non-causal information, while outperforming the \gls{mdp}-based solution. It is worth mentioning that the latter result is due to the fact that deep reinforcement learning enables a finer discretization step compared to the \gls{mdp}-based solution, thanks to its much lower computational complexity.

\begin{table}[t!]
	\renewcommand{\arraystretch}{1.5}
	\caption{Performance of deep reinforcement learning online policy for a point-to-point link with $m=10$ in comparison with the \gls{mdp}-based solution and with the offline solution. The deep reinforcement learning uses a discretized action space with step $10^{-3}$, while the \gls{mdp} uses a discretization with step $10^{-2}$, due to its higher complexity.}
	\label{table_P2P_mean10}
	\centering
	\begin{tabular}{|c|c|c|c|c|}
		\hline
		\begin{tabular}[x]{@{}c@{}}Variance\\ (v)\end{tabular} &  \begin{tabular}[x]{@{}c@{}}Offline Policy\\  (nats/s)\end{tabular} & \begin{tabular}[x]{@{}c@{}}DQN Policy\\ (Percentage )\end{tabular} & \begin{tabular}[x]{@{}c@{}}MDP Policy\\ (Percentage )\end{tabular}\\
		\hline
		$1$  &  2.0434 &  95.56\% & 83.32\%  \\ 
		\hline
		$2$  &  2.0375 & 95.24\% & 83.60\%   \\
		\hline
		$3$  &   2.0372 & 98.11\% & 83.32\%  \\
		\hline
		$4$  &  2.0347 & 96.54\% & 83.37\%  \\
		\hline
		$5$  &  2.0310 & 95.28\% & 83.29\%  \\
		\hline
		$6$  &  2.0284 & 98.18\% & 83.21\%  \\
		\hline
	\end{tabular}
\end{table}

\begin{figure}[!t]
\centering
\includegraphics[width=\columnwidth]{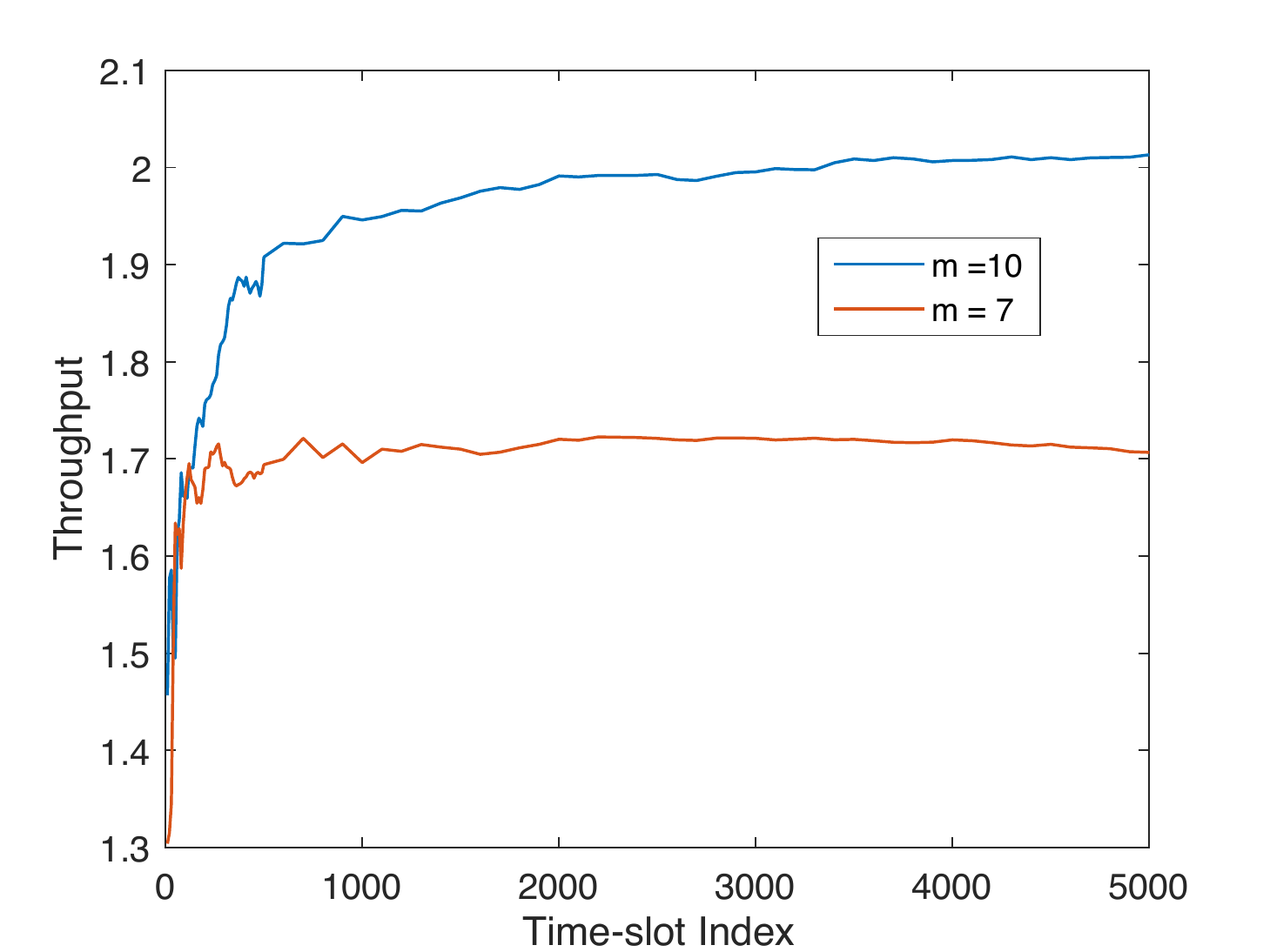}
\caption{Convergence of the deep reinforcement learning algorithm in terms of time-slots.}
\label{Fig:DRL_Convergence}
\end{figure}

Finally, Figure \ref{Fig:DRL_Convergence} shows the convergence of the considered deep reinforcement learning method in terms of number of time slots until the value of the system throughput stabilizes, for $m=7$ and $m=10$. It is seen that a few thousands of time-slots are required to reach convergence.

\section{Conclusions and future research directions}\label{Sec:Conclusions}
The complexity of future wireless communication networks makes deep learning an indispensable design tool. Moreover, recent technological advancements in the area of computer processing units and distributed data storage make the use of deep learning more practical than ever. Nevertheless, research in this field has just started, and a great deal of open problems must be solved before \gls{ann}-based wireless communication networks can be deployed. 

The first challenge to be overcome is represented by the large amount of data that \gls{ann} need in order to ensure satisfactory performance. As remarked in Section \ref{Sec:MLandDL}, deep learning outperforms other machine learning techniques in the large data regime. However, while this might not be an obstacle in other fields of science, the acquisition of large datasets in wireless networks requires measurement campaigns that could be too expensive and/or not practical. In addition, wireless networks are very dynamic, especially in outdoor environments, and it may be difficult to gather new accurate data within the coherence time of the channel of the environment itself \cite{EURASIP_RIS}.

As shown in this work, the most promising approach to overcome this challenge is the joint use of data-driven and model-based approaches. The transfer learning approach developed in Section \ref{Sec:Applications} demonstrates how even approximate mathematical models contain useful prior information that, if successfully transferred into deep learning techniques, can significantly reduce the amount of data required to achieve the desired performance. Nevertheless, this represents only the tip of the iceberg, and many open issues remain to be investigated. As far deep transfer learning is concerned, it is not clear how to set the hyperparameters (e.g. the amount of model-based data, the number of \gls{ann} layers and neurons, etc.) to prevent a negative transfer, i.e. that the \gls{ann} tuned with empirical data provides worse performance than the model-based \gls{ann}. Moreover, other transfer learning techniques remain to be explored, as well as other ways of embedding expert knowledge into \glspl{ann}, based for example on the deep unfolding and deep reinforcement learning methods. As an example, embedding some prior information into a deep reinforcement learning algorithm could potentially speed up its convergence. In addition, a research direction that could provide guidance to achieve a cross-fertilization between mathematical models and deep learning is aimed at deriving a theoretical explanation of how \glspl{ann} work and how to configure them to perform a certain task. Opening the \emph{black box} of \glspl{ann} in order to understand the information-theoretic principles that regulate their behavior is surely a major topic for future investigation. A recent contribution in this direction is \cite{Wolchover17}, which employs the so-called information bottleneck approach. 

The second challenge to be overcome is the integration of \gls{ann} into future wireless network architectures. As motivated in this work, deep learning should be implemented in a distributed fashion. However, this poses several issues that need to be overcome in the next years. Integrating \gls{ai} into distributed wireless networks will not only affect the transmission technologies, but it will also significantly impact the way the network is controlled through feedback signals to avoid instability and malfunctioning. A distributed network in which each node has its own \gls{ann}, that is trained based on a dataset acquired from local measurement and experience, inevitably leads to different nodes having different learning capabilities. Each distributed dataset might differ in size, since different nodes might have different measurement and storage capabilities, as well as quality, since different nodes might experience different data perturbations due to the non-ideality of the measurement sensors. This could potentially lead to instabilities and, in the worst case, cause the wireless network to collapse. Moreover, another issue to be addressed in distributed setups is the possibility for each node to optimize its own performance, rather than the system-wide utility, which might cause a device to learn how to cheat for individual gain. Thus, security mechanisms must be put in place to ensure the correct evolution of a distributed, \gls{ann}-based wireless  communication network.

A third challenge to be overcome is to make deep learning robust against corrupted data. Indeed, due to inevitable errors over feedback channels or in the storage process of data into memory banks, the datasets used to train \glspl{ann} might be corrupted and possibly lead to undesirable training results. Techniques that are able to make the training process robust to these events are warranted, especially in light of the distributed implementation of \gls{ann}-based wireless networks, which makes the overall network highly prone to inconsistencies and failures. 

In conclusion, it is apparent that deep learning is a promising tool to \enquote{make things work}. However, lots of data (for deep learning) or time (for reinforcement learning) is needed to achieve the desired performance. Compared with other fields of research, wireless is unique, since decades of research allowed us to gain deep expert knowledge. This prior information can be used to \enquote{initialize} deep learning, in order to reduce the amount of data, the computational complexity, the energy, and the overhead that are needed to achieve these gains. Communications theory still has a fundamental role in the era of deep learning.

\bibliographystyle{IEEEtran}
\bibliography{DeepLearning,FracProg,MDR}

\end{document}